\pdfoutput=1
\documentclass[cernpreprint]{na61doc}
\usepackage{hyperref}
\usepackage{color}
\usepackage{nicefrac}
\usepackage{multicol}
\usepackage{rotating}
\usepackage{overpic}
\usepackage[capitalize]{cleveref} 
\usepackage{wrapfig}
\usepackage{subfigure}
\usepackage{upgreek}
\usepackage{cite}
\usepackage{rviewport}
\usepackage{diagbox}
\usepackage[titletoc]{appendix}
\usepackage{tikz}
\usetikzlibrary{shapes,arrows}
\usetikzlibrary{matrix,fit}
\usetikzlibrary{arrows,decorations.markings}
\usepackage{relsize}
\usepackage{siunitx}
\usepackage{booktabs}
\usepackage[english]{babel}

\newcommand{\eV}{\ensuremath{\mbox{e\kern-0.1em V}}\xspace}
\newcommand{\GeV}{\ensuremath{\mbox{Ge\kern-0.1em V}}\xspace}
\newcommand{\MeV}{\ensuremath{\mbox{Me\kern-0.1em V}}\xspace}
\newcommand{\GeVc}{\ensuremath{\mbox{Ge\kern-0.1em V}\kern-0.1em/\kern-0.05em c}\xspace}
\newcommand{\GeVcc}{\ensuremath{\mbox{Ge\kern-0.1em V}\kern-0.1em/\kern-0.05em c^2}\xspace}
\newcommand{\AGeV}{\ensuremath{A\,\mbox{Ge\kern-0.1em V}}\xspace}
\newcommand{\AGeVc}{\ensuremath{A\,\mbox{Ge\kern-0.1em V}\kern-0.1em/\kern-0.05em c}\xspace}
\newcommand{\MeVc}{\ensuremath{\mbox{Me\kern-0.1em V}\kern-0.1em/\kern-0.05em c}\xspace}
\newcommand{\MeVcc}{\ensuremath{\mbox{Me\kern-0.1em V}\kern-0.1em/\kern-0.05em c^2}\xspace}

\newcommand{\pT}{\ensuremath{p_\text{T}}\xspace}
\newcommand{\pp}{\ensuremath{p}\xspace}

\newcommand{\logp}{\ensuremath{\lg p}\xspace}

\newcommand{\minv}{\ensuremath{m_\text{inv}}\xspace}
\newcommand{\dedx}{\ensuremath{\mathrm{d}E/\mathrm{d}x}\xspace}

\newcommand{\mdedx}{\ensuremath{\langle\mathrm{d}E/\mathrm{d}x\rangle}\xspace}

\newcommand{\UrqmdLong}{{\scshape U}r{\scshape qmd\,1.3.1}\xspace}

\newcommand{\Corsika}{{\scshape Corsika}\xspace}

\newcommand{\VenusLong}{{\scshape Venus\,4.12}\xspace}

\newcommand{\Geant}{{\scshape Geant}\xspace}

\newcommand{\Epos}{{\scshape Epos}\xspace}
\newcommand{\EposLong}{{\scshape Epos\,1.99}\xspace}
\newcommand{\QGSJetLong}{{\scshape QGSJet\,II-04}\xspace}
\newcommand{\QGSJetOldLong}{{\scshape QGSJet\,II-03}\xspace}
\newcommand{\DPMJetLong}{{\scshape DPMJet\,3.06}\xspace}
\newcommand{\SibyllLong}{{\scshape Sibyll\,2.1}\xspace}
\newcommand{\SibyllNewLong}{{\scshape Sibyll\,2.3c}\xspace}
\newcommand{\EposLHCLong}{{\scshape Epos\,LHC}\xspace}

\newcommand{\CernVM}{\textsc{Cern\-\kern-0.05emVM}\xspace}

\newcommand{\cmc}{\ensuremath{\mbox{$C_\text{sim}$}}\xspace}

\def \pions{$\uppi^\pm$\xspace}
\def \kaons{K$^\pm$\xspace}
\def \proton{p\xspace}
\def \antiproton{$\bar{\text{p}}$\xspace}

\def \protons{p($\bar{\text{p}}$)\xspace}
\def \protonpm{p$^\pm$\xspace}
\def \lamb{$\Lambda$\xspace}
\def \antilamb{$\bar{\Lambda}$\xspace}
\def \lambs{$\Lambda(\bar{\Lambda})$\xspace}
\makeatletter
\newcommand{\oset}[3][0ex]{%
  \mathrel{\mathop{#3}\limits^{
    \vbox to#1{\kern-2\ex@
    \hbox{$\scriptstyle#2$}\vss}}}}
\makeatother

\def \kzeros{K$^{0}_\text{S}$\xspace}

\def \vzero{$\text{V}^0$\xspace}
\def \vzeros{$\text{V}^0$s\xspace}
\def \p{$p$\xspace}

\def \piminus{$\uppi^-$\xspace}
\def \rhozero{$\uprho^0$\xspace}
\def \Tint{\ensuremath{{\text{T}_\text{int}}}\xspace}
\def \Tbeam{\ensuremath{\text{T}_\text{beam}}\xspace}
\def \hmC{\ensuremath{\text{h}^-\text{-\,C}}\xspace}
\def \pimC{\ensuremath{\uppi^-\text{-\,C}}\xspace}
\def \pipC{\ensuremath{\uppi^+\text{-\,C}}\xspace}
\def \pimp{\ensuremath{\uppi^-\text{-\,C}}\xspace}
\def \recn{\ensuremath{\hat{n}}\xspace}
\def \recN{\ensuremath{\hat{N}}\xspace}
\def \genn{\ensuremath{n_\text{gen}}\xspace}
\def \genN{\ensuremath{N_\text{gen}}\xspace}

\newcommand{\dedxfig}[2]{
\begin{figure}[!#2]
  \centering
  \def\dedxheight{0.235}
  \includegraphics[clip, rviewport=0 0 1 1,height=\dedxheight\textheight]{Plots/Dedx_#1_0}\quad\includegraphics[clip, rviewport=0 0 1 1,height=\dedxheight\textheight]{Plots/Dedx_#1_1}
  \caption{Energy deposit vs momentum of negatively and positively
    charged tracks for the #1 \GeVc data set. The dashed lines
    indicate the average energy deposit for each particle type
    \ifnum#1=158 (see \cref{fig:dedx:bb350} in the appendix for
    $p_\text{beam} =350~\GeVc$)\fi.}
  \label{fig:dedx:bb#1}
\end{figure}
}

\newcommand{\betafig}[1]{
\begin{figure}[!ht]
  \centering
  \begin{overpic}[clip, rviewport=0 0.143 1 1,width=0.45\textwidth]{Plots/fac_#1_All_beta_c0_p1}
    \put(20,53){$\pi^+$}
  \end{overpic}
  \begin{overpic}[clip, rviewport=0 0.143 1 1,width=0.45\textwidth]{Plots/fac_#1_All_beta_c1_p1}
    \put(20,53){$\pi^-$}
  \end{overpic}

  \begin{overpic}[clip, rviewport=0 0.143 1 1,width=0.45\textwidth]{Plots/fac_#1_All_beta_c0_p2}
    \put(20,53){K$^+$}
  \end{overpic}
  \begin{overpic}[clip, rviewport=0 0.143 1 1,width=0.45\textwidth]{Plots/fac_#1_All_beta_c1_p2}
    \put(20,53){K$^-$}
  \end{overpic}

  \begin{overpic}[clip, rviewport=0 0 1 1,width=0.45\textwidth]{Plots/fac_#1_All_beta_c0_p3}
    \put(20,63){p$^+$}
  \end{overpic}
  \begin{overpic}[clip, rviewport=0 0 1 1,width=0.45\textwidth]{Plots/fac_#1_All_beta_c1_p3}
    \put(20,63){p$^-$}
  \end{overpic}
  \begin{overpic}[clip, rviewport=0 0 1 1.1,width=0.99\textwidth]{Plots/beta#1}
    \put(0,22){\lamb}
    \put(33,22){\antilamb}
    \put(67,22){\kzeros}
  \end{overpic}
  \caption{$\beta$ correction factors for the #1 \GeVc data set \ifnum#1=158 (see \cref{fig:correction:beta350} in the appendix for $p_\text{beam} =350~\GeVc$)\fi.}
  \label{fig:correction:beta#1}
\end{figure}
}

\newcommand{\sysfig}[1]{
\begin{figure}[p]
  \centering
  \def\sysfigwh{0.45}
  \def\sysfigwe{0.328}
  \def\sysfigh{0.265}
  \begin{overpic}[clip, rviewport=0 0 1 1,width=\sysfigwh\textwidth]{Plots/syst_#1_c0_p1_std}
    \put(2,68){$\pi^+$}
  \end{overpic}\qquad
  \begin{overpic}[clip, rviewport=0 0 1 1,width=\sysfigwh\textwidth]{Plots/syst_#1_c1_p1_std}
    \put(2,68){$\pi^-$}
  \end{overpic}

  \begin{overpic}[clip, rviewport=0 0 1 1,width=\sysfigwh\textwidth]{Plots/syst_#1_c0_p2_std}
    \put(2,68){K$^+$}
  \end{overpic}\qquad
  \begin{overpic}[clip, rviewport=0 0 1 1,width=\sysfigwh\textwidth]{Plots/syst_#1_c1_p2_std}
    \put(2,68){K$^-$}
  \end{overpic}

  \begin{overpic}[clip, rviewport=0 0 1 1,width=\sysfigwh\textwidth]{Plots/syst_#1_c0_p3_std}
    \put(2,68){p$^+$}
  \end{overpic}\qquad
  \begin{overpic}[clip, rviewport=0 0 1 1,width=\sysfigwh\textwidth]{Plots/syst_#1_c1_p3_std}
    \put(2,68){$\bar{\text{p}}$}
  \end{overpic}

  \begin{overpic}[clip, rviewport=0.03 0 0.97 1,height=\sysfigh\textwidth]{Plots/syst_#1_h0}
    \put(20,16){\lamb}
  \end{overpic}
  \begin{overpic}[clip, rviewport=0.1 0 0.97 1,height=\sysfigh\textwidth]{Plots/syst_#1_h1}
    \put(15,16){\antilamb}
  \end{overpic}
  \begin{overpic}[clip, rviewport=0.12 0 0.97 1,height=\sysfigh\textwidth]{Plots/syst_#1_h2}
    \put(13,16){\kzeros}
  \end{overpic}

  \caption{Systematic uncertainties of the single-differential spectra
    $\text{d}n/\text{d}\pp$ for the charged hadrons (top three rows) and
    \vzeros (bottom row) as a function of momentum
    for the data set recorded with $p_\text{beam} = #1 \GeVc$ \ifnum#1=158 (see \cref{fig:syst350} in the appendix for $p_\text{beam} =350~\GeVc$)\fi.}
  \label{fig:syst#1}
\end{figure}
}

\ShineTitle{Measurement of Hadron Production \\
            in \pimC Interactions at 158 and 350 \GeVc\\
            with \NASixtyOne at the CERN SPS}

\PreprintIdNumber{CERN-EP-2022-194}

\definecolor{rossoCP3}{cmyk}{0,.88,.77,.40}
\definecolor{darkBlue}{rgb}{0, 0, 0.8}

\hypersetup{
    bookmarksopen=true,     
    bookmarksopenlevel=1,
    unicode=false,          
    pdftoolbar=true,        
    pdfmenubar=true,        
    colorlinks=true,        
    linkcolor=darkBlue,     
    citecolor=darkBlue,        
    filecolor=darkBlue,      
    urlcolor=darkBlue           
}

\ShineJournal{Phys.~Rev.~C}
\ShineAbstract{
We present a measurement of the momentum spectra
of \pions, \kaons, \protonpm, \lamb, \antilamb and \kzeros produced in
interactions of negatively charged pions with carbon nuclei at beam
momenta of 158 and 350\,\GeVc. The total production cross sections are
measured as well.  The data were collected with the large-acceptance
spectrometer of the fixed target experiment \NASixtyOne at the CERN
SPS.  The obtained double-differential \p-\pT spectra provide a unique
reference data set with unprecedented precision and large phase-space
coverage to tune models used for the simulation of particle production
in extensive air showers in which pions are the most numerous
projectiles.}

\begin{document}
\maketitle
\section{Introduction}

When cosmic rays collide with the nuclei of the atmosphere, they
initiate a cascade of secondary particles called air shower.  The
interpretation of cosmic-ray data from air-shower arrays such as
KASCADE-Grande~\cite{Navarra:2004hi}, IceTop~\cite{IceCube:2012nn},
Telescope Array~\cite{AbuZayyad:2012kk} or the Pierre Auger
Observatory~\cite{Abraham:2004dt} relies to a large extent on the
understanding of these particle cascades in the atmosphere,
specifically on the correct modeling of hadron-air interactions that
occur during shower development.  However, it is a
well-established fact that air shower simulations using current
state-of-the-art models of high-energy hadronic interactions produce
significantly less muons than observed in data~\cite{HiresMia2000,Aab:2014pza,Aab:2014dua,Aab:2016hkv,KASCADE-Grande:2017wfe,Bogdanov:2018sfw,TelescopeArray:2018eph,Bellido:2018toz,Gesualdi:2020ttc,PierreAuger:2020gxz, wgicrc21,
  Albrecht:2021cxw,IceCube:2022zhn}.

The majority of muons in air showers are created in decays of charged
pions when the energy of the pion is low enough such that its decay
length is smaller than its interaction length in air.  The projectiles
creating these pions are typically produced at equivalent beam
energies below a TeV~\cite{Drescher:2002vp,Meurer:2005dt,Maris:2009uc}
which is well within the reach of current accelerators.  However, only a
very limited amount of data exists on the interactions of the most numerous
projectile in air showers, the $\uppi$-meson~\cite{previousdata}.

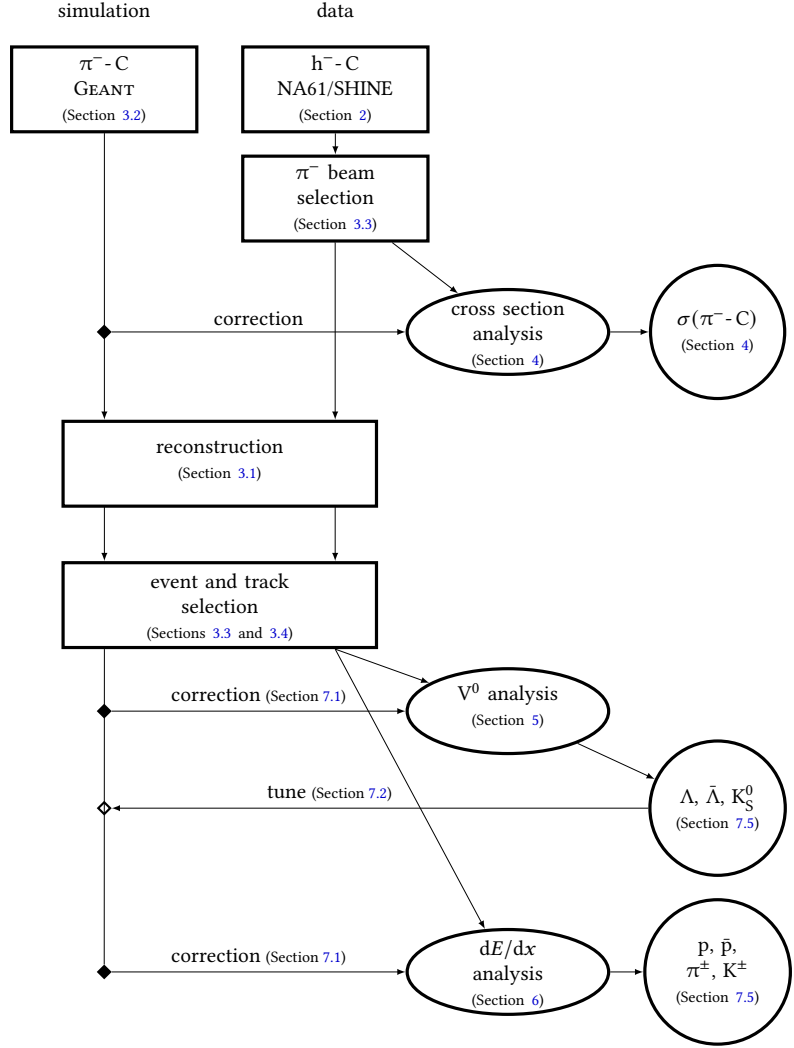
\begin{figure}[t]
\begin{center}
  \smaller[2]
  \def\secref#1{{\smaller[2](\cref{#1})}}
  \begin{tikzpicture}[auto,
    correction/.style={diamond, draw=black, thick, fill=black,
    text width=0.1em, text badly centered,
    inner sep=1pt, font=\sffamily\small},
    tune/.style={diamond, draw=black, thick, fill=white,
    text width=0.1em, text badly centered,
    inner sep=1pt, font=\sffamily\small},
    analysis/.style ={ellipse, draw=black, very thick, fill=white,
      text width=6em, text centered,anchor=center,
      minimum height=4em},
    result/.style ={circle, draw=black, very thick, fill=white,
      text width=5em, text centered,anchor=center,
      minimum height=4em},
    block/.style ={rectangle, draw=black, very thick, fill=white,
      text width=8em, text centered,anchor=center,
      minimum height=4em},
    multiblock/.style ={rectangle, draw=black, very thick, fill=white,
      text width=14em, text centered,anchor=center,
      minimum height=4em},
    empty/.style ={rectangle, draw=none, very thick, fill=none,
      text width=.1em, text centered},
    line/.style ={draw, ultra thin, -latex, shorten >=0pt}]
    \matrix (table) [ matrix of nodes,
      nodes in empty cells,
      column sep=2em,row sep=1em] {
      simulation & data \\
      \node [block] (simDetection) {\pimC \\\Geant \\ \secref{sec:simu}}; &
      \node [block] (dataDetection) {\hmC \\\NASixtyOne \\ \secref{sec:experiment}}; & \\
     & \node [block] (dataBeamSel) {\piminus beam\\ selection\\ \secref{sec:beamselection}}; & \\
      \node [correction] (C1){};&&&&&\node [result] (xs) {$\sigma(\pimC)$\\ \secref{sec:xsana}};\\
      &&\\
      &&\\
      &&\\
      &&\\
      &&\\
      &&\\
      &&\\
      &&\\
      \node [correction] (C2){};&&&&&\\
     \node [tune] (tune){}; &&&&&\node [result] (v0) {\lamb, \antilamb, \kzeros\\\secref{sec:v0results}};\\
      \node [correction] (C3){};&&&&&\node [result] (dedx) {\proton, \antiproton, \pions, \kaons \\ \secref{sec:dedxresults}};\\
    };
   \node[fit=(table-6-1)(table-6-2), multiblock] (reco) {reconstruction \\ \secref{sec:reco}};
   \node[fit=(table-10-1)(table-10-2), multiblock] (sel) {event and track \\ selection \\ \secref{sec:eventselection,sec:trackselection}};
    \node[fit=(table-4-3)(table-4-4), analysis] (xsAna) {cross~section analysis \\ \secref{sec:xsana}};
    \node[fit=(table-13-3)(table-13-4), analysis] (v0Ana) {\vzero analysis\\ \secref{sec:v0ana}};
    \node[fit=(table-15-3)(table-15-4), analysis] (dedxAna) {\dedx analysis \\ \secref{sec:dedx}};
    \begin{scope}[every path/.style=line]
      \path (dataDetection) -- (dataBeamSel);
      \path (dataBeamSel) -- (xsAna);
      \path (xsAna) -- (xs);
      \path (C1) -- node[anchor=south] {correction} (xsAna);
      \path (simDetection.south) -| ([xshift=-5.4em]reco.north);
      \path (dataBeamSel.south)  -| ([xshift=+5.4em]reco.north);
      \path ([yshift=.1em]reco.south) -| ([xshift=-5.4em]sel.north);
      \path ([yshift=.1em]reco.south)  -| ([xshift=+5.4em]sel.north);
      \path ([xshift=5.4em]sel.south)  -- (dedxAna);
      \path ([xshift=+5.4em]sel.south)  -- (v0Ana);
      \path ([xshift=-5.4em]sel.south)  -- (C3)-- node[anchor=south] {correction \secref{sec:correction}} (dedxAna);
      \path (v0Ana)  -- (v0);
      \path (v0)  -- node[anchor=south] {\hspace*{-1.4cm} tune \secref{sec:correction:fd}}  (tune);
      \path (dedxAna)  -- (dedx);
      \path (C2) -- node[anchor=south] {correction \secref{sec:correction}} (v0Ana);
    \end{scope}
  \end{tikzpicture}
\end{center}
\caption{Schematic view of the analysis presented in this
  paper.}
\label{fig:analysisflow}
\end{figure}

In this paper, we present new data from the \NASixtyOne experiment at
the CERN SPS on the particle production in interactions of pion beams
at 158 and 350\,\GeVc with a thin carbon target (used as a proxy for
nitrogen, the most abundant nucleus in air).  After a brief
introduction to the experiment in \cref{sec:experiment}, we will
describe the various data analysis steps that lead to the results
presented in this paper.  These steps are sketched in
\cref{fig:analysisflow} for a better orientation of the flow of the
analysis and the corresponding sections in which they are described in
this article.  The processing and selection of data and simulations
are introduced in \cref{sec:dataProcAndSel}.  The three main analyses
of the cross section, \vzero decays, and identified charged particles
are explained in \cref{sec:xsana,,sec:v0ana,,sec:dedx}, respectively.
The calculation of the particle spectra and the estimation of their
uncertainties is outlined in~\cref{sec:spectra}, where the measured
production spectra of \pions, \kaons, \protonpm, \lamb, \antilamb and
\kzeros are presented.  In \cref{sec:discussion} we conclude by
comparing our measurements to the predictions of hadronic interaction
models used for the modeling of air showers.

\section{Experimental Setup and Data Taking}
\label{sec:experiment}

\begin{figure*}[t]
\centering
\includegraphics[clip,rviewport=0 0 0.9 1,width=0.95\textwidth]{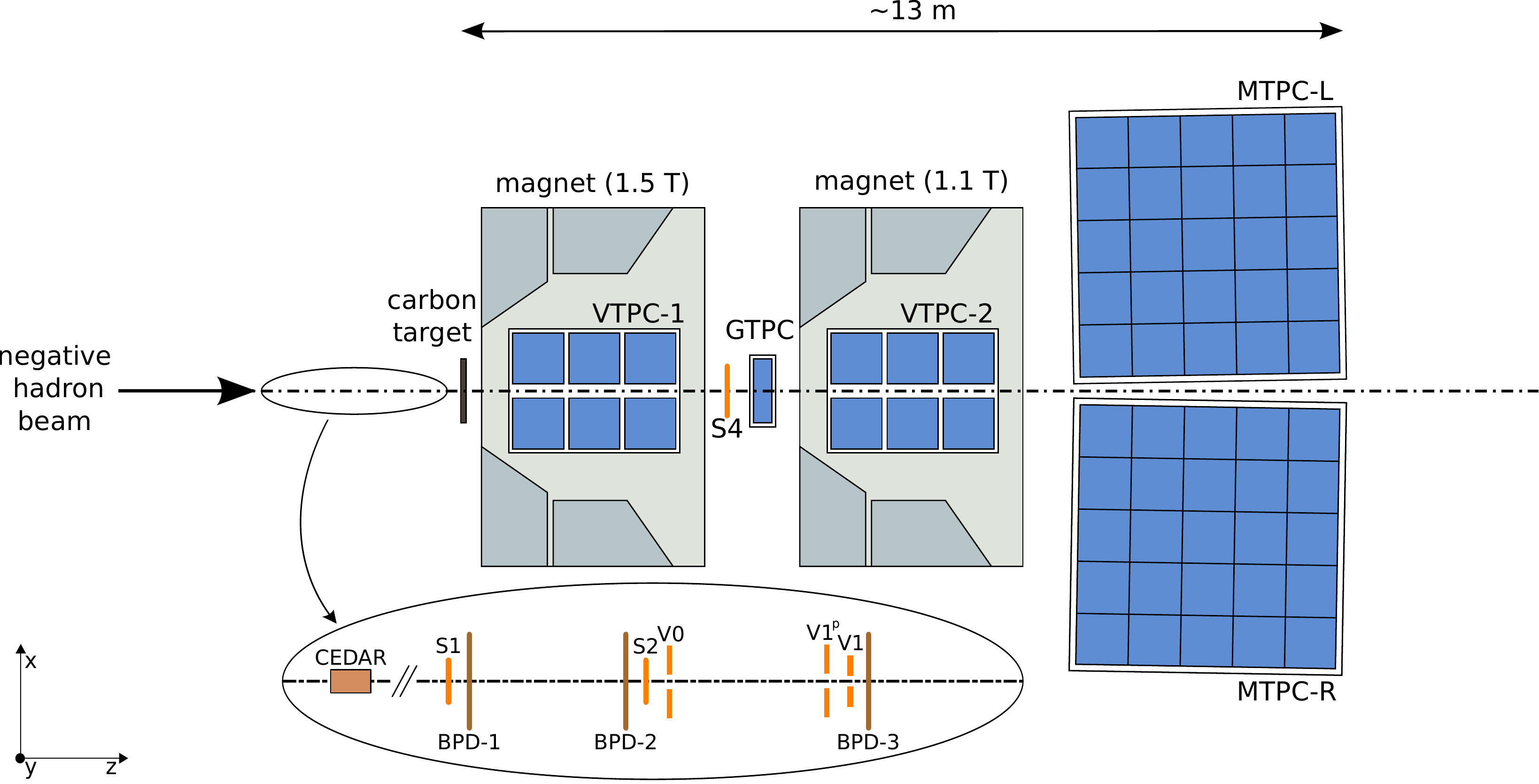}
\caption{Experimental Setup of the NA61/SHINE
  experiment~\cite{Abgrall:2014fa} (configuration for the \pimC data
  taking).  The coordinate system used in this paper is indicated on
  the lower left.  The incoming beam direction is along the $z$ axis.
  The magnetic field bends charged particle trajectories in the
  $x$-$z$ (horizontal) plane.  The drift direction in the TPCs is
  along the $y$ (vertical) axis.  A zoomed view of the beam and
  trigger instrumentation is shown as an elliptical inset at the
  bottom.}
\label{fig:na61Layout}
\end{figure*}

The data reported in this paper were taken in 2009 with the \NASixtyOne
instrument, a wide-acceptance hadron spectrometer at the CERN SPS on
the H2 beam line of the CERN North Area~\cite{Abgrall:2014fa}.  The
experimental setup used to record \pimC interactions is shown in
\cref{fig:na61Layout}.

The main part of the detector consists of five Time Projection
Chambers (TPCs) which were inherited from NA61's predecessor, the NA49
experiment~\cite{Afanasev:1999iu}.  Two Vertex TPCs
(VTPC\nobreakdash-1 and VTPC\nobreakdash-2) are located inside the
magnetic field produced by two superconducting dipole magnets.  For
the measurements presented in this paper the two magnets were operated at
full electric current providing a field of 1.5 and 1.1\,T,
respectively.  Two Main-TPCs are located downstream of the VTPCs to
measure particles bent in the left and right hemispheres
(MTPC\nobreakdash-L and MTPC\nobreakdash-R).  An additional small TPC
is placed between VTPC\nobreakdash-1 and VTPC\nobreakdash-2, covering
the very-forward region, and is referred to as the gap\nobreakdash-TPC
(GTPC).  The combined bending power of the magnets is 9\,T\,m and the
coordinates on a track are measured with a precision of a few
100\,$\upmu$m.  The resolution of the measurement of particle momenta
in the TPCs depends on the track topology, i.e.\ on the overall track
length and the number of position
measurements~\cite{Gluckstern:1963ng}.  Typical values for the
momentum resolution are $\sigma(p)/p^2 =
7{\times}10^{-4}\,(\text{GeV}/c)^{-1}$ for low-momentum tracks
measured only in the VTPC\nobreakdash-1 ($p\lesssim 8$\,\GeVc) and
${\sim}3{\times}10^{-3}\,(\text{GeV}/c)^{-1}$ for tracks traversing
the full detector up to the MTPCs ($p\gtrsim 8$\,\GeVc).

For the study reported in this paper, the SPS delivered a secondary
hadron beam originating from interactions of 400\,\GeVc primary
protons impinging on a 10-cm-long beryllium target.  The negatively
charged hadrons ($\text{h}^-$) produced in these interactions were
transported through the H2 beam line to the \NASixtyOne experiment.  A
beam momentum of 158\,\GeVc was requested for the first part of data
taking and 350\,\GeVc for the second part.  At these momenta the
negatively charged beam particles are mostly $\uppi^-$ mesons.  They are
identified by a differential ring-imaging Cherenkov detector
(CEDAR)~\cite{Bovet:1975bx} and the fraction of pions was measured to
be ${\sim}95\%$ for 158\,\GeVc and ${\sim}100\%$ for 350\,\GeVc (see
Fig.\,2 in~\cite{Aduszkiewicz:2017anm}).  The CEDAR signal is recorded
during data taking and then used as an offline selection cut.

Downstream of the CEDAR three proportional chambers are used as Beam
Position Detectors (BPDs) to measure the trajectories of the incoming
particles.  Two scintillation counters (S1 and S2) and three veto
counters (V0, V1 and V1$^\prime$) define the beam trigger with the
coincidence
\begin{equation}
  \text{T}_\text{beam} =
    \text{S1} \land \text{S2} \land \overline{\text{V0}} \land \overline{\text{V1}} \land \overline{\text{V1}^\prime},
\end{equation}
see inset in \cref{fig:na61Layout}.
The beam trigger is a \emph{zero bias trigger}.
Furthermore, an {interaction trigger},
\begin{equation}
  \text{T}_\text{int} = \text{T}_\text{beam} \land \overline{\text{S4}},
\end{equation}
is defined as the anti-coincidence of the incoming beam particle and
S4, a scintillation counter with a diameter of 2\,cm placed between
the VTPC\nobreakdash-1 and VTPC\nobreakdash-2 along the beam
trajectory at about 3.7\,m from the target.  If an inelastic
interaction occurs then the produced particles typically have momenta
considerably lower than the beam momentum and are thus bent away from the
beam trajectory such that no particle reaches S4.  The
anti-coincidence with S4, therefore, serves as a \emph{minimum-bias interaction
  trigger}.

During data taking a prescaled fraction of the \Tbeam and \Tint
signals can trigger the data acquisition system to read out the TPCs
and write a raw event to disk.  For most of the data taking period,
the pre-scaling was 0.4\% for the zero-bias beam trigger and 100\% for
the minimum-bias interaction trigger.

The target used for this study was an isotropic graphite plate with a
thickness along the beam axis of 2\,cm and a density of
$\rho=1.840$\,g/cm$^3$, equivalent to about 4\% of a nuclear
interaction length.  90\% of data was recorded with the target
inserted and 10\% with the target removed.  The latter data was
used to subtract interactions that took place in the material upstream
and downstream of the target. In total, 5.5 million events were recorded with
the target inserted at a beam momentum of 158\,\GeVc and 4.5 million
events at 350\,\GeVc.

\section{Data Processing and Selection}
\label{sec:dataProcAndSel}

\subsection{Reconstruction}
\label{sec:reco}

The raw data recorded during data taking are processed with the
standard \NASixtyOne reconstruction chain (based on tried-and-tested
algorithms developed by the NA49 collaboration) to obtain high-level
physics information such as the charge and momentum of the produced
particles.  Firstly, charge clusters are identified in the raw data of
the TPCs and their three-dimensional positions are reconstructed from
the centroids in drift time and in position on the TPC readout pads.
A pattern recognition combines these clusters to form local track segments
in each TPC separately.  The local track segments are matched to
\emph{global tracks} for which the track parameters (track position at
reference plane, charge and three-momentum) are fitted.

The beam trajectory before the target is determined using the position
measurements from the BPDs.  The position of the nominal \emph{main interaction
  vertex} is estimated as the intersection of the reconstructed beam
trajectory with a plane located at the center $z$-position of the
2\,cm-carbon target.  Global tracks compatible with this position are
then re-fitted with this vertex hypothesis leading to
\emph{main-vertex track candidates} with track parameters at the
nominal interaction vertex.  Furthermore, the $z$-position of the main
interaction vertex is reconstructed by fitting a common origin of
these global tracks along the beam trajectory.  Depending on the track
multiplicity, the resolution along the $z$-axis for the vertex position obtained
in this way exceeds the target thickness, therefore better
main-vertex-constrained track-momenta can be obtained by using the
nominal interaction vertex in the middle of the target instead.  Yet, the
reconstructed vertex position is useful for the rejection of obvious
out-of-target interactions during the event selection (see below).

In addition to the main-vertex hypothesis, each combination of
positively and negatively charged global tracks within one event is
investigated for a common origin downstream of the target from the
decay of a long-lived neutral particle (so-called \vzero events).  All
pairs with a distance of closest approach $\leq 1$\,cm anywhere along
their trajectory downstream of the target are re-fitted with the
constraint to originate from a common vertex resulting in \emph{\vzero
  candidates} that will be analyzed in \cref{sec:v0ana}.

The results of the reconstruction are stored in a dedicated
\textsc{Root}-based~\cite{Antcheva:2009zz} output
format~\cite{Sipos:2012hs} for further processing during data
analysis.

\subsection{Simulation}
\label{sec:simu}

Through this analysis, we will use simulations of the measurement
to correct the raw-data spectra for various distortions originating from the detector
acceptance, re-interactions with the detector material and within the
target, feed-down from weak decays etc.  For this purpose we simulated
\pimC interactions at both beam momenta with the hadronic event
generators \EposLong~\cite{Pierog:2006qv} and
\QGSJetLong~\cite{Ostapchenko:2004ss} with the \textsc{crmc}
program~\cite{crmc}.  The 1.99 version of the \Epos model was used
rather than its newer ``LHC'' variant as the former is better tuned to
interactions at SPS energies \cite{Tanguy_Epos}.  The particles
produced by these generators are then passed to a simulation of the
passage of particles through the material and magnetic field of the
\NASixtyOne setup using the \textsc{Geant}\,3.21
package~\cite{Geant3}.  The hits generated in the active detector
volumes are digitized to produce the same raw information as for real data
and an interaction trigger is simulated by checking whether any of the
charged particles hit the S4 counter.  The simulated information is then
processed with the same reconstruction algorithms discussed in the
previous section and the results are stored in the same output format
as the reconstructed data.

\subsection{Event Selection}
\label{sec:beamselection}
\label{sec:eventselection}

We apply the following event-selection criteria\footnote{The event
  selection criteria used here are identical to the ones described in
  Ref.~\cite{Aduszkiewicz:2017anm} for the same data set discussed in
  this paper.  We therefore refer to Sections 2 and 3 of that paper
  for a detailed description of the event selection.} to obtain a set
of high-quality interaction triggers.

Pion projectiles are selected with the CEDAR (see previous section)
and a pile-up of interactions is avoided by rejecting events in which
the S1 scintillator detected another beam particle within $\pm2\,\upmu$s
of the time of the interaction trigger.  Furthermore, it is required
that the direction of the beam was well measured with the three beam
position detectors.  These three cuts select high-quality
$\uppi^-$ projectiles based on measurements from the beam detectors
upstream of the target.

Further event selection criteria define the subset of
$\uppi^-$-projectiles with an interaction in the target.  Here we
analyze the particles produced in events recorded with the
minimum-bias trigger.  Furthermore, we reject events with an
interaction vertex reconstructed far from the center of the target
($|\Delta z| >17$\,cm), since such events mostly originate from
interactions outside of the target.

With these criteria we select $2.8{\times}10^6$ and $2.6{\times}10^6$
minimum-bias triggers recorded with an inserted C-target at 158 and
350\,\GeVc, respectively.  By construction, due to the restriction to
events with a reconstructed vertex close to the target position, only
very few events recorded with a removed target survive the selection
(${\lesssim}7{\times}10^3$ events in each data set).  The sum of
simulated events with an inserted target is $7.2{\times}10^6$ and
$6.0{\times}10^6$ for beam momenta of 158 and 350\,\GeVc respectively.

\subsection{Track Selection}
\label{sec:trackselection}

\begin{figure}[t]
\def\fh{0.56}
\centering
\includegraphics[clip,height=\fh\linewidth]{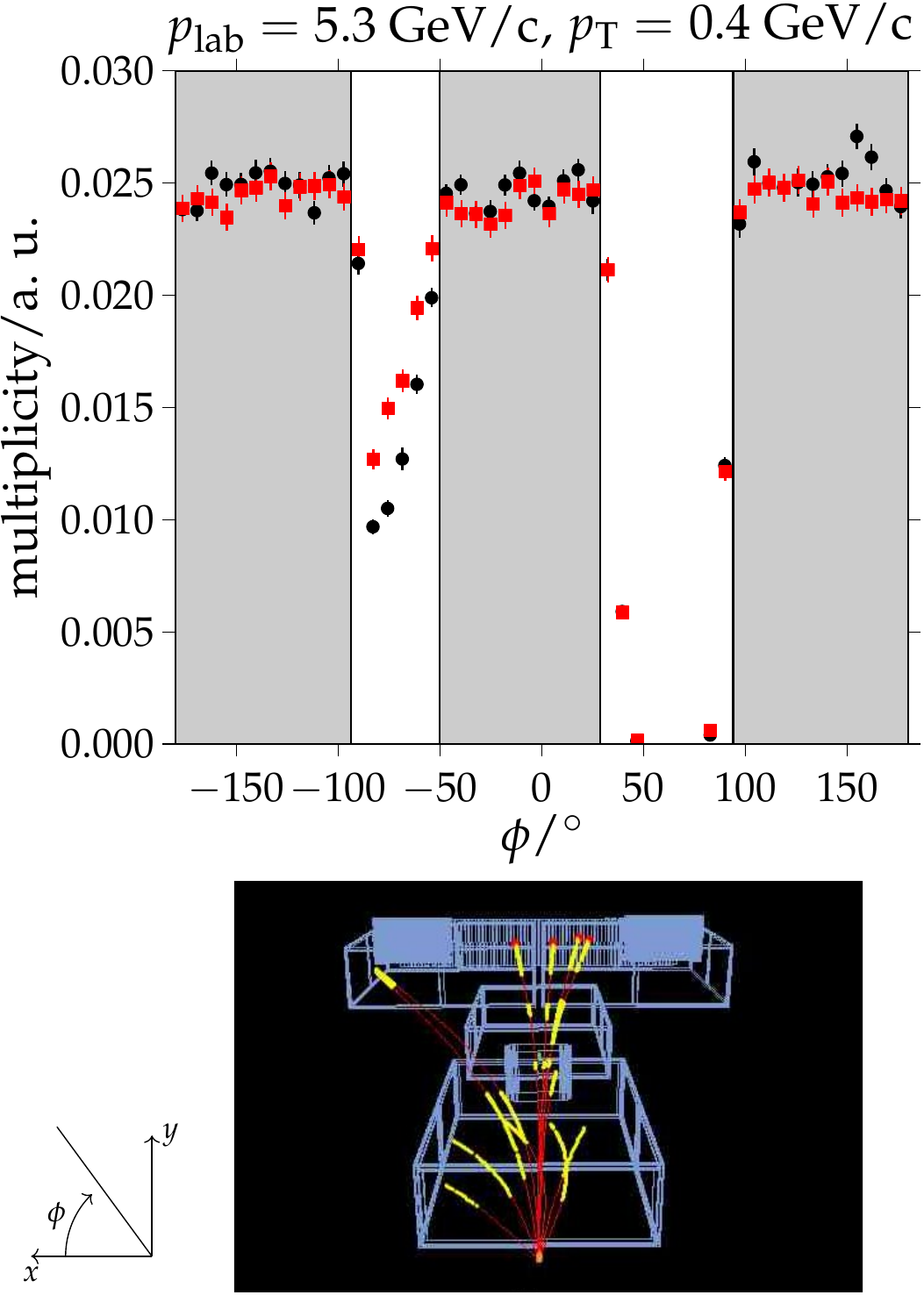}\hfill
\includegraphics[height=\fh\linewidth]{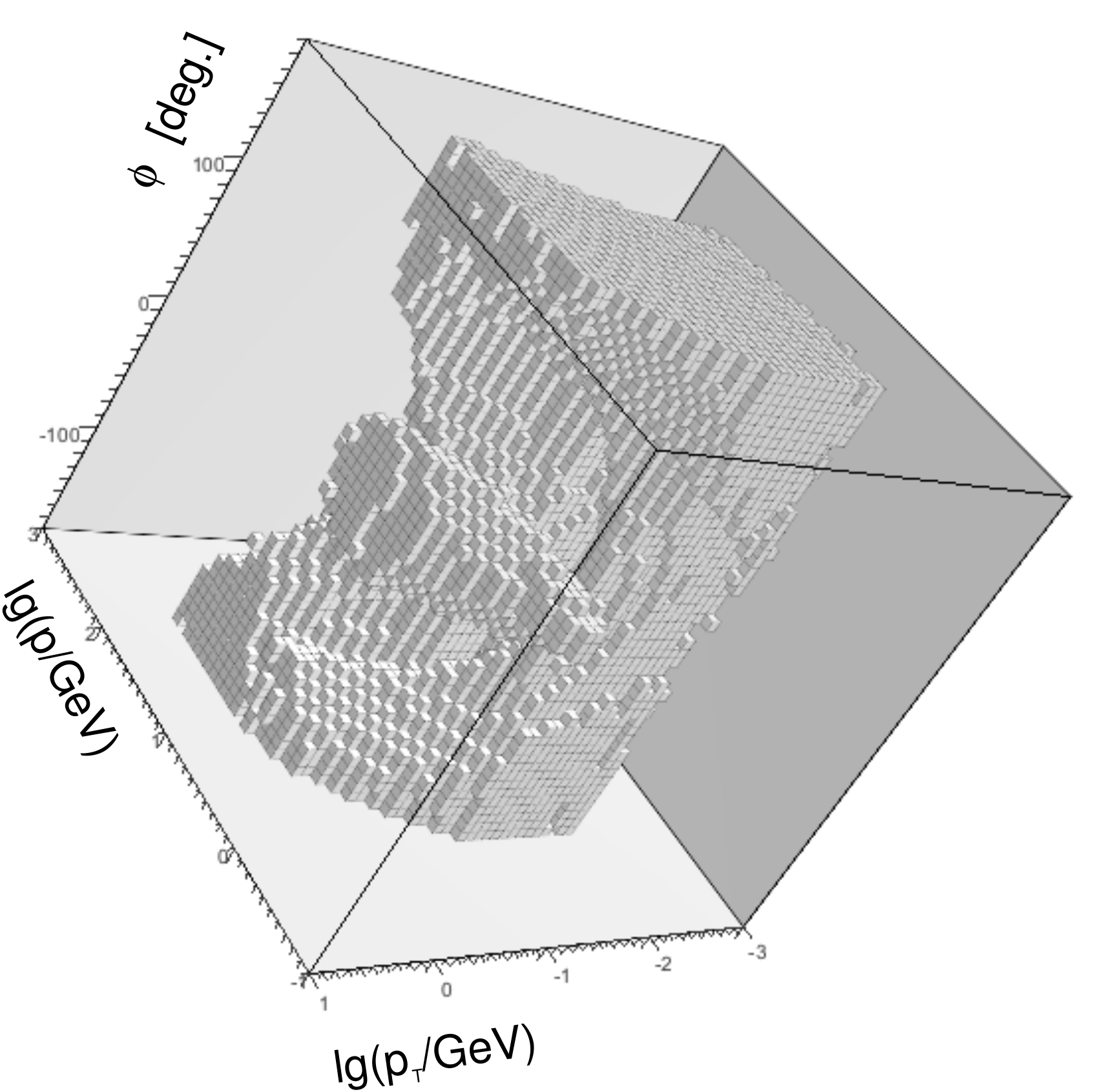}
\caption{\emph{Top left:} Example of the charged particle multiplicity
  at $(\pp,\pT) = (5.3, 0.4)$\,\GeVc as a function of track
  angle $\phi$.  Red squares are simulated data and black circles are
  measurements.  \emph{Bottom left:} Event display illustrating the
  aspect-ratio of the TPC chambers (blue boxes) and showing
  reconstructed tracks (red lines), clusters (yellow points) and
  the primary interaction vertex (orange point).
  \emph{Right:} 3D view of the acceptance definition
  for positively charged tracks.  Each ($\phi$, \pp, \pT) bin which
  does not satisfy the acceptance criteria is empty.}
\label{fig:acc}
\end{figure}

The set of selection criteria given below is applied to the tracks
measured in the TPCs.  They are constructed to assure good quality of the
momentum measurements and to select regions of the detector with a solid
understanding of the detection efficiency.
\paragraph{Track Quality}
The total number of clusters on the track must be $\geqslant 25$ and the sum of clusters in both VTPCs must be $\geqslant 12$,
or the number of clusters in the GTPC must be $\geqslant 6$.  These cuts
assure a good momentum determination in either the VTPCs within the
magnetic field, or a large lever arm for forward tracks measured with
a combination of GTPC and MTPC~\cite{MartinThesis}.
\paragraph{Fiducial Acceptance}
For each particle charge, we restrict the analysis to bins in azimuthal
angle, total momentum and transverse momentum, ($\phi$, \pp, \pT), in which the track selection efficiency is larger than 90\%
for simulated tracks.\footnote{Since the acceptance is only a property
  of the track momentum and direction (not of the beam momentum or
  track multiplicity), it can be determined with high statistical
  accuracy averaging all available simulation sets.  A total of
  $2.1{\times}10^7$ simulated events were used to construct the
  acceptance map.  This number is higher than the sum of simulated
  events mentioned in \cref{sec:eventselection}, since for the
  acceptance study we also included events generated with the
  \DPMJetLong model, which were not used otherwise for this study.}
This cut mainly removes tracks at the edges of the detector for which
the efficiency drops rapidly, as illustrated in the example shown in
the top-left panel of \cref{fig:acc}.  Since the TPCs have a larger
width than height (cf.\ lower left panel in \cref{fig:acc}), the
acceptance is not uniform in $\phi$ and especially poor along
$\pm\pi/2$.  But since neither the beam nor the target are polarized, the
particle yields are independent of $\phi$ and we can thus restrict the
analysis to $\phi$ regions where the detection efficiency is near
100\%.  The \emph{fiducial volume} selected by the acceptance cut
therefore leads to a near-geometric acceptance that is given
by the number of accepted $\phi$-bins~\cite{MartinThesis}.  The
acceptance for positively charged tracks is visualized in
\cref{fig:acc}.
\paragraph{Origin at Primary Vertex}
Furthermore, we require that the distance between the position of the
intercept of the extrapolation of the track (reconstructed without
vertex-constraint) to the interaction plane and the position of the
main interaction vertex must be smaller than 4~cm in both horizontal
and vertical direction. This cut removes out-of-target interactions
and tracks from particle decays (``feed-down'').

Furthermore, we require that the distance between the extrapolation of
the track (reconstructed without vertex-constraint) to the interaction
plane and the interaction point must be smaller than 4\,cm in both
horizontal and vertical direction.  This cut removes out-of-target
interactions and tracks from particle decays (``feed-down'').

After the track selection, the measured tracks are split into two
subsets called right-side tracks (RSTs) and wrong-side tracks (WSTs).
The former group is defined as the tracks that bend away from the beam
axis, while the latter as the tracks that bend towards the beam
axis\footnote{The terminology of wrong- and right-side tracks dates
  back to NA49 and reflects the fact that the tilt angle of the pads
  in the VTPCs were optimized to measure the right-side topologies.},
i.e.\ $p_x q > 0$ (RST) and $p_x q < 0$ (WST), where $p_x = p \sin
\theta \cos\phi$ and $p$ and $q$ are the momentum and charge of the
track. These conditions simplify to $q\cos\phi>0$ for RSTs and
$q\cos\phi<0$ for WSTs for the usual case of forward production in the
laboratory frame, $\theta>0$.  The distinction between the two track
topologies is motivated by the fact that a right- and a wrong-side track
with the same \pp and \pT, cross different regions of the detector,
which has important implications for the particle
reconstruction and identification step based on the \dedx measurements
(see~\cref{sec:dedx}).  Furthermore, this subdivision defines two
independent data sets that can be compared to estimate the systematic
uncertainties of the measured particle multiplicities, see
\cref{sec:spec:syst}.

\section{Cross Section Analysis}
\label{sec:xsana}
\label{sec:xs}

The measurement of the production cross section\footnote{The
  production cross section is defined by $\sigma_\text{prod} =
  \sigma_\text{tot} - \sigma_\text{el} - \sigma_\text{qe}$, where
  $\sigma_\text{tot}$ denotes the total cross section,
  $\sigma_\text{el}$ the coherent elastic cross section and
  $\sigma_\text{qe}$ the quasi-elastic cross section. Quasi-elastic interactions are processes in which no new particles are produced, but the target nucleus is fragmented.} in \pimC interactions
closely follows the analysis procedure detailed in Ref.~\cite{Abgrall:2015hmv}.
The experimental interaction cross section is measured by counting the
number of interaction triggers, $N(\Tbeam \wedge \Tint)$, within the
recorded zero-bias beam triggers, $N(T_\text{beam})$, to obtain the
interaction trigger probability
\begin{equation}
  P_\text{\Tint} = \frac{N(\Tbeam \wedge \Tint)}{N(\Tbeam)}.
\label{eq:ptint}
\end{equation}
The interaction probability in the carbon target is then obtained by
correcting for out-of-target interactions via
\begin{equation}
  P_\text{int} = \frac{P_{\Tint}^\text{I} - P_{\Tint}^\text{R}}{1 - P_{\Tint}^\text{R}},
\label{eq:pint}
\end{equation}
where the trigger probabilities measured with the target removed are
denoted by a superscript R and the ones with the target inserted with a
superscript I.

The interaction trigger cross section is given by
\begin{equation}
  \sigma_\text{trig} = \frac{m_A}{L\,\rho\,N_\text{A}} \ln\left(\frac{1}{1 - P_\text{int}} \right),
\label{eq:sigma_trig_anal}
\end{equation}
where $N_\text{A}$ is Avogadro's number and $\rho$, $A$ and $L$ are
the density, molar mass and length of the target, respectively.  The
logarithmic term, $\ln\left(\frac{1}{1 - P_\text{int}}\right) =
P_\text{int} + \frac{1}{2}P_\text{int}^2 + \frac{1}{3}P_\text{int}^3 +
\cdots$, accounts for the exponential attenuation of the beam inside
the target.

The experimentally accessible interaction trigger cross section can be
related to the production cross section by correcting for the residual
contributions to $\sigma_\text{trig}$ originating from elastic
and quasi-elastic scattering ($\sigma_\text{el}$ and
$\sigma_\text{qe}$).  Furthermore, a correction for inelastic
interactions, to which the interaction trigger is not sensitive to, is
needed
\begin{equation}
  \sigma_\text{prod} =
    \left(\sigma_\text{trig} - \sigma_\text{el}\, f_\text{el} - \sigma_\text{qe}\,f_\text{qe}\right) \frac{1}{f_\text{prod}},
\label{eq:prodFromTrig}
\end{equation}
where $f_\text{el}$, $f_\text{qe}$, and $f_\text{prod}$ are the
fractions of elastic, quasi-elastic, and production events that lead
to an \Tint trigger.  $f_\text{el}$ and $f_\text{qe}$ thus
give the fraction of false-positive interaction triggers from
(quasi-)elastic interactions and $1-f_\text{prod}$ is the fraction of
false-negative production interactions.

\begin{table}
\centering
\caption{List of correction factors used in \cref{eq:prodFromTrig} to
  convert the trigger cross section to the production cross section.
  The quoted uncertainties are systematic uncertainties compared to
  which the statistical uncertainties are negligible. The
  fractions $f$ are given for a trigger radius of 0.9 and
  0.6\,cm (see text) for the beam energies of 158 and 350\,\GeVc,
  respectively.}
\label{tab:xscorr}
\sisetup{
table-number-alignment=center,
separate-uncertainty=true,
table-figures-integer = 1,
table-figures-decimal = 2}
\medskip
\begin{tabular}{
  c
  S[separate-uncertainty,table-figures-uncertainty=1]
  S[separate-uncertainty,table-figures-uncertainty=1]
  }
\toprule
beam momentum & \multicolumn{1}{c}{158 \GeVc} & \multicolumn{1}{c}{350 \GeVc} \\
\midrule
$\sigma_\text{el}$/mb     & 35.1 \pm 0.8 & 36.4 \pm 0.9\\
$\sigma_\text{qe}$/mb     & 12.5 \pm 0.3 & 12.0 \pm 0.3\\
\midrule
$f_\text{el}$ & {0.0012}  & {0.00} \\
$f_\text{qe}$ & 0.27 \pm 0.03 & 0.06 \pm 0.02 \\
$f_\text{prod}$ & 0.91 \pm 0.02 & 0.87 \pm 0.02 \\
\bottomrule
\end{tabular}
\end{table}

We derived model predictions for $\sigma_\text{el}$ and
$\sigma_\text{qe}$ by performing a Glauber
calculation~\cite{Glauber:1970jm} of \pimC interactions using a fit to
previously measured cross sections of \pimp~\cite{Patrignani:2016xqp}
as an input.  The resulting cross sections are listed in
\cref{tab:xscorr} where the quoted systematic uncertainty originates
from different assumption on the inelastic screening in \pimC
interactions~\cite{explainInelastic}.

The fractions $f$ depend on the chosen interaction trigger condition.
We use the FTFP\_BERT physics list of
\Geant\,4.9.4.p01~\cite{Agostinelli:2002hh} to estimate $f_\text{el}$
and $f_\text{qe}$.  For systematic checks, we assumed that the angular
distribution of quasi-elastic scattering in \pimC is very similar to
free pion-nucleon scattering and can thus be modeled using the elastic
slope $B_\text{ela}$ in \pimp scattering.  $f_\text{prod}$ was
estimated by generating interactions with the hadronic models
\textsc{Fluka}\,2011.2.9~\cite{Ferrari:2005zk},
\EposLong~\cite{Pierog:2006qv},
\textsc{QGSJet01}~\cite{Kalmykov:1997te},
\QGSJetOldLong~\cite{Ostapchenko:2004ss},
\VenusLong~\cite{Werner:1993uh}, \SibyllLong~\cite{Ahn:2009wx}
\UrqmdLong~\cite{Bleicher:1999xi,Uzhinsky:2011ir} using the
\texttt{INTTEST} mode of \Corsika~\cite{Heck:1998vt}.  The arithmetic
mean of the different predictions of $f_\text{prod}$ is used to
correct the data, and the maximum and minimum values as an estimate of
the systematic uncertainty.

In previous cross-section analyses within \NASixtyOne we used the
absence of a signal in the S4 counter to define an interaction.  But,
as can be seen in \cref{pic:fractions} in the appendix, especially for
the 350\,\GeVc data studied in this paper, the radius of 1\,cm of the
S4 would lead to a large model-dependent correction for
$\sigma_\text{prod}$ with \cref{eq:prodFromTrig}.  Therefore, we
instead use the GTPC to define an offline interaction trigger by
requiring the absence of a track within a radius $r_\text{trig}$ from
the beam extrapolation at 3.7\,m downstream of the target (i.e.\ at
the $z$-position of the S4 plane, located before the GTPC).  We choose
the trigger radius which minimizes the quadratic sum of statistical and
systematic uncertainty of the measured cross section leading to an
$r_\text{trig}$ of 0.9 and 0.6\,cm for beam energies of 158 and
350\,\GeVc respectively, see Ref.~\cite{MichaelThesis}.  The analysis
is performed on the zero-bias beam trigger data with
$4.8\,(2.0)\,{\times}10^5$ and $6.5\,(2.2)\,{\times}10^5$ events at
158 and 350\,\GeVc, respectively, where the number in brackets refers
to the data taken with a removed target.

With the above definition of the offline interaction trigger, we find
an interaction probability of
\begin{equation}
  P_\text{int} =  0.0293 \pm 0.0003 \, (\text{stat.}),
  \quad
  p_\text{beam} = 158\,\GeVc
\end{equation}
and
\begin{equation}
  P_\text{int} =  0.0284 \pm 0.0003 \, (\text{stat.}),
  \quad
  p_\text{beam} = 350\,\GeVc
\end{equation}
and the cross section fractions as listed in \cref{tab:xscorr}.

\begin{figure}[t]
\centering
\includegraphics[width=0.7\linewidth]{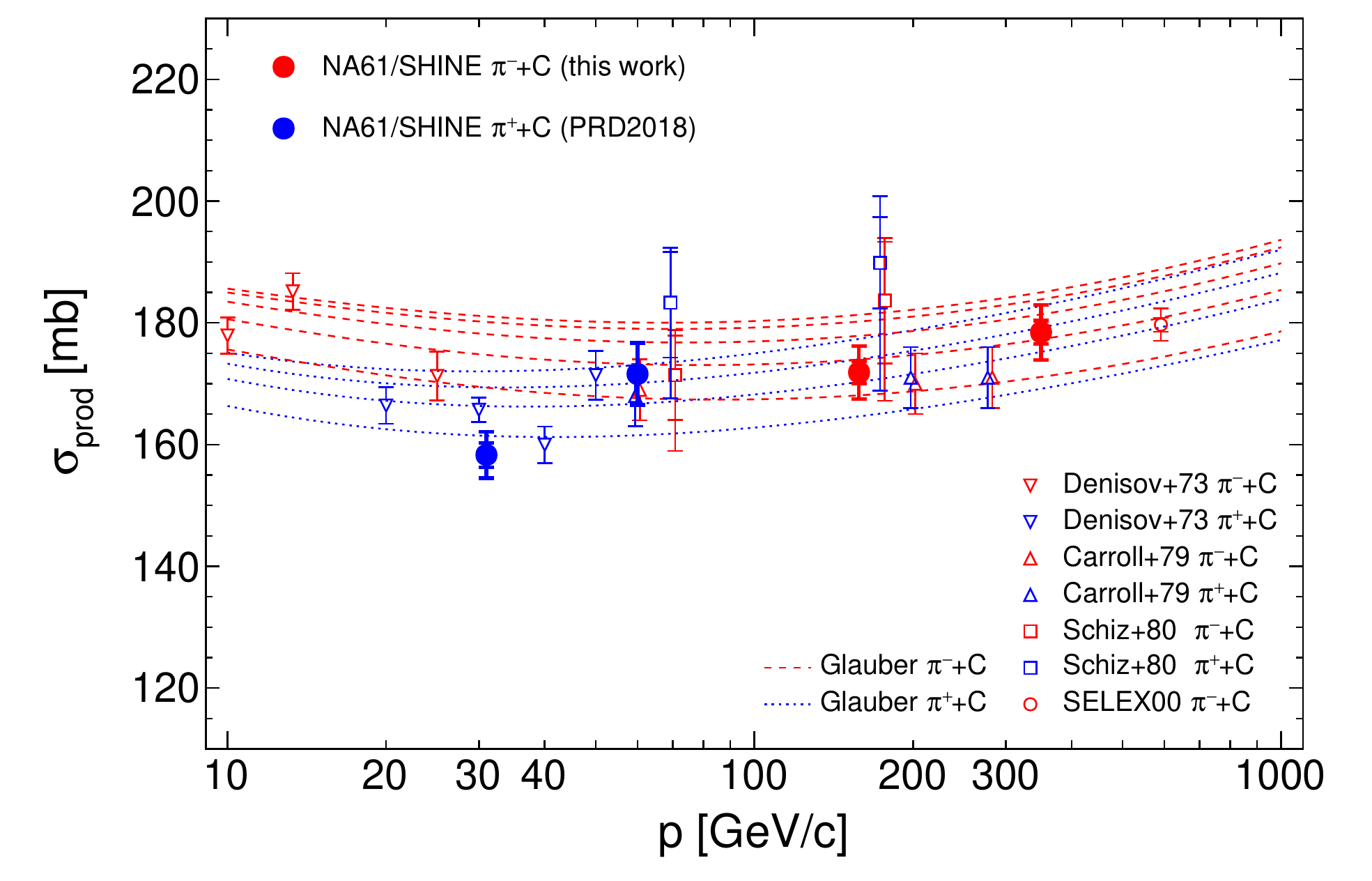}
\caption{Production cross section in \pimC interactions (red) and
  \pipC interactions (blue).  Measurements by \NASixtyOne are shown as
  filled circles, previous measurements from
  \cite{Carroll:1978hc,Denisov:1973zv,Schiz:1979qf,Dersch:1999zg} are
  indicated by open symbols.  Predictions from the Glauber model with
  different values of the inelastic screening are shown as dashed
  lines ($\lambda=0.1$, 0.3, 0.5, 0.7 and 0.9 from top to bottom).}
\label{fig:crosssection}
\end{figure}

Evaluating \cref{eq:prodFromTrig} with the trigger cross section
derived via \cref{eq:sigma_trig_anal} from these measurements as well
as with the correction factors listed in \cref{tab:xscorr}, leads to
our estimates of the production cross section in \pimC interactions of
\begin{equation}
  \sigma_\text{prod} =  \left(172 \pm 2 \, (\text{stat.}) \, \pm 4 \, (\text{sys.})\right) \text{mb},
  \quad
  p_\text{beam} = 158\,\GeVc
\end{equation}
and
\begin{equation}
  \sigma_\text{prod} =  \left(178 \pm 2 \, (\text{stat.}) \, \pm 5 \, (\text{sys.})\right) \text{mb},
  \quad
  p_\text{beam} = 350\,\GeVc.
\end{equation}

The systematic uncertainty includes contributions from the
inefficiency of the GTPC, uncertainties of the detector simulation, and
the uncertainties of the correction factors quoted in
\cref{tab:xscorr}.  The largest contribution of about 3.5\,mb
originates from the model uncertainty of $f_\text{prod}$.  In
\cref{sec:xsapp} we provide the efficiency map corresponding to the
two beam energies such that in the future it will be possible to
recalculate $f_\text{prod}$ with a different set of models, and
possibly smaller systematic uncertainty.

The obtained cross sections are presented in \cref{fig:crosssection}
along with the theoretical prediction using the Glauber
theory~\cite{Glauber:1970jm} for several assumptions on the inelastic
screening parameter $\lambda$~\cite{explainInelastic} (from top to
bottom $\lambda=0.1$, 0.3, 0.5, 0.7 and 0.9) and previous
measurements~\cite{Carroll:1978hc,Denisov:1973zv,Schiz:1979qf,Dersch:1999zg,Aduszkiewicz:2018uts}.\footnote{Here
  we use Glauber predictions of $\sigma_\text{prod}$,
  $\sigma_\text{tot}$ and $\sigma_\text{inel}$ to scale all
  measurements to $\sigma_\text{prod}$.  The measurements of
  $\sigma_\text{tot}$ from Refs.~\cite{Schiz:1979qf} and
  \cite{Dersch:1999zg} are multiplied by
  $(\sigma_\text{prod}/\sigma_\text{tot})_\text{Glauber} \sim 0.77$
  and the measurement of $\sigma_\text{inel}$ from
  Ref.~\cite{Denisov:1973zv} is multiplied by
  $(\sigma_\text{prod}/\sigma_\text{inel})_\text{Glauber} \sim 0.91$.}
As can be seen, the only previous data set on the production cross
section in \pimC interactions from Ref.~\cite{Carroll:1978hc} agrees
well with our measurement and the total uncertainty of the cross
sections presented here matches the statistical precision of this old
measurement (no systematic error was quoted in this reference).
Furthermore, our measurement agrees well with the Glauber predictions
for a broad range of inelastic screening assumptions, but small values
are disfavored and the preferred range is within $0.5 < \lambda <
0.9$.

\section{\vzero analysis}
\label{sec:v0ana}

As a first step of the analysis of particle spectra, we investigate
the production of the neutral weakly-decaying particles \lamb,
\antilamb and \kzeros.  The production spectra of these strange
particles provide important constraints for the tuning of hadronic
interaction models, see e.g. Ref.~\cite{Liu:2002gw}.  Moreover, a good
knowledge of these spectra is also important to distinguish particles
created directly in $\pimC$ interactions from particles originating
from weak decays, see \cref{sec:correction:fd}.

Neutral weakly decaying particles with an average decay length ($c\tau$)
of the order of a few to tens of centimeters can be detected by \NASixtyOne
experiment through their charged decay products.  This type of
particle is traditionally called \vzero, because of its neutral charge
and the V-shaped decay topology.  Although the \vzero itself does not
create a track in the TPCs, the products of its decay may do, allowing
us to reconstruct the position of the decay vertex and, by using the
momenta of the daughter tracks, reconstruct the properties of the
parent particle.

In~\cref{tab:vzero:part} we list the three \vzero particles studied in
this work together with the branching ratio of the decay channels
investigated here.
\begin{table}[!t]
  \begin{center}
    \caption{List of the \vzero particles measured in this work,
      together with their mass and decay length.  The last two columns
      give the decay channel used for the reconstruction as well as
      its branching ratio~\cite{Patrignani:2016xqp}.  }\medskip
    \begin{tabular}{cccr@{+}lc}
    \toprule
      \vzero  & mass [GeV/$c^2$] & $c\tau$ [cm]& \multicolumn{2}{c}{decay} & branching ratio \\
      \midrule
      \lamb            & 1.1157 & 7.89 & \proton   & $\uppi^-$      & 63.9$\%$ \\
      \antilamb        & 1.1157 & 7.89 & $\uppi^+$ & \antiproton  & 63.9$\%$ \\
      \kzeros          & 0.4976 & 2.68 & $\uppi^+$ & $\uppi^-$    & 69.2$\%$ \\
      \bottomrule
    \end{tabular}
    \label{tab:vzero:part}
  \end{center}
\end{table}
\vzero candidates are selected by calculating the distance of closest
approach (dca) for each combination of one positively and one negatively charged track.
To assure a good momentum resolution, both tracks should in total have more
than 30 clusters each and more than 15 clusters in the VTPCs.
Each combination with a $\text{dca}<2$\,cm downstream of the main
vertex is re-fitted under a common vertex hypothesis.  To increase the
signal-to-background ratio, the reconstructed \vzero momentum is then
extrapolated to the vertex plane and the radial impact parameter, $b_r
= \sqrt{\smash[b]{(b_x/2)^2 + b_y^2}}$, is required to be $\leq2$\,cm.
Here $b_x$ and $b_y$ denote the coordinates of the impact point in the
target plane with respect to the main vertex and the factor
$\nicefrac{1}{2}$ accounts for the fact that the resolution in the
$xz$ bending plane is approximately twice as large as in the $yz$-plane.
The efficiency of this cut is $\geq90\%$ in all of the phase space
bins studied here.

A further reduction of the background is possible by selecting events
with a reconstructed decay vertex far from the target, as the
background arises mostly from combinations involving non-\vzero tracks
from the main vertex.  A large distance cut, $d_\text{min}$, improves
the purity of the sample, but also diminishes the selection efficiency,
$\varepsilon = \exp(-\gamma c \tau / d_\text{min})$ for an ideal
detector.  Therefore an optimal selection distance that minimizes the
uncertainty of the extracted signal is chosen, depending on the \vzero
particle type and momentum as described in Ref.~\cite{RaulPhD}.

\begin{figure}[!t]
\def\vheight{0.32}
\centering
\begin{overpic}[clip,rviewport=0.02 0 0.98 1,height=\vheight\textheight]{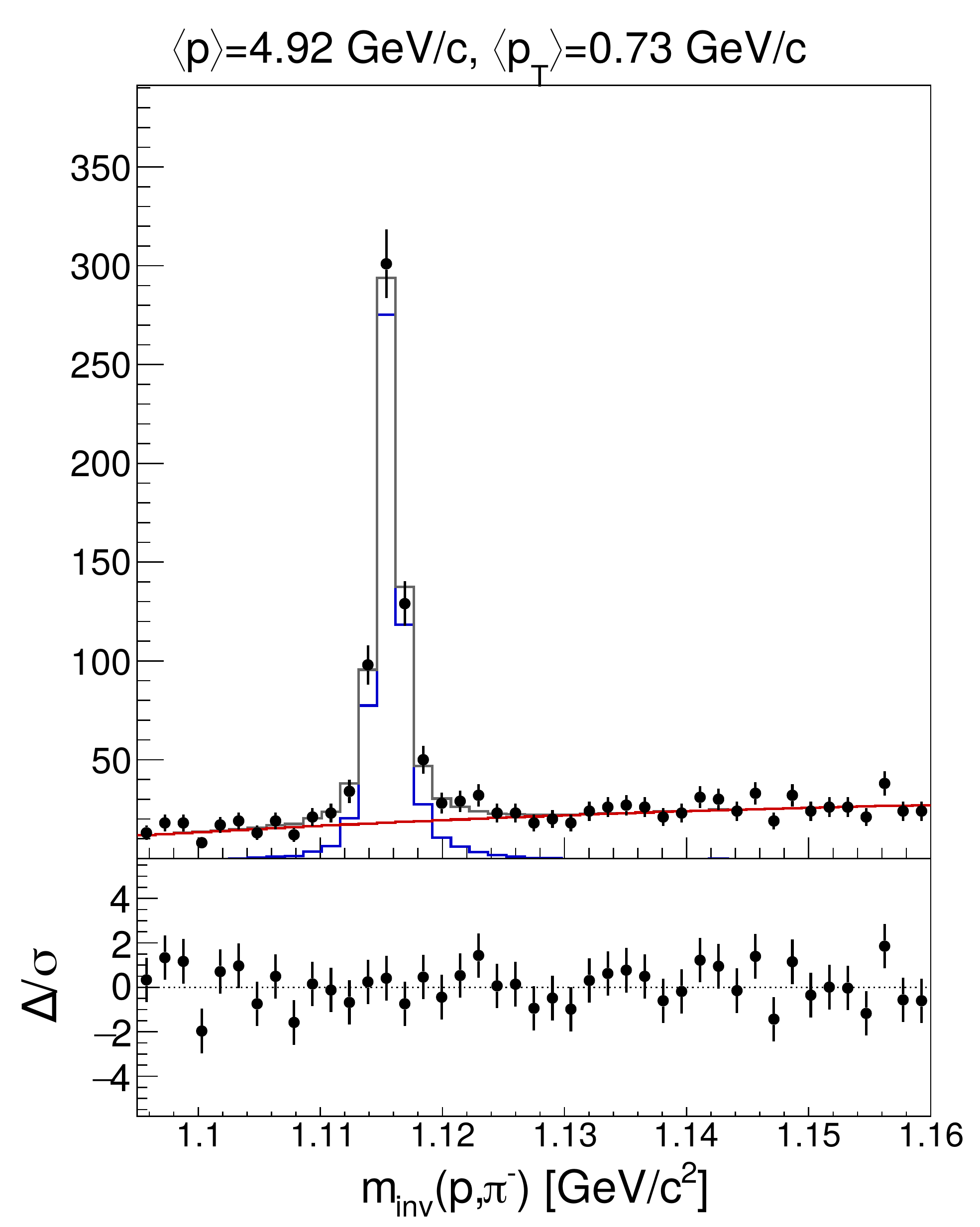}
  \put(16,86){\lamb}
  \put(-2,50){\rotatebox{90}{\scalebox{0.8}{entries}}}
\end{overpic}
\begin{overpic}[clip,rviewport=0.07 0 0.98 1,height=\vheight\textheight]{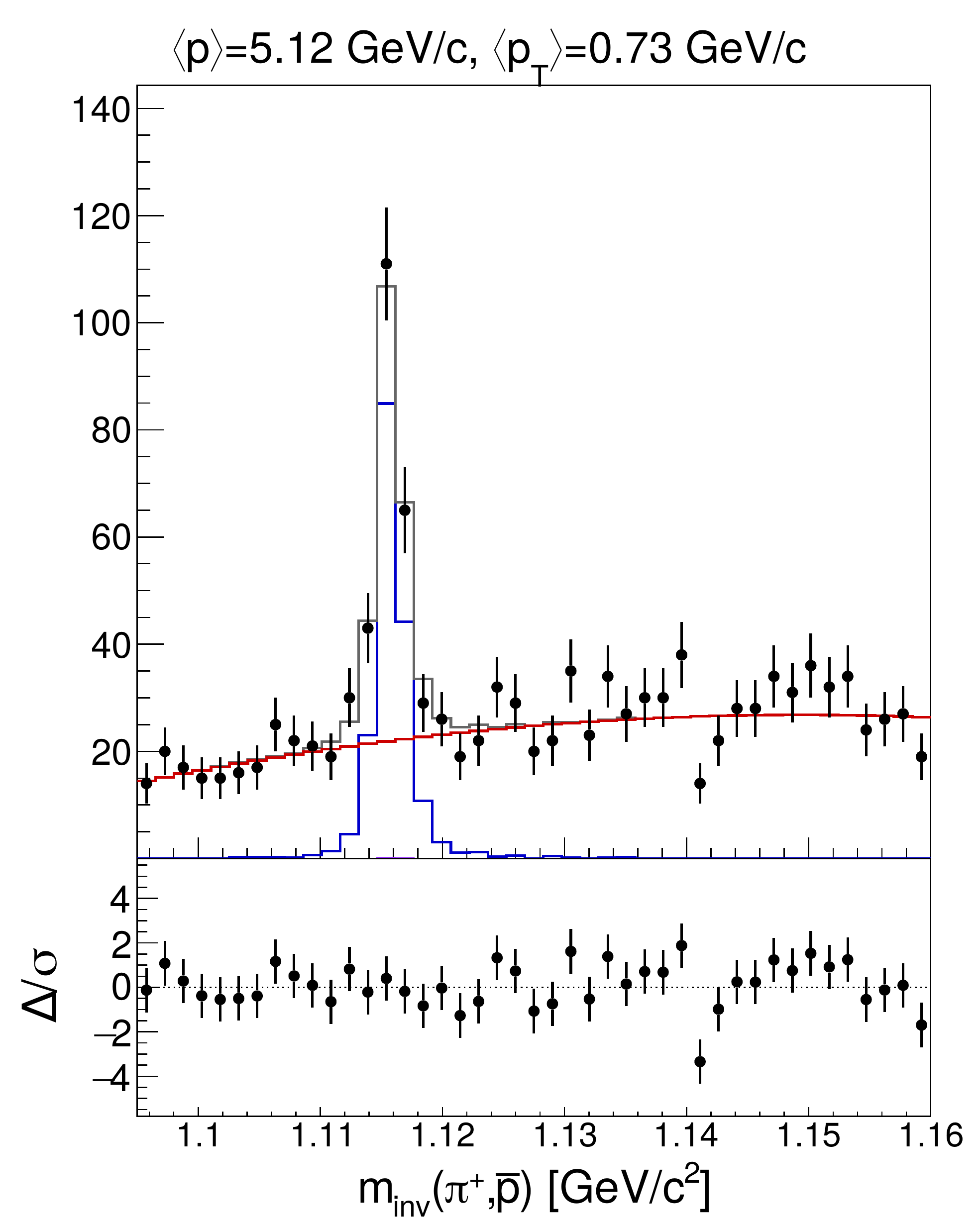}
  \put(13,86){\antilamb}
\end{overpic}
\begin{overpic}[clip,rviewport=0.07 0 0.98 1,height=\vheight\textheight]{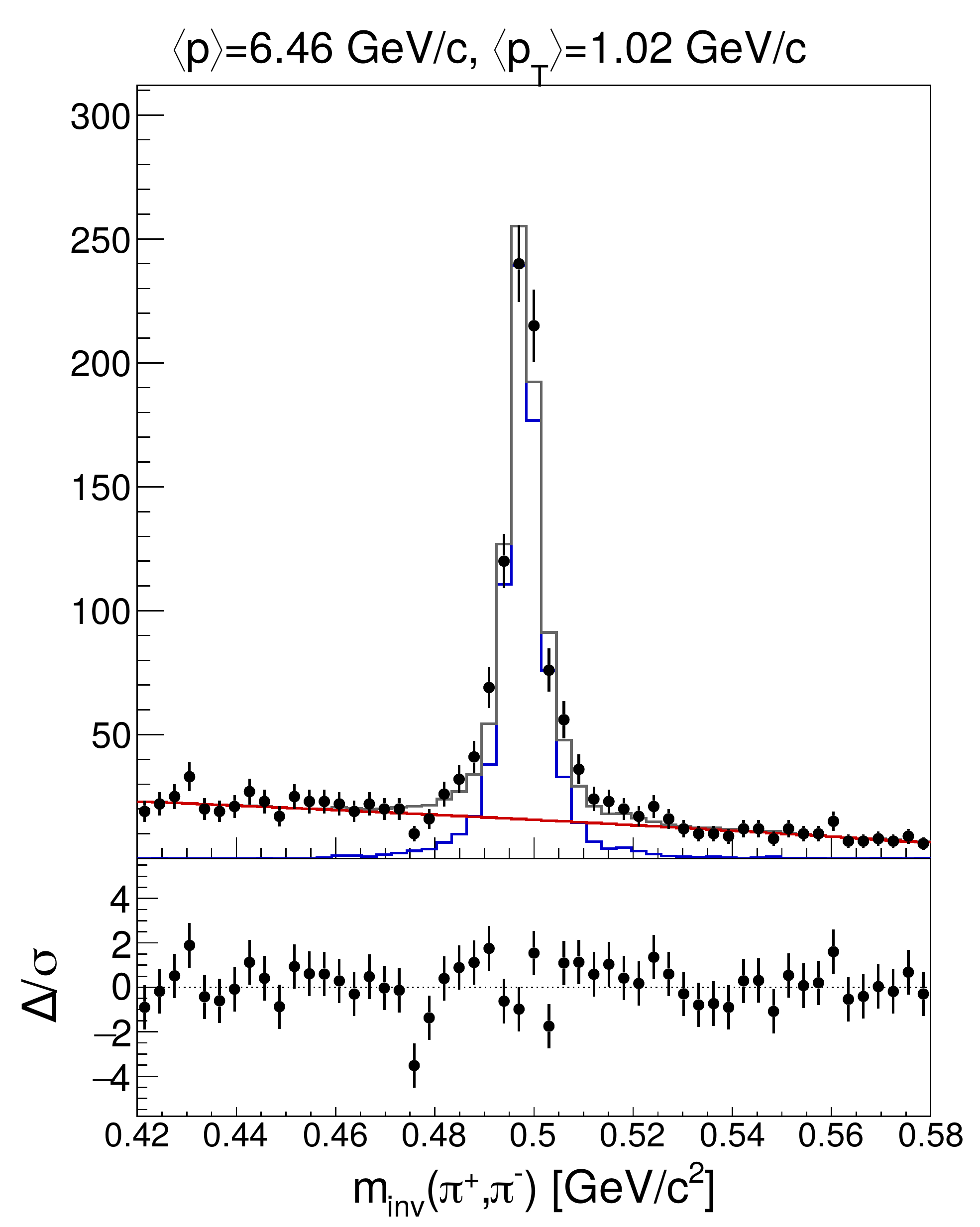}
  \put(13,86){\kzeros}
\end{overpic}
\caption{Examples of the fitted \minv distributions for the 158\,\GeVc
  data set with target inserted for \lamb (left), \antilamb (middle)
  and \kzeros (right).  The black markers show the measured \minv
  distributions.  The curves show the results of the fit with the
  signal in blue, the background in red and the total in gray.  On the
  bottom of each plot we show the residual distributions, i.e.\ the
  difference $\Delta$ between the observed and fitted number of entries in
  units of the uncertainties $\sigma$ of the observed number.  The
  $\langle\pp\rangle$ and $\langle\pT\rangle$ of the phase space bin
  are indicated on the top of each panel.}
\label{fig:vzero:signal:dist:158:in}
\end{figure}

For the measurement of the multiplicity of \vzero particles, we study
the distribution of the invariant mass \minv of combinations of candidate
tracks,
\begin{equation}
  \minv^2 = (\mathbf{p_+} + \mathbf{p_-})^2 =  (m_+^2 + m_-^2)\,c^4 + 2\,(E_+E_--\vec{p}_+\vec{p}_-\,c^2),
  \label{eq:minv}
\end{equation}
in bins of \pp and \pT (cf.\ \cref{sec:binning}). Here the subscripts
$+$ and $-$ refer to the positively and negatively charged daughter
particles,  $\mathbf{p_\pm}$ denote their four-momenta and $\vec{p}_\pm$ the reconstructed three-momenta of
candidate tracks.  \cref{eq:minv} is evaluated for three mass
combinations $m_\pm$ corresponding to the main \lamb, \antilamb and
\kzeros decays listed in \cref{tab:vzero:part} and the energies
$E_\pm$ are calculated accordingly via $E_\pm =
\sqrt{\smash[b]{m_\pm^2c^4 + |\vec{p}_\pm|^2c^2}}$.  Examples of
invariant mass distributions are shown in
\cref{fig:vzero:signal:dist:158:in}.  As can be seen, the ``real''
combinations peak around the mass of the \vzero whereas the background
of unrelated track combinations exhibits a broad and nearly flat
distribution.

The number of \vzeros is extracted in a Poissonian likelihood
fit~\cite{Baker:1983tu} describing the invariant mass distribution as
the sum of the signal template and the background function.  For this
purpose, the signal distribution is modeled using the shape of the
reconstructed \vzero as predicted by the detector simulation and for
the background distribution a polynomial function is used.  We found
that a second-degree polynomial provides a satisfactory description of
the data in the chosen \minv interval.  A third-degree polynomial was
used to estimate the systematic uncertainties of the background shape
description (see~\cref{sec:spec:syst}).  These signal and background
templates are illustrated as blue and red histograms in
\cref{fig:vzero:signal:dist:158:in}.  The number of \vzeros produced
in the fitted \pp-\pT bin is then inferred from the normalization of
the fitted distributions.

\section{\dedx analysis}
\label{sec:dedx}

The spectra of identified particles produced at the main vertex are
obtained by extracting the average particle yields in each kinematic bin
from the measured distributions of the energy loss of tracks in the
TPCs.  For this purpose, we calculate the truncated
mean~\cite{Blum:1993nw}, $\mdedx_j = \nicefrac{2}{N_j}
\sum_{i=1}^{N_j/2} \dedx_{ij}$, of the charges $\dedx_{ij}$ of the
50\% of clusters with the lowest charge among the $N_j$ clusters
detected along each particle track $j$.  As in the Bethe
formula~\cite{Bethe:1930ku}, \mdedx depends on $p/m = \beta \gamma$
and can thus be used to identify particles of different masses $m$ at
a particular momentum $p$.  The two-dimensional distribution of energy
deposit and momenta of selected main-vertex tracks at a beam momentum
of 158\,\GeVc are displayed in \cref{fig:dedx:bb158} along with a
Bethe-inspired parametrization of the mean energy deposit for the
particles considered here, i.e.\ electrons, pions, kaons, protons, and
deuterons and their anti-particles.

\dedxfig{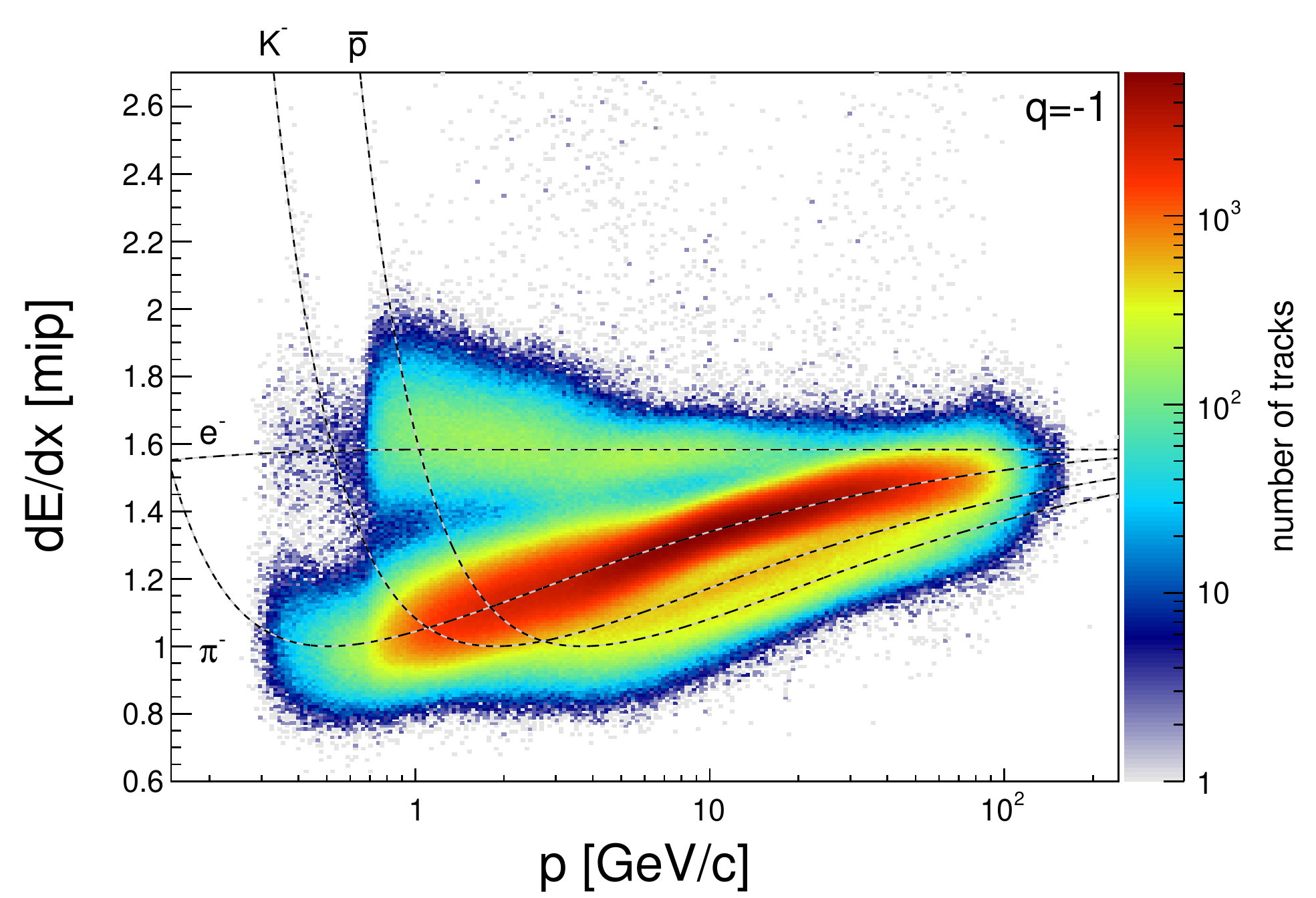}{t}

\begin{figure}[!t]
\centering
\begin{overpic}[clip,rviewport=0 0 1 0.95,width=0.49\textwidth]{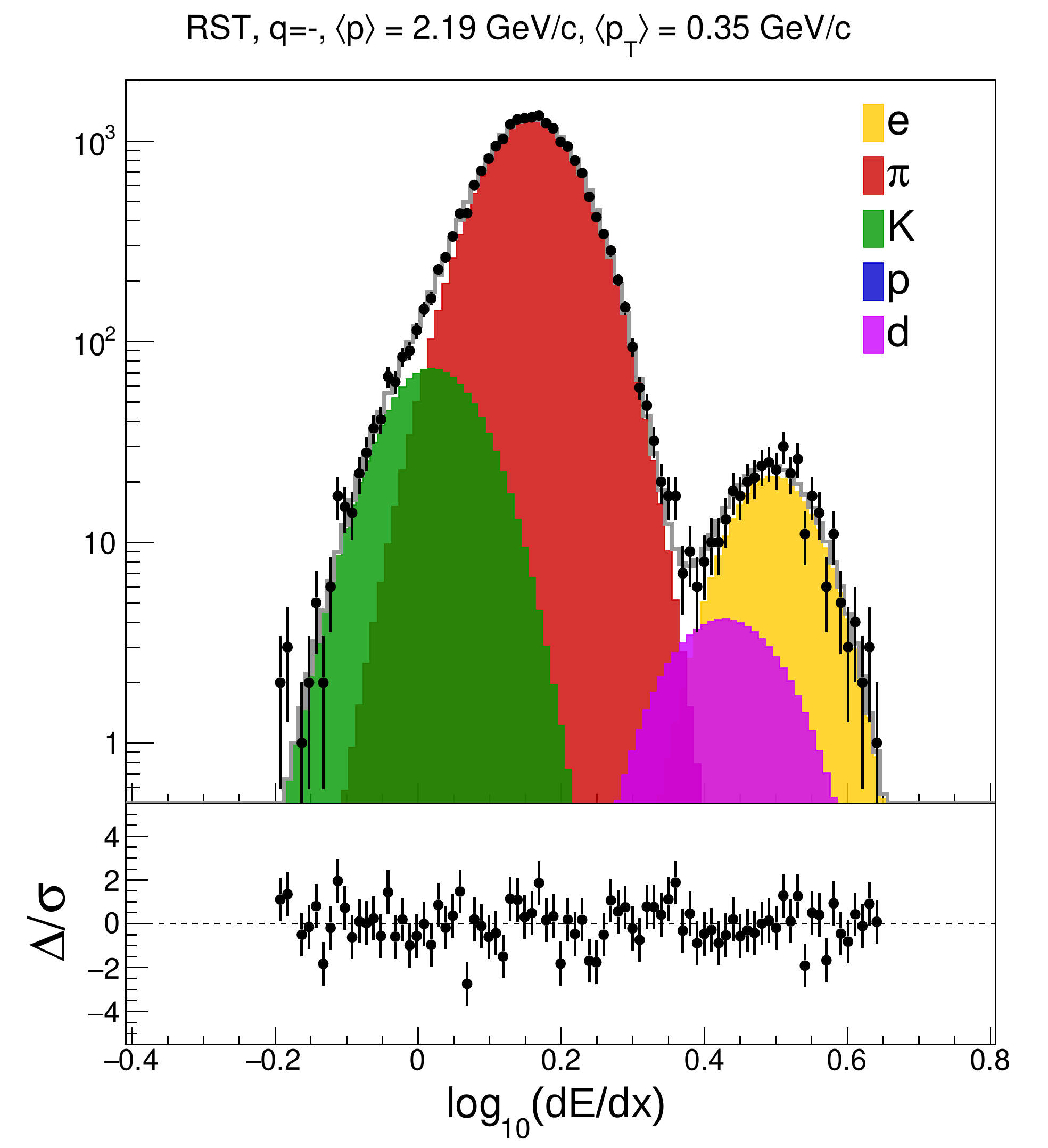}
  \put(16,91){$q=-1$}
  \put(2,50){\rotatebox{90}{\scalebox{1}{entries}}}
\end{overpic}
\begin{overpic}[clip,rviewport=0 0 1 0.95,width=0.49\textwidth]{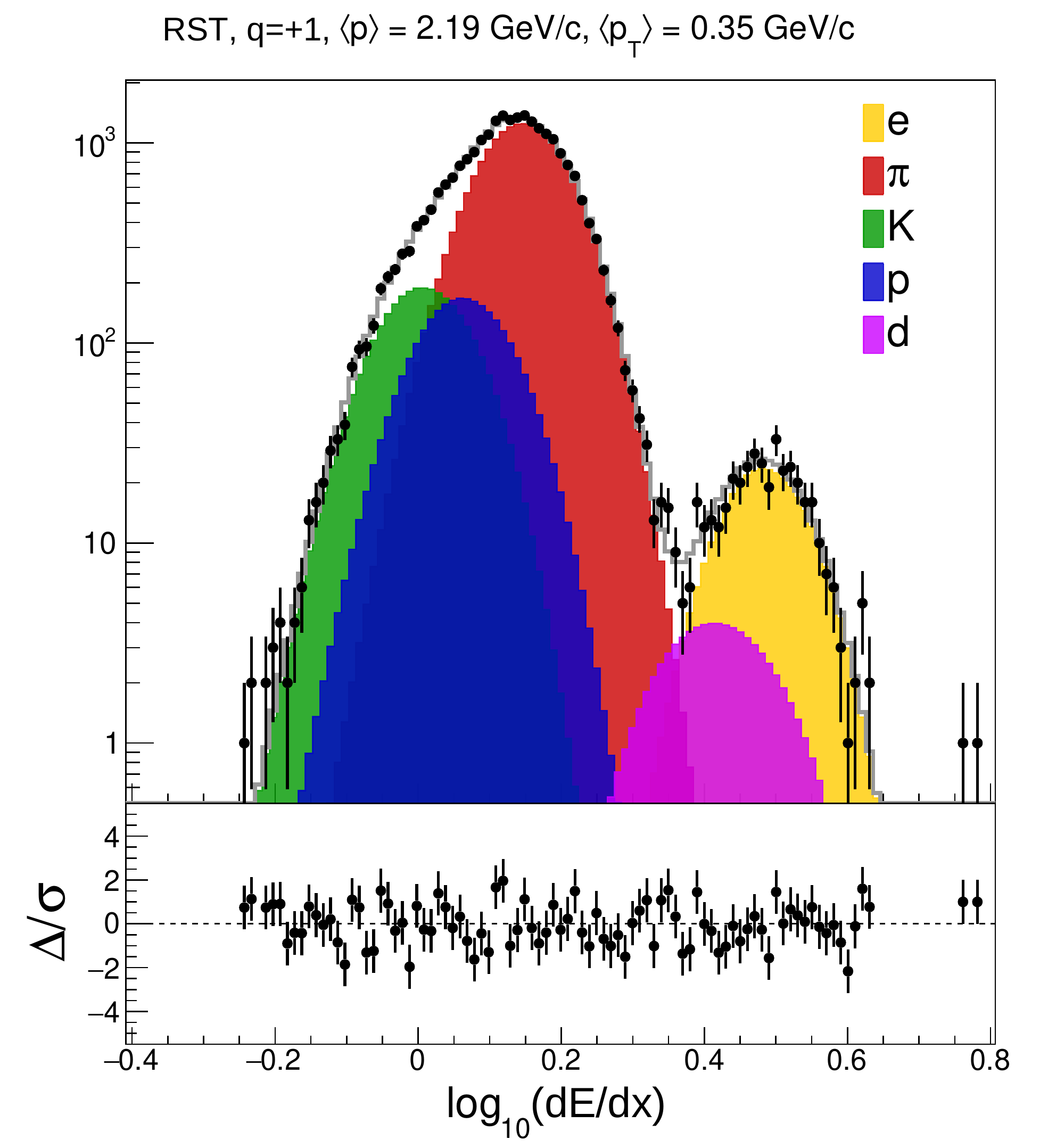}
  \put(16,91){$q=+1$}
  \put(2,50){\rotatebox{90}{\scalebox{1}{entries}}}
\end{overpic}
\caption{Example of the \dedx distributions for one phase space bin
  ($\langle p \rangle = 2.19\,\GeVc$ and $\langle p_\text{T}\rangle =
  0.35\,\GeVc$) of the 158\,\GeVc data set.  The black markers show
  the measured distributions and the colored areas show the result of
  the \dedx fit for different particles.  Negatively and positively
  charged particles are shown on the left and right, respectively.
  Fit residuals are shown in the lower panels in units of the
  statistical uncertainty of the data distribution.}
\label{fig:hadron:dedx}
\end{figure}

An example of the distribution of the \mdedx of tracks in a particular
\pp and \pT bin is shown in \cref{fig:hadron:dedx} for negatively and
positively charged particles.  The production yields of different
particle types are determined by fitting the distribution with the sum
of \mdedx-templates for electrons, pions, kaons, protons, and
deuterons.  The shapes of these templates are based on previous studies
from \NAFortyNine and
\NASixtyOne~\cite{Marco_fit,VeresPhD,NA61SHINE:2017fne} which describe
the distribution of energy deposit of each particle type by the sum of
asymmetric Gaussians taking into account the distribution of $N_j$
(number of clusters per track) and the distribution of momenta within
the bin.

The fitting is performed using a binned maximum-likelihood
method.  In the general case, 10 particle fractions (5 particle types
and 2 charges) and 10 model parameters are taken as free parameters of
the fit.  Most of these model parameters are nuisance parameters that
allow the mean and width of the distributions to stray away from the
global \dedx parametrization.  In that way, residual systematic
offsets of the calibration of cluster charges in different parts of
the detector can be corrected.

As illustrated in \cref{fig:hadron:dedx}, in general the \dedx-fits lead
to a very satisfactory description of the data.  The example
shown here is close to a momentum where the mean \dedx values of
protons and kaons as well as the one of electrons and deuterons
overlap (cf.\ \cref{fig:dedx:bb158}).  In case of a near-complete
degeneracy between the fitting templates of different masses in
certain momentum bins (``Bethe crossings''), the fitted yields of
these ambiguous particles are excluded from the final results.

\section{Derivation of the Particle Spectra}
\label{sec:spectra}

The \vzero and \dedx analyses described in the previous two sections
result in the number $\hat{n}$ of identified tracks (\pions, \kaons,
\proton, \antiproton, \lamb, \antilamb, and \kzeros) in bins of \pp
and \pT for each of the two data taking modes, i.e.\ with the target
inserted and removed.  For the choice of binning, see
\cref{sec:binning}.

Given these measurements, the double-differential particle production
spectra can be computed for each particle as
\begin{equation}
  \frac{1}{N_\text{prod}} \frac{\mathrm{d}^2n}{\mathrm{d}\pp\,\mathrm{d}\pT} =
    \frac{\cmc}{\Delta\pp\,\Delta\pT}\, \frac{\recn_\text{I}-b\,\recn_\text{R}}{\recN_\text{I}-b\,\recN_\text{R}}.
\label{eq:spec}
\end{equation}
This quantity is the track multiplicity per production event in a
certain \pp-\pT bin, sometimes also referred to as \emph{average
  multiplicity}.  Here $\Delta \pp$ and $\Delta \pT$ denote the widths
of the phase-space bin and the transverse component \pT of the
momentum vector $\mathbf p$ is defined with respect to the beam
direction ${\mathbf u}_\text{b}$, $\pT = \sqrt{{\mathbf p}^2 -
  ({\mathbf u}_\text{b}\, {\mathbf p})^2}$.  \cmc is a correction
factor derived from simulations that will be discussed in the next
section.  The indexes I and R refer to target inserted and removed,
respectively, and $\recN$ is the number of selected minimum-bias
interaction-trigger events.  The factor $b$ is the so-called
target-removed factor and it normalizes both target inserted and
removed data set to the same number of beam particles.  Given the
number of events, $N_\Tint$, and the probability of a beam particle to
produce an interaction trigger, $P_\Tint$, the number of beam
particles for a given data set can be estimated as $N_\text{beam} =
N_\Tint/P_\Tint$.  $P_\Tint$ is measured from zero-bias beam triggers,
see \cref{eq:ptint}.  The target-removed factor is given by $b =
N_\text{beam}^\text{I}/N_\text{beam}^\text{R}$ and its value is about
five, corresponding to the time spent in target-removed configuration
during data taking.

\subsection{Correction Factors}
\label{sec:correction}

The selected number of events $\recN$ and the estimated number of
particles $\recn$ in a given phase space bin are biased estimators of
the true number of production events $N_\text{prod}$ and the true
number of produced particles $n$.  These biases are corrected for with
the help of simulations for which it is easy to determine the ratio
of generated and measured spectra,
\begin{equation}
  \cmc = \left(\frac{\genn}{\genN}\right)\big/\left(\frac{\recn}{\recN}\right),
\label{eq:correction:cmc}
\end{equation}
where $n$ and $N_\text{prod}$ are the values from the event generator,
\recn and $N_\Tint$ are obtained after the detector simulation,
event reconstruction, and event and track selection.  Two simulated
data sets generated with different hadronic interaction models
(see~\cref{sec:dataProcAndSel}) are used for systematic studies.

To gain further insights into the different contributions to the
correction factor it is useful to split \cmc into event- and
particle-contribution factors, referred to as $\alpha$ and $\beta$ in the
following,
\begin{equation}
  \cmc = \left( \frac{\recN}{\genN} \right) \big/ \left( \frac{\recn}{\genn} \right) = \alpha / \beta.
\label{eq:correction:cmc:2}
\end{equation}
The $\alpha$ factor is a property of the data set as a whole and thus
depends only on the beam energy, whereas the $\beta$ factor depends
on the phase space bin and on the particle type.  We get $\alpha= 0.872
\pm 0.004$ and $0.732 \pm 0.012$ for beam momenta of 158 and
350\,\GeVc, respectively.  These values were determined from the
arithmetic average of the two simulated data sets, with the upper
uncertainty range corresponding to the result predicted by \EposLong
and the lower range to the one from \QGSJetLong.  These values give
the product of the efficiency of the vertex-$z$ cut
(cf.\ \cref{sec:eventselection}) and the efficiency of the
minimum-bias interaction trigger (cf.\ \cref{sec:experiment}).  The
former amounts to approximately 0.975 with a good agreement between
real and simulated data.  Therefore, $\alpha$ is dominated by the
trigger efficiency and is thus mainly a model-dependent correction,
since the efficiency depends on the fraction of events with a
high-momentum charged particle triggering the S4 scintillator.
Overall, the two hadronic generators used here show a good overall
agreement in their predictions of $\alpha$.  In the following, we will
use the average value for the correction and the difference for
systematic uncertainty.  A slightly larger uncertainty would result if
the standard deviation of efficiencies of all the models shown in
\cref{pic:fractions} of the appendix would be used as an estimate of
the systematic uncertainty.  Note that for the majority of phase-space
bins, the total correction for the trigger efficiency is small, as it
affects both \recN and \genN.  The correction is mostly relevant at
high momenta when the trigger efficiency affects \genN but not \recN
since the conservation of the beam momentum does not allow for the
simultaneous presence of a high-momentum particle in the TPCs and
another high-momentum particle hitting the S4 scintillator.

The $\beta$ correction factors as a function of \pp and \pT are given
by the ratio of the generated and measured number of tracks.  They are
shown in \cref{fig:correction:beta158} for \pions, \kaons, \protonpm,
\lamb, \antilamb, and \kzeros.  Note that the $\beta$ factors for the
\vzeros also include the effect of the \vzero cuts, presented
in~\cref{sec:v0ana}.

\betafig{158}

The geometrical acceptance of the detector is the dominant
contribution to the $\beta$ factor at large \pT.  Most of the overall
structure visible in the $\beta$ plots is due to the acceptance and
reflects the aspect ratio of the TPCs (rectangular in the $xy$ plane)
and the bending of particles in the magnetic field.  At particle
momenta of ${\sim}5$\,\GeVc, the TPCs provide full coverage for
$\pT\leq0.2$\,\GeVc increasing up to $\pT\leq1$\,\GeVc at
$\sim$30\,\GeVc.  Due to the requirement on the number of clusters in
the different TPCs (cf.\ \cref{sec:trackselection}), the coverage
decreases at higher momenta due to the forward gap between the TPCs
and at lower momenta due to the strong bending of particles in the
magnetic field.  The fiducial acceptance cuts assure that these
effects are taken into account by a purely geometrical correction and
the corresponding efficiency is simply given by the number of fiducial
$\phi$ bins in the map of the ($\phi$, \pp, \pT) acceptance.

Other corrections subsumed in $\beta$ are related to the efficiencies
of the event selection which, as mentioned above, partially cancels
out with the $\alpha$ correction. By construction, the reconstruction
and selection efficiency within the fiducial acceptance is $>90\%$,
but typically $>95\%$.  $\beta$ also includes the effects of
bin migration due to the finite momentum resolution (\genn is counted
in a bin of generated \pp and \pT whereas \recn in a bin of
reconstructed momenta).  In most of the bins, this correction is at
the sub-percent level, as the bin width is much larger than the
momentum resolution.  Only at $\pp\gtrsim 40$~\GeVc this correction
becomes significant but stays $\leq 5\%$ over the whole phase space
studied here.

Finally, $\beta$ also includes corrections for ``feed-down'',
i.e.\ the fact that a fraction of the measured particles are not
produced in the main target interaction, but instead, in processes
like interactions of secondary particles inside the target or in the
detector material, or the decay of unstable particles which are
produced in the main interaction.  Therefore, in contrast to the
corrections discussed before, which are efficiency corrections
($\recn/\genn<1$), the feed-down correction gives rise to
$\recn/\genn>1$.  The particles most affected by feed-down are protons
and anti-protons for which a substantial fraction originates from weak
decays, most prominently the decays of \lamb and \antilamb, leading to
$\beta$ factors $>1$ in \pp-\pT regions where the other
efficiency-type corrections are near unity (see the p$^\pm$ panels in
\cref{fig:correction:beta158}).  Due to the potentially large model
dependence of this correction, it is treated with special care as
detailed in the next section.

\subsection{Feed-down from weak decays}
\label{sec:correction:fd}

According to the two hadronic generators used here, the contribution
of weak decays is negligible for \kaons, but typically several percent
for \pions and up to 20\% for \protonpm, depending on the phase-space
bin.  \kzeros decays are responsible for $\gtrsim 70\%$ of the
feed-down to charged pions and \lamb and \antilamb decays dominate the
feed-down to \proton and \antiproton respectively.  The \lambs and
\kzeros production varies substantially among different hadronic event
generators and therefore the model dependence on the $\beta$
correction for \pions and \protonpm could be very large.  To avoid the
corresponding large systematic uncertainties, we use here the measured
spectra of \lamb, \antilamb, and \kzeros to correct the feed-down
contribution from these particles.

\begin{figure}[!t]
\centering
\begin{overpic}[clip, rviewport=0 0 1 1,width=0.31\textwidth]{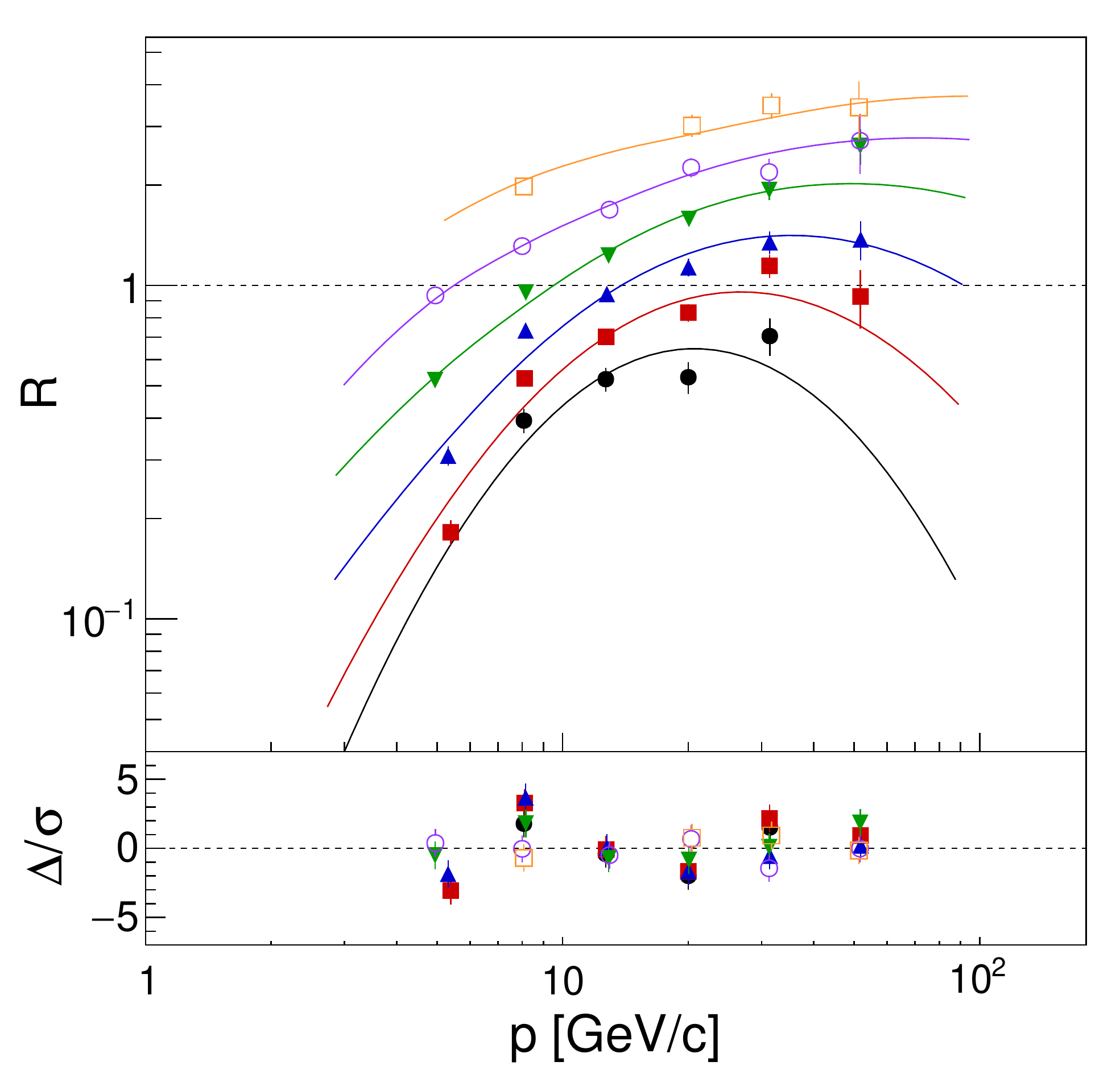}
  \put(18,84){\lamb}
\end{overpic}\quad
\begin{overpic}[clip, rviewport=0 0 1 1,width=0.31\textwidth]{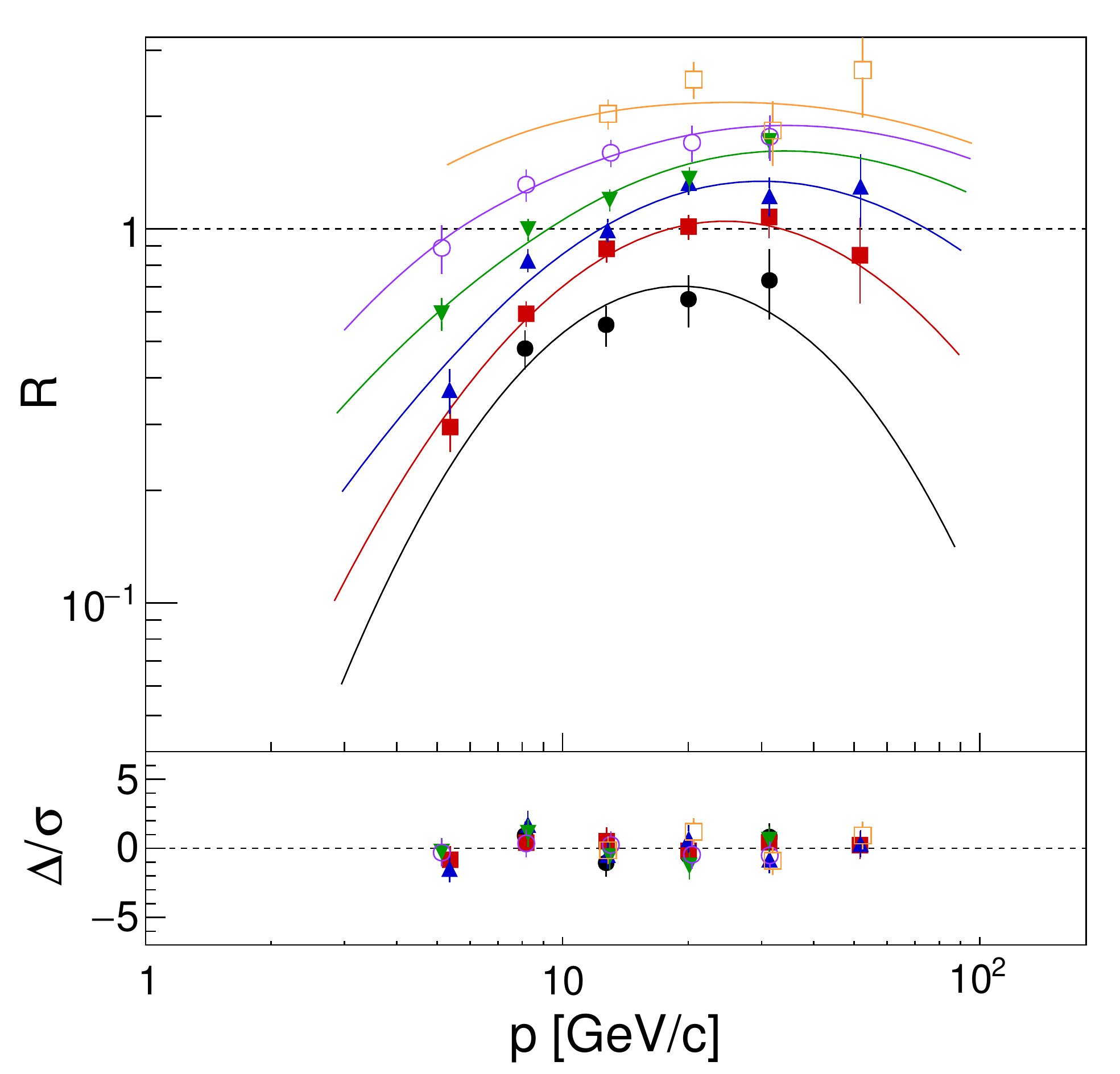}
  \put(18,84){\antilamb}
\end{overpic}\quad
\begin{overpic}[clip, rviewport=0 0 1 1,width=0.31\textwidth]{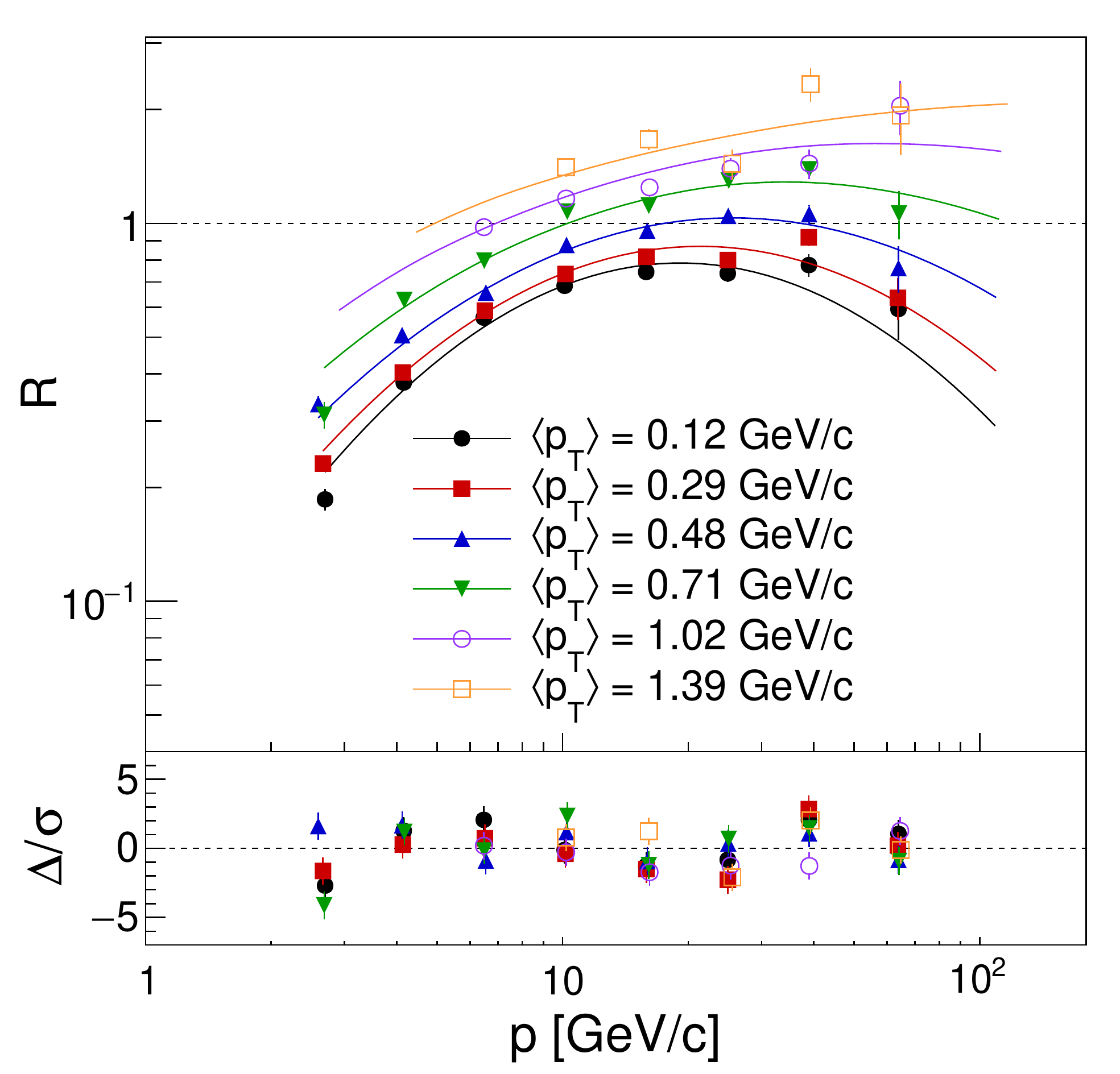}
  \put(18,84){\kzeros}
\end{overpic}
\caption{Ratio between the measured and generated spectra, $R$, as a
  function of \pp (\EposLong) for \lamb, \antilamb, and \kzeros
  particles.  The markers show different \pT bins and the colored
  lines show the result of the parametrization.  On the bottom of each
  panel the differences $\Delta$ between the observed value of $R$
  and the parametrization divided by its uncertainty $\sigma$ are
  shown.}
\label{fig:correction:beta:ratio}
\end{figure}

The procedure adopted for this purpose is based on a re-weighting of
the simulated particles which are produced from the decay of \lamb,
\antilamb, and \kzeros, see also Ref.~\cite{Abgrall:2015hmv}.  This
weight is determined by the ratio, $R$, between the reconstructed
spectra of these spectra in data and simulation.
In~\cref{fig:correction:beta:ratio} we show examples of the ratios for
the three \vzero particles for the model \EposLong and beam momentum
of 158\,\GeVc.  A fit of $R$ as a function of \pp and \pT is performed
by using a log-normal function in \pp which parameters are
interpolated as a function of \pT by a second-degree polynomial
function.  This parametrization is shown as colored lines
in~\cref{fig:correction:beta:ratio}.  The weight given for the
simulated particles is then computed by using the parametrization of
the ratio between measured and generated spectra. Note that since $R$
is calculated from the ratio of {\itshape reconstructed} \vzero
spectra without feed-down correction from heavier baryons such as
$\Omega$ and $\Xi$, it also adjusts the feed-down contribution from
\vzeros produced in decays of these baryons.

Concerning the weak feed-down for the particle spectra of \vzeros, the
correction is negligible for \kzeros, but can reach up to 25\% for
\lamb and \antilamb, in which case the decaying particles are mostly
charged and neutral $\Xi$ baryons as well as $\Omega^\pm$.  Since we
did not measure the production spectra of these particles, this
correction is fully model-dependent with correspondingly larger
uncertainties than achieved for \pions and \protonpm from the primary
interaction.

\subsection[Integration over \pT]{\boldmath Integration over \pT}
\label{sec:spec:int}

For some purposes it is useful to calculate the single-differential
spectra, $1/N\,\text{d}n/\text{d}\pp$, by integrating the
double-differential spectra given by~\cref{eq:spec} over \pT.  Since
the measured spectra do not cover the full \pT range, an extrapolation
is needed to perform this integration.  We therefore fit the
double-differential spectra as a function of \pT for each \pp bin and
then use the integral of the fitted function to extrapolate the
measured spectra to full phase space in \pT.

We found that a Gaussian function convoluted with an exponential one
gives a very satisfactory description of the spectra. For the purpose
of evaluating the systematic uncertainty of the extrapolation we also
used an exponential in transverse mass, $m_\text{T} = \sqrt{\smash[b]{\pT^2\,c^2 +
  m^2\,c^4}}$, which describes the data equally well. The \pT-integrated
spectra are computed by summing the measured spectra over all the
available \pT bins and adding the integral of the fitted function over
the remaining \pT range.  Single-differential spectra are calculated
only for \pp bins where the fraction of the extrapolation is smaller
than 5\% of the total for the charged hadrons and 20\% for the \vzero
particles.

\subsection{Uncertainties}
\label{sec:spec:uncert}
\sysfig{158}
\subsubsection{Statistical uncertainties}
\label{sec:spec:stat}
The statistical uncertainties of the measured spectra, \cref{eq:spec},
are dominated by the statistical uncertainties of $\recn_\text{I}$
originating from the \dedx-fit in case of charged hadrons and from the
invariant-mass-fit for \vzero particles.  These are in general larger
than the simple Poisson uncertainties,
$\sigma(\recn_\text{I})\gtrsim\sqrt{\recn_\text{I}}$.  Since the
number of target-removed tracks is substantially smaller than the
target-inserted ones, the statistical uncertainties on $b\,n^\text{R}$
can be neglected.  Furthermore, the statistical uncertainty of \cmc
due to the limited number of simulated tracks is taken into account,
but it constitutes only a minor contribution to the overall
uncertainty.
\subsubsection{Systematic uncertainties}
\label{sec:spec:syst}
The contributions of different sources to the overall systematic
uncertainty of the charged hadron and \vzero spectra as a function of
\pp are displayed in \cref{fig:syst158} and are listed in the
following with the corresponding label in the figure given in
brackets.
\paragraph{Modeling of the energy loss distributions (\dedx)}
The uncertainties related to the \dedx model used for the fit to
establish the fractions of \pions, \kaons, and \protonpm are estimated
by repeating the fit with different configurations of the model of the
\dedx distribution including different assumptions of the center
position of the Gaussian constraints on the mean value of the \dedx of
the six particles (moved separately by two standard deviations in both
directions) and different assumptions of the shape and signal
dependence of the distribution (see Ref.~\cite{RaulPhD} for more
details).  The resulting spectra obtained with these model variations
are compared to the standard spectra and the size of the uncertainties
is taken as the differences between the extreme cases and the standard
one.
\paragraph{Minimum Bias Interaction Trigger Efficiency (T2)}
Distortions of the spectra due to the minimum bias interaction trigger
are corrected for with the \cmc correction factor, see
\cref{eq:correction:cmc:2}.  Since both, the number of events and the
number of tracks are affected by the trigger, the model differences
were estimated by combining both $\alpha$ and $\beta$ factors.  To
compute the systematic uncertainties, we first isolate the
contributions of the trigger efficiency to the correction factor,
$\alpha_{T2}$ and $\beta_{T2}$, and then compute the factor
$\alpha_{T2}/\beta_{T2}$ for the two event generators separately.  The
relative differences between the extreme values of
$\alpha_{T2}/\beta_{T2}$ and the average one are used to define the
relative systematic uncertainties.
\paragraph{Main-Vertex Cut (vtx Z)}
The model dependence of \cmc due to the cut on the $z$ position of the
main vertex in the event selection is evaluated analogously to the
systematics of the interaction trigger.
\paragraph{Feed Down (FD)}
The systematic uncertainties of the feed-down correction are estimated
by comparing the differences between the corrections predicted by the two
different event generators.  The corresponding systematic
uncertainties for the feed down to \vzeros are substantial, but since
the predictions of the feed down to \pions, \kaons, and \protonpm from
\lamb, \antilamb, and \kzeros are constrained to the \vzero data, the
feed-down-related systematics for the charged hadron spectra are small.
\paragraph{Event Topology (RST/WST)}
The data set is subdivided into two statistically independent subsets
based on the sign of the product $q \cos\phi$, where $q$
denotes the charge and $\phi$ is the azimuthal angle (see
\cref{sec:trackselection}).  Since neither the beam nor the target is
polarized, we expect identical spectra for the two data sets.
Differences between the particle spectra derived for the two data sets
are small ($\lesssim 3\%$ in most of the phase space bins), but larger
than the statistical uncertainties.  We interpret these differences as
an evaluation of the residual disagreement between the measured data and
the idealized detector simulation originating from e.g.\ calibration
uncertainties not present in the simulated events and add them in
quadrature to the other contributions.
\paragraph{Selection of \vzero candidates (\vzero sel.)}
The minimum number of clusters required for the daughter tracks of the
\vzero selection is changed from 30 to 20 and both, the signal
extraction and the calculation of \cmc, are repeated.  This cut
variation results in slightly different \vzero spectra and the
differences are conservatively added to the overall systematic
uncertainty of the \vzero spectra.
\paragraph{\minv Background Model (BG)}
The shape of the background used for the \minv fit of \vzero candidates
is changed from a second-degree to a third-degree polynomial.  The
systematic uncertainty related to the background subtraction is then
estimated as the relative difference between the particle multiplicity,
obtained by the new background function, and the standard one.

\def\specfigw{0.99}
\begin{figure}[!p]
\centering
\includegraphics[page=1,width=\specfigw\linewidth]{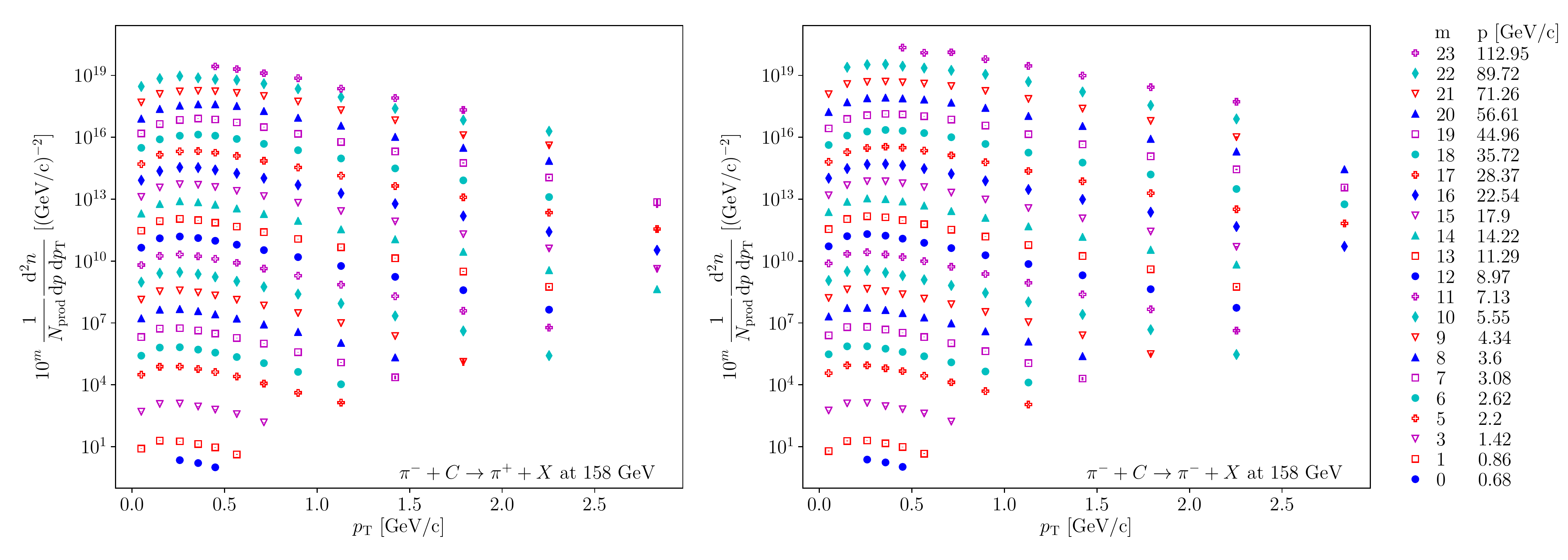}\\
\includegraphics[page=2,width=\specfigw\linewidth]{Plots/all_figures_color}\\
\includegraphics[page=3,width=\specfigw\linewidth]{Plots/all_figures_color}\\
\caption{Production spectra of \pions, \kaons and \protons in \pimC interactions at $p_\text{beam} =
  158$\,\GeVc.  For each bin in momentum \p, the spectrum was
  multiplied by $10^{m}$ with the value of $m$ shown on the right. Error bars show the statistical uncertainties and are most of the times smaller than the marker size.}
\label{fig:dndpdpt158}
\end{figure}
\subsection{Results}
\label{sec:results}
\label{sec:v0results}
\label{sec:dedxresults}
\begin{figure}[!p]
\centering
\includegraphics[page=4,width=\specfigw\linewidth]{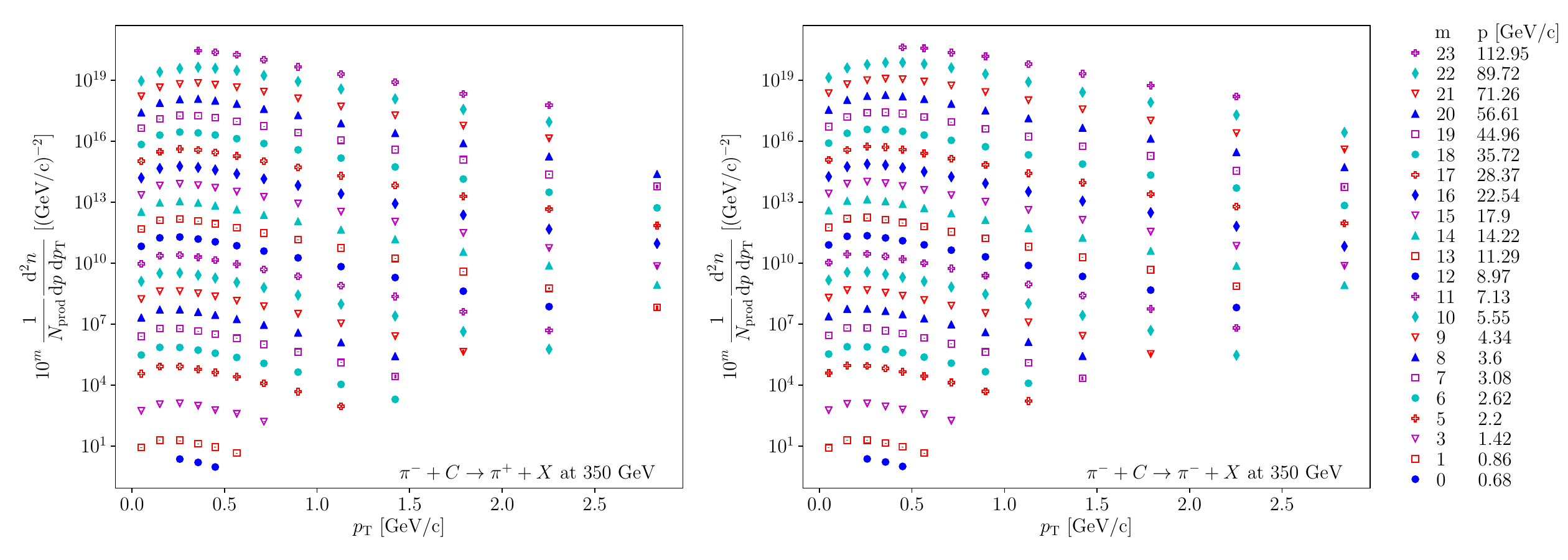}\\
\includegraphics[page=5,width=\specfigw\linewidth]{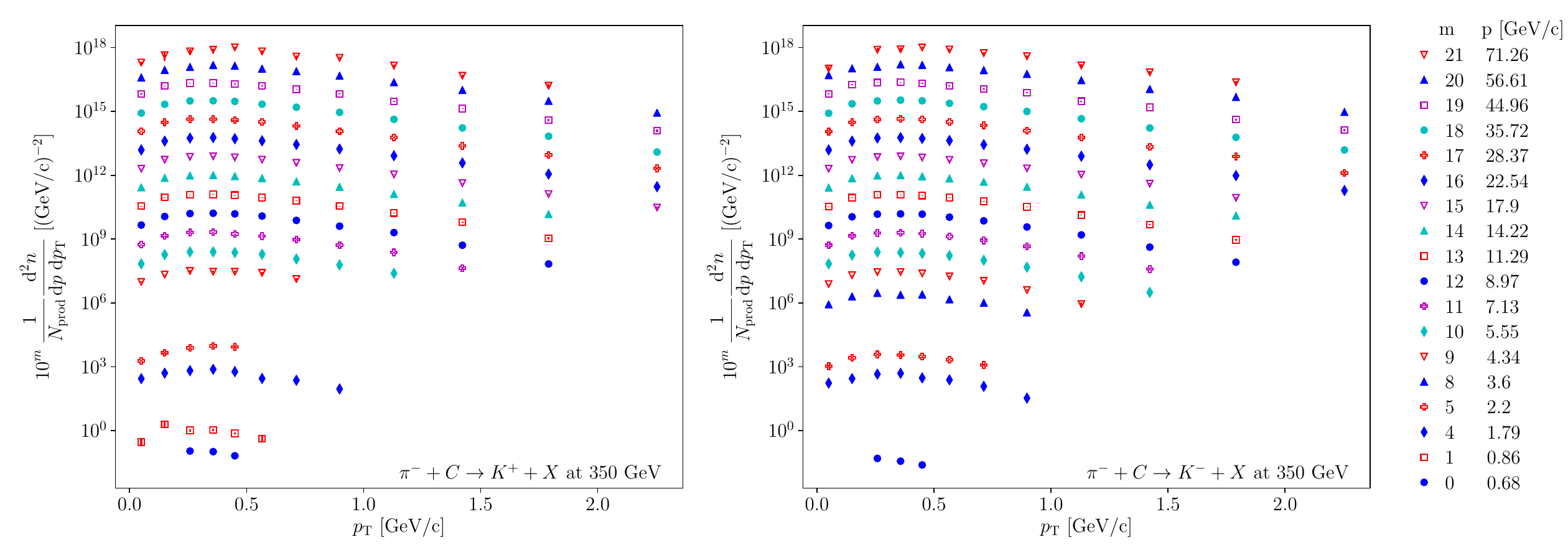}\\
\includegraphics[page=6,width=\specfigw\linewidth]{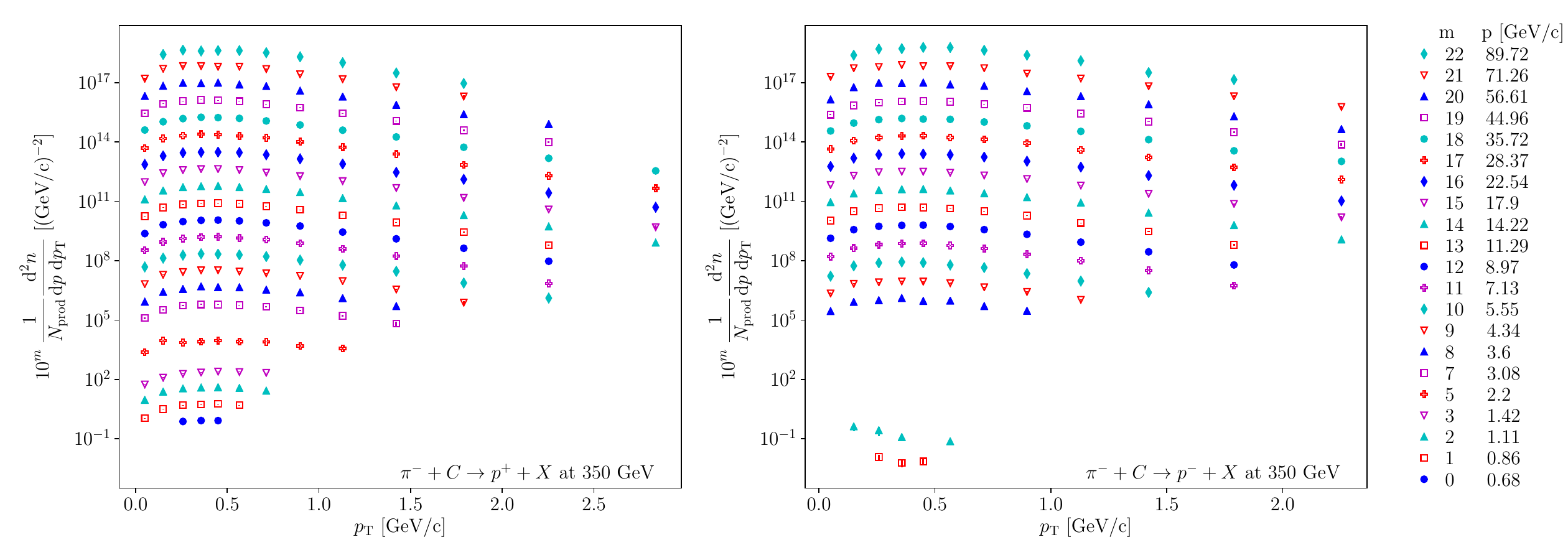}\\
\caption{Production spectra  of \pions, \kaons and \protons in \pimC interactions at $p_\text{beam} =
  350$\,\GeVc.  For each bin in momentum \p, the spectrum was
  multiplied by $10^{m}$ with the value of $m$ shown on the right. Error bars show the statistical uncertainties and are most of the times smaller than the marker size.}
\label{fig:dndpdpt350}
\end{figure}

\begin{figure}[!p]
\centering
\includegraphics[page=7,width=\specfigw\linewidth]{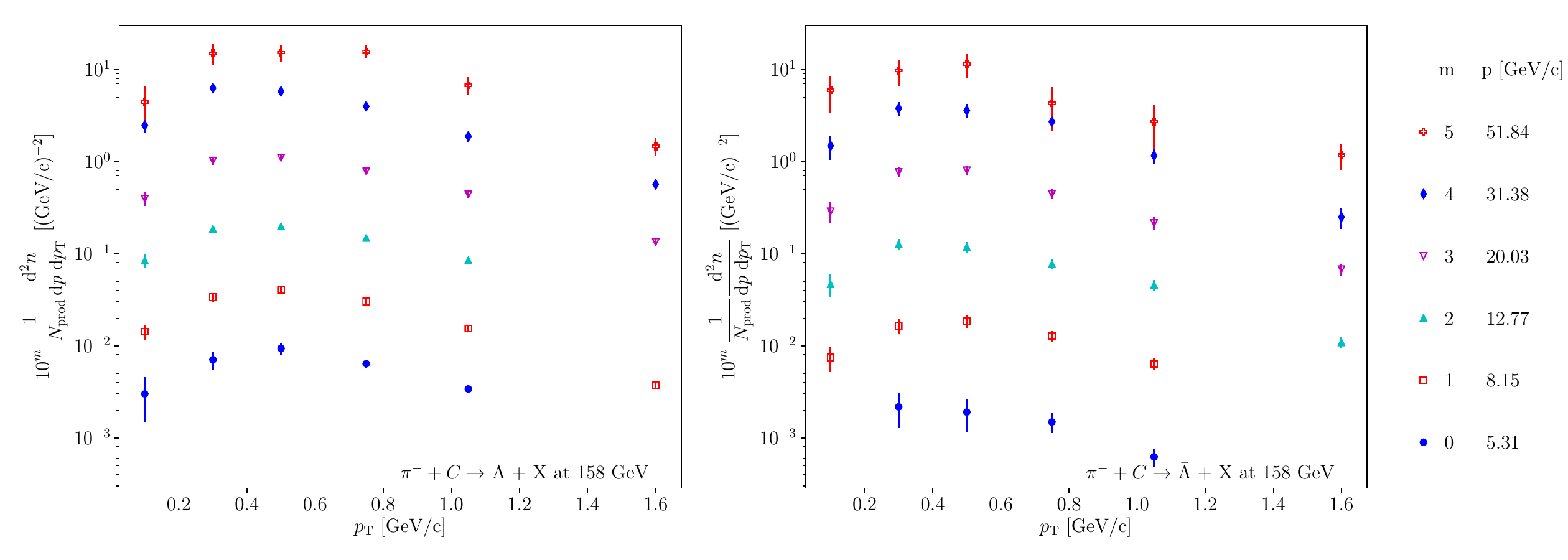}\\
\includegraphics[page=8,width=\specfigw\linewidth]{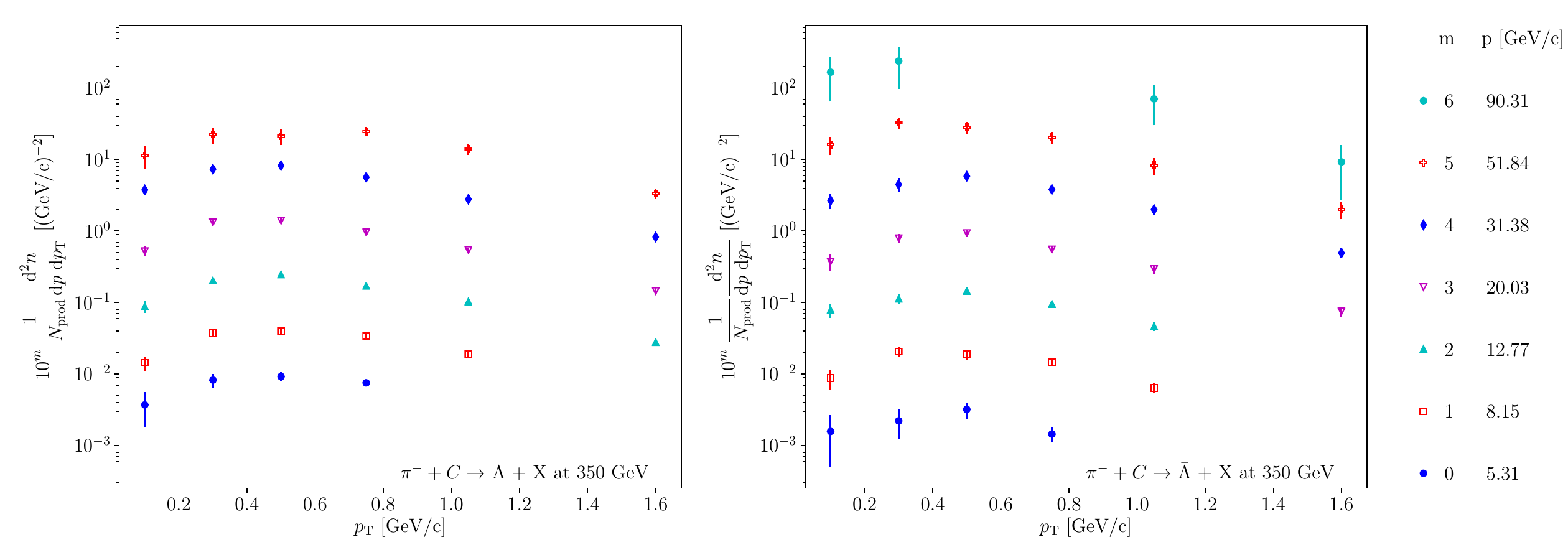}\\
\includegraphics[page=9,width=\specfigw\linewidth]{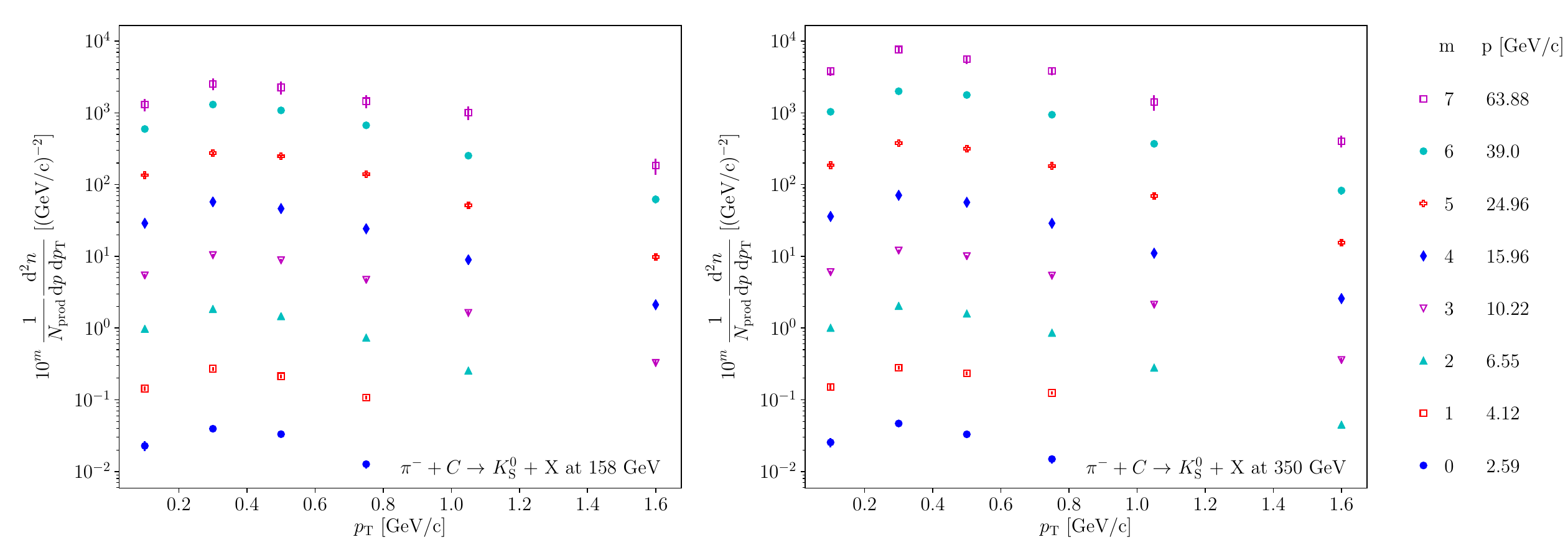}\\
\caption{Production spectra of \lamb, \antilamb and \kzeros in \pimC interactions at $p_\text{beam} =
  158$ and $350$\,\GeVc.  For each bin in momentum \p, the spectrum was
  multiplied by $10^{m}$ with the value of $m$ shown on the right. Error bars show the statistical uncertainties.}
\label{fig:dndpdptv0}
\end{figure}

The measured double-differential spectra of \pions, \kaons, \protonpm,
\lamb, \antilamb, and \kzeros spectra in \pimC interactions at 158 and
350\,\GeVc are shown in~\cref{fig:dndpdpt158,fig:dndpdpt350,fig:dndpdptv0}.
The single-differential,
\pT-integrated spectra are displayed in~\cref{fig:pt1,fig:pt2} and
will be discussed in more detail in the next Section. Tables of the measured
spectra can be downloaded at~\cite{dataDownload}.

\section{Discussion}
\label{sec:discussion}

\def\rfigw{0.43}
\def\rffigw{0.382}
\begin{figure}[!p]
\centering
\includegraphics[rviewport=0.025 0.125 0.97 1,clip,width=\rfigw\linewidth]{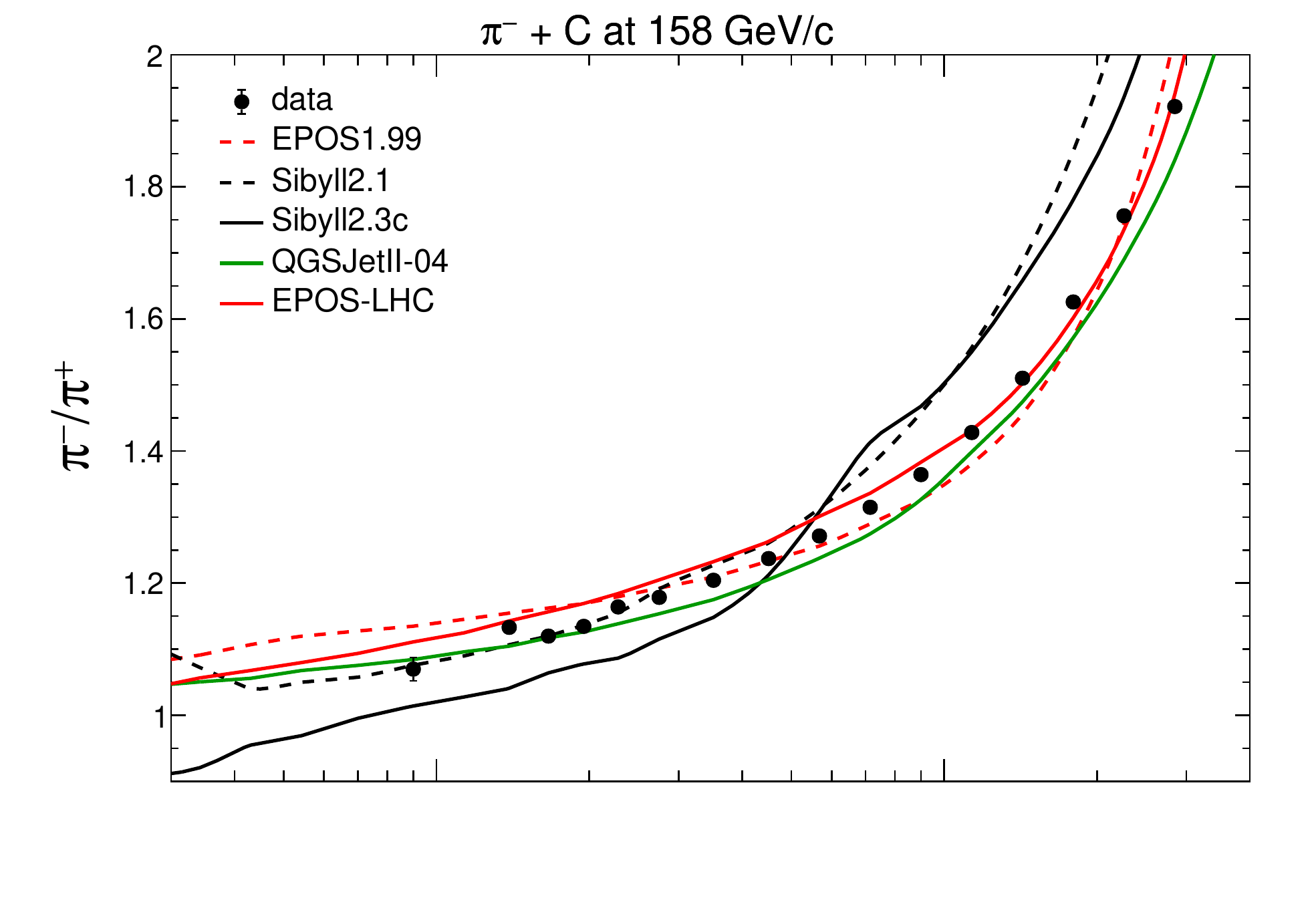}%
\includegraphics[rviewport=0.13 0.125 0.97 1,clip,width=\rffigw\linewidth]{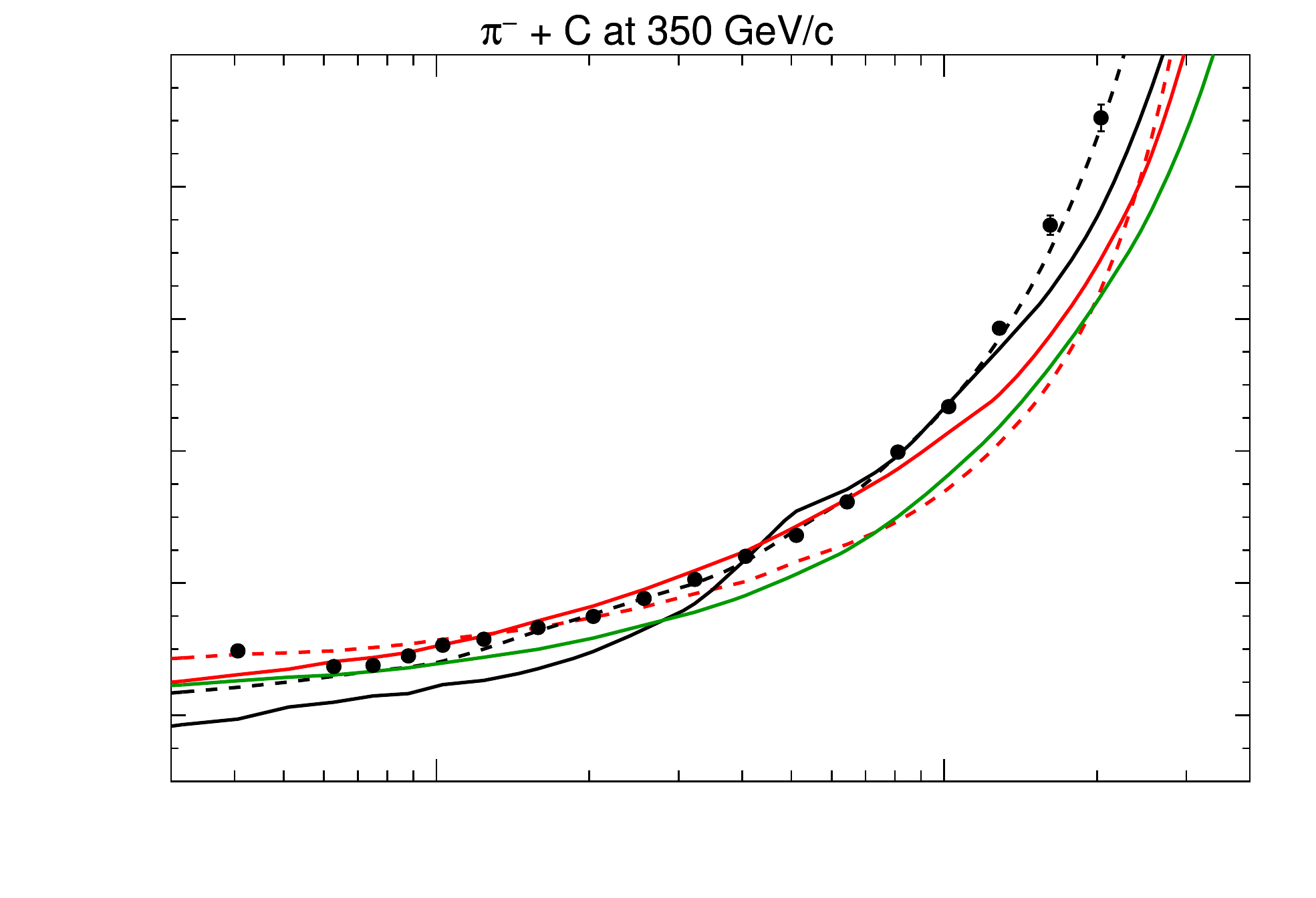}\\
\includegraphics[rviewport=0.025 0.125 0.97 0.95,clip,width=\rfigw\linewidth]{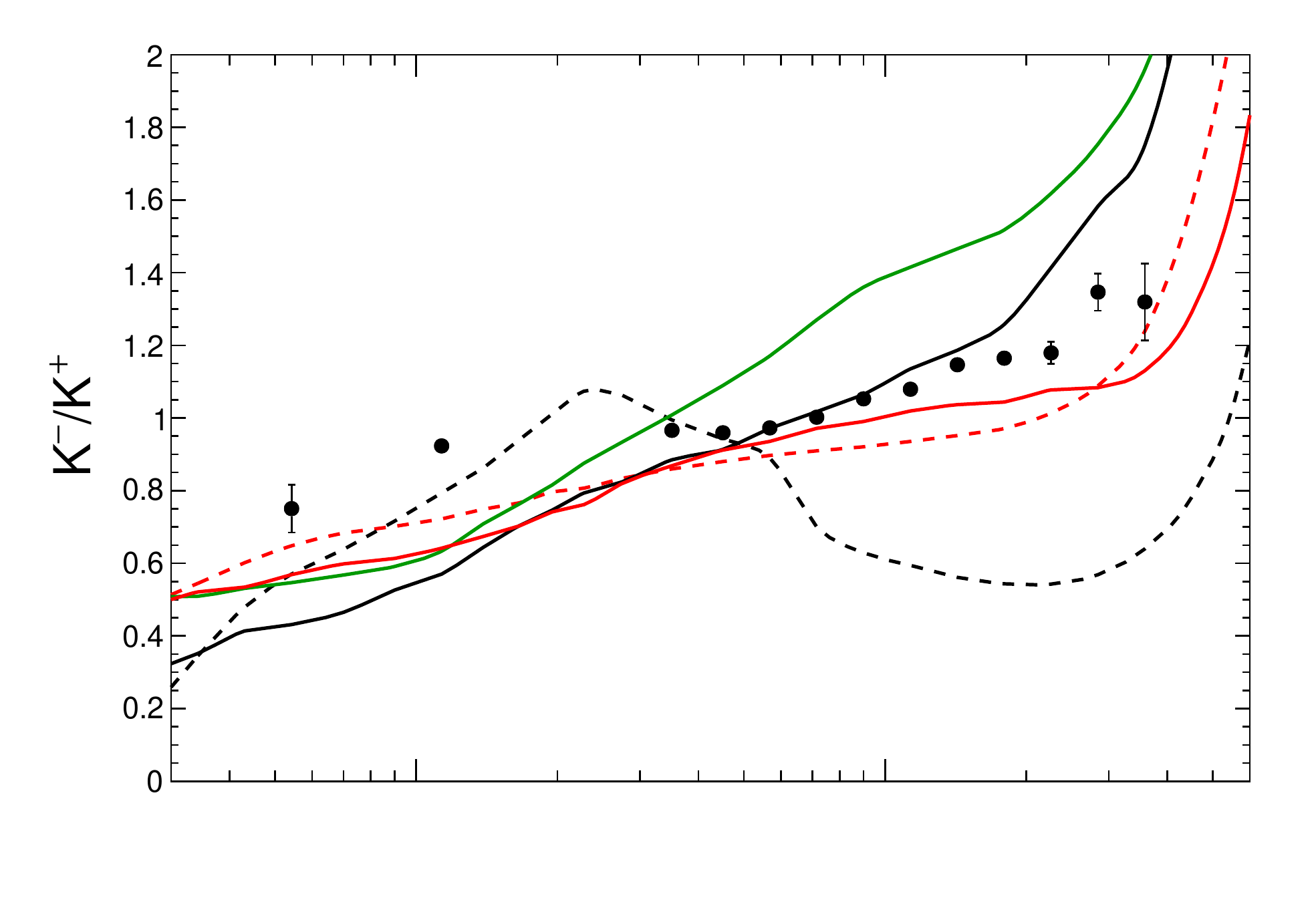}%
\includegraphics[rviewport=0.13 0.125 0.97 0.95,clip,width=\rffigw\linewidth]{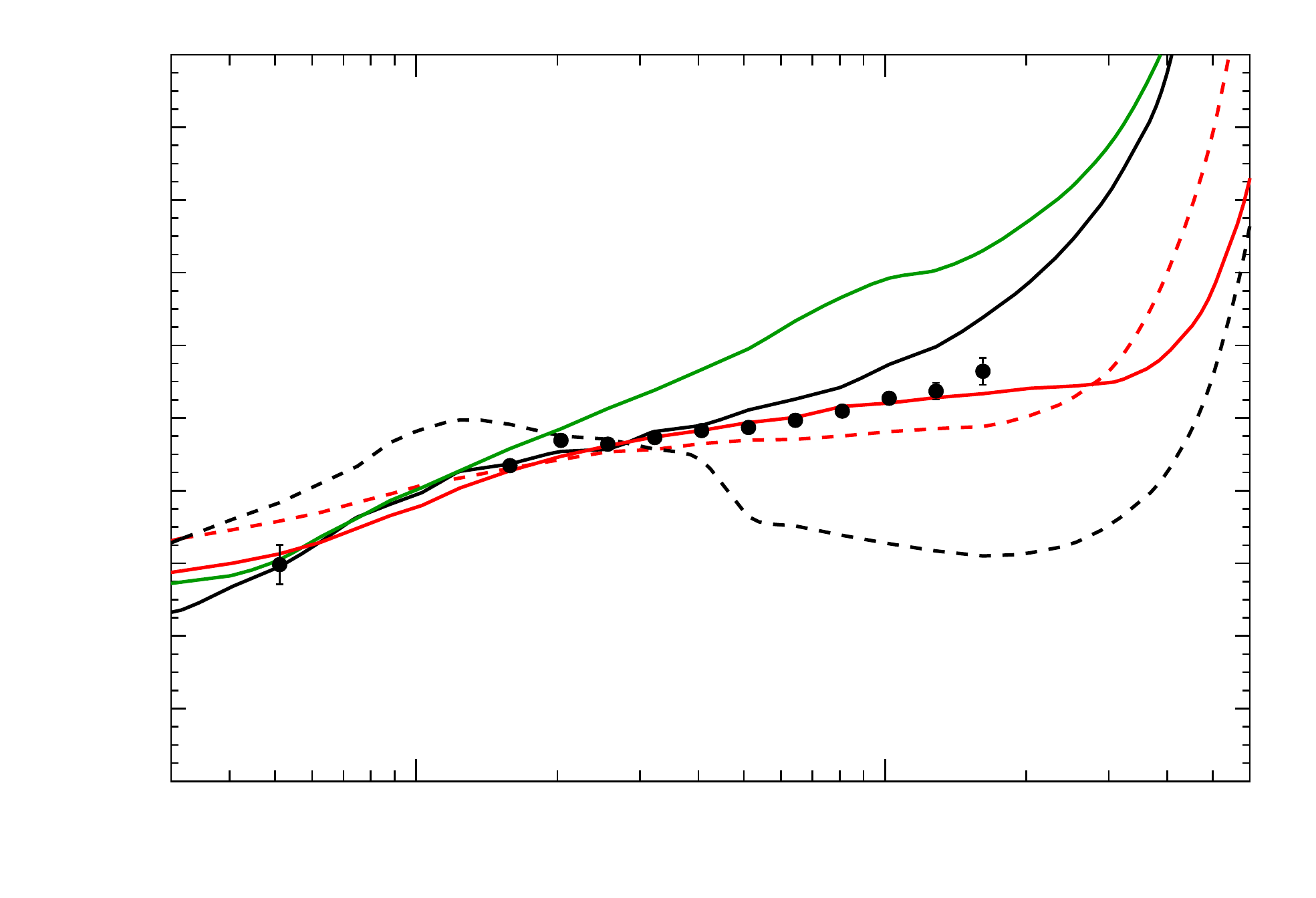}\\
\includegraphics[rviewport=0.025 0.125 0.97 0.95,clip,width=\rfigw\linewidth]{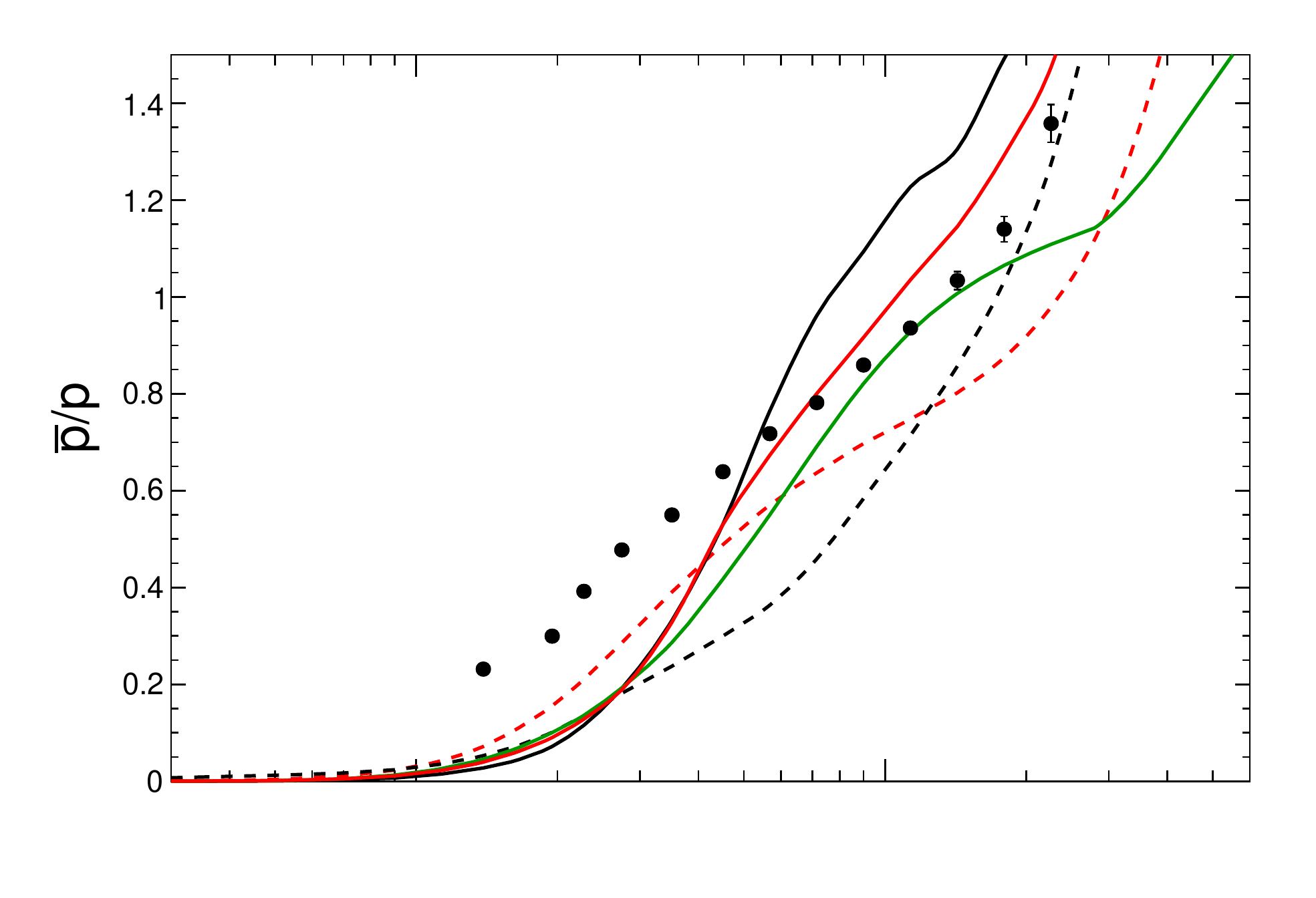}%
\includegraphics[rviewport=0.13 0.125 0.97 0.95,clip,width=\rffigw\linewidth]{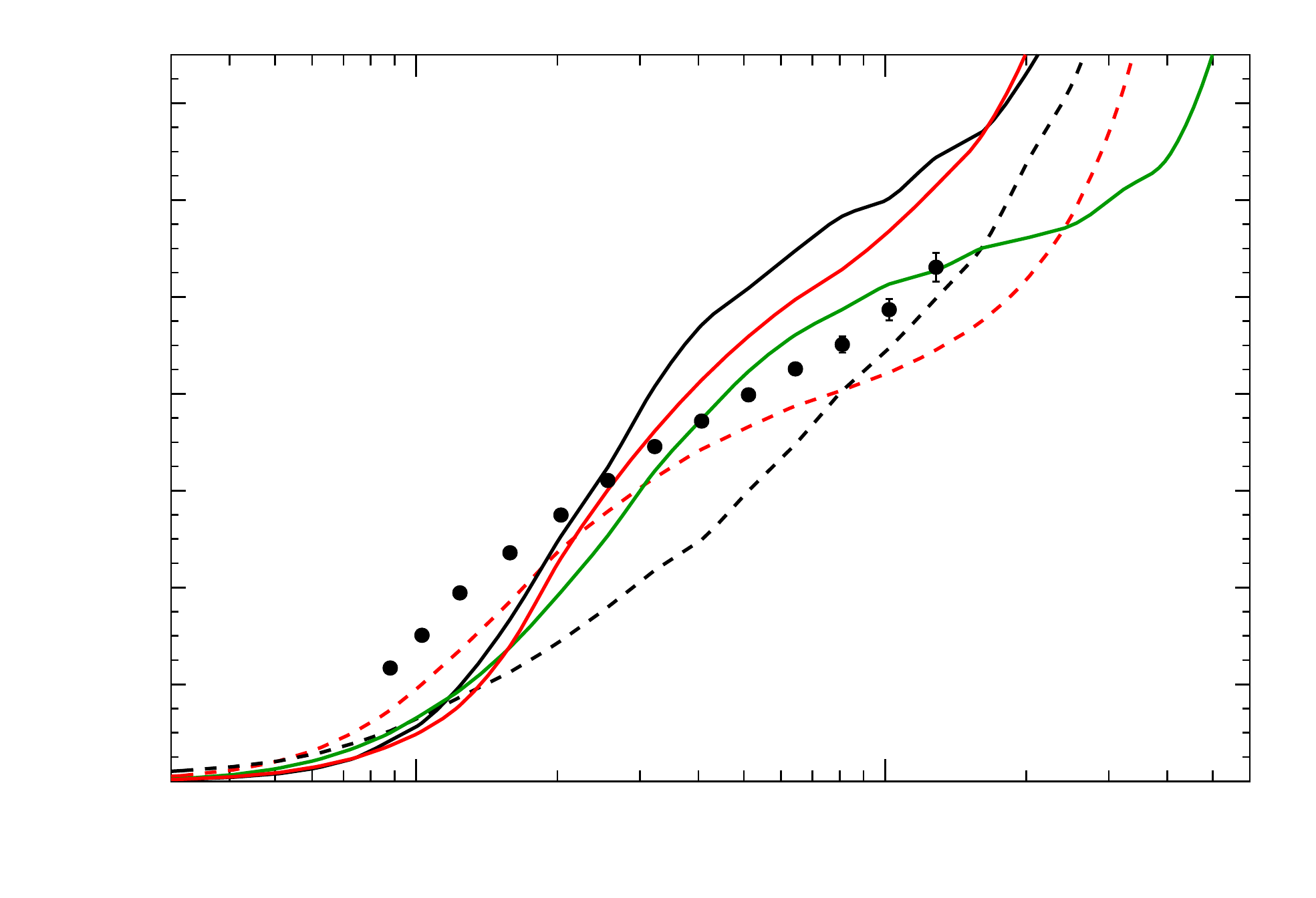}\\
\includegraphics[rviewport=0.025 0.125 0.97 0.95,clip,width=\rfigw\linewidth]{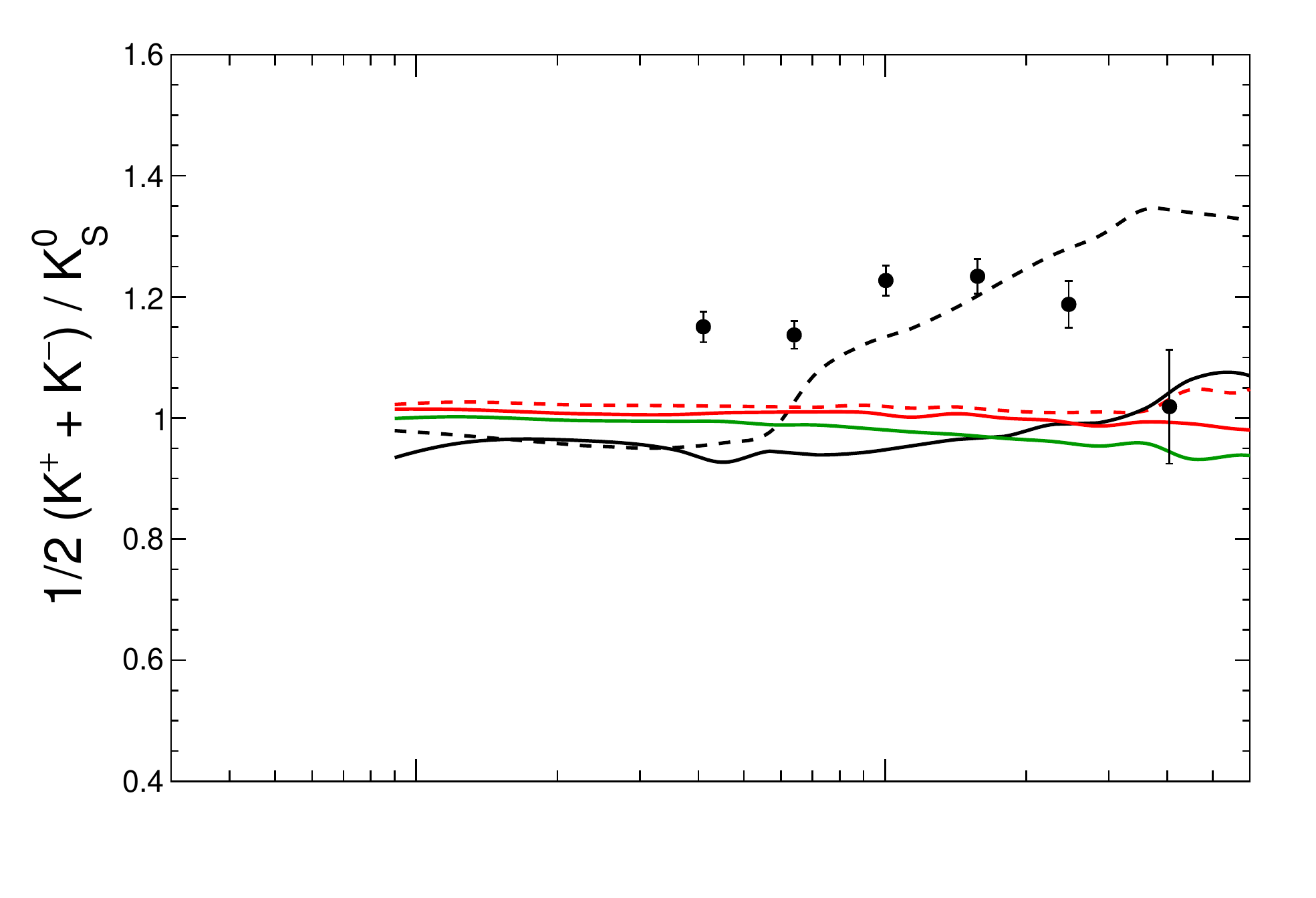}%
\includegraphics[rviewport=0.13 0.125 0.97 0.95,clip,width=\rffigw\linewidth]{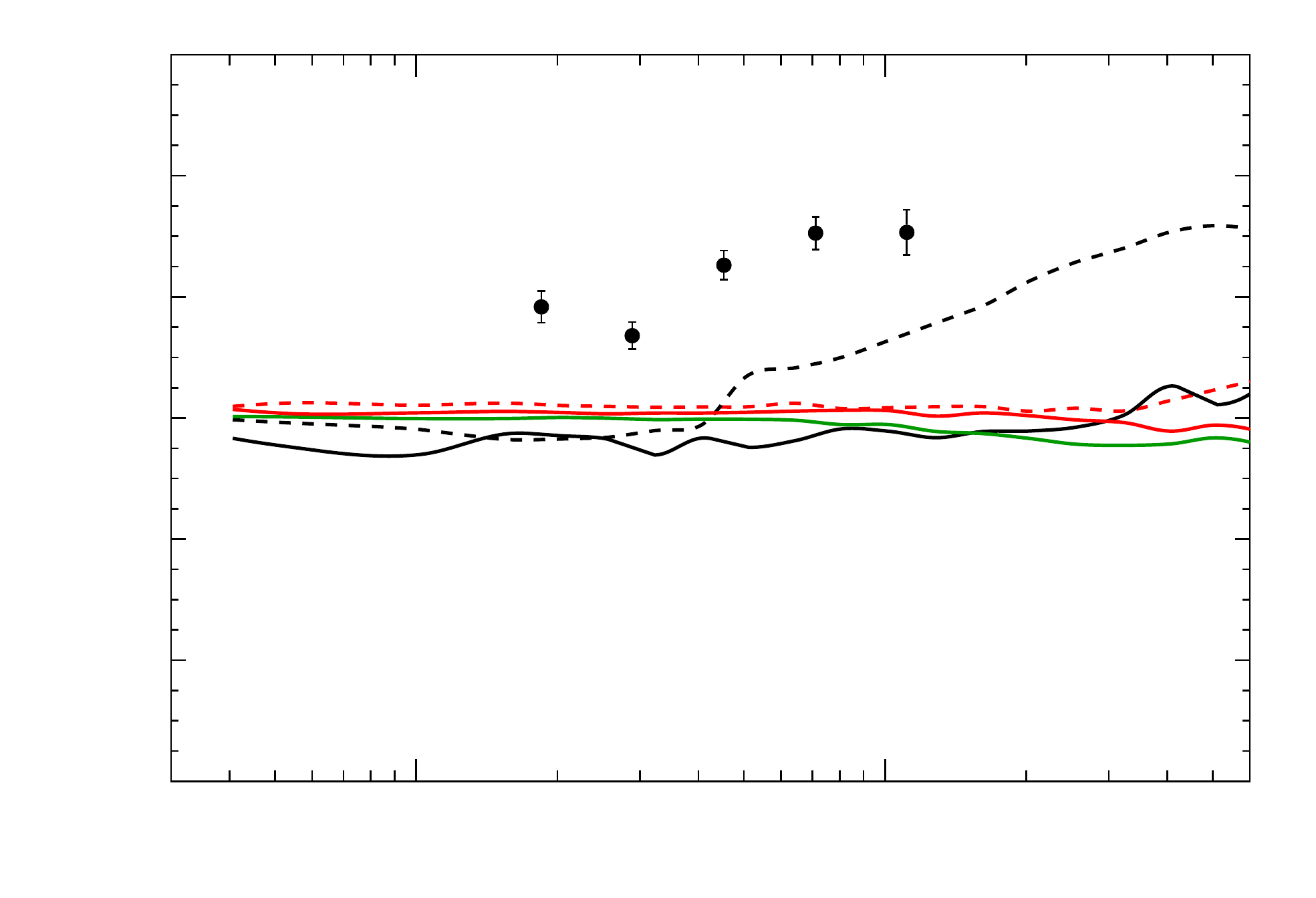}\\
\includegraphics[rviewport=0.025 0 0.97 0.95,clip,width=\rfigw\linewidth]{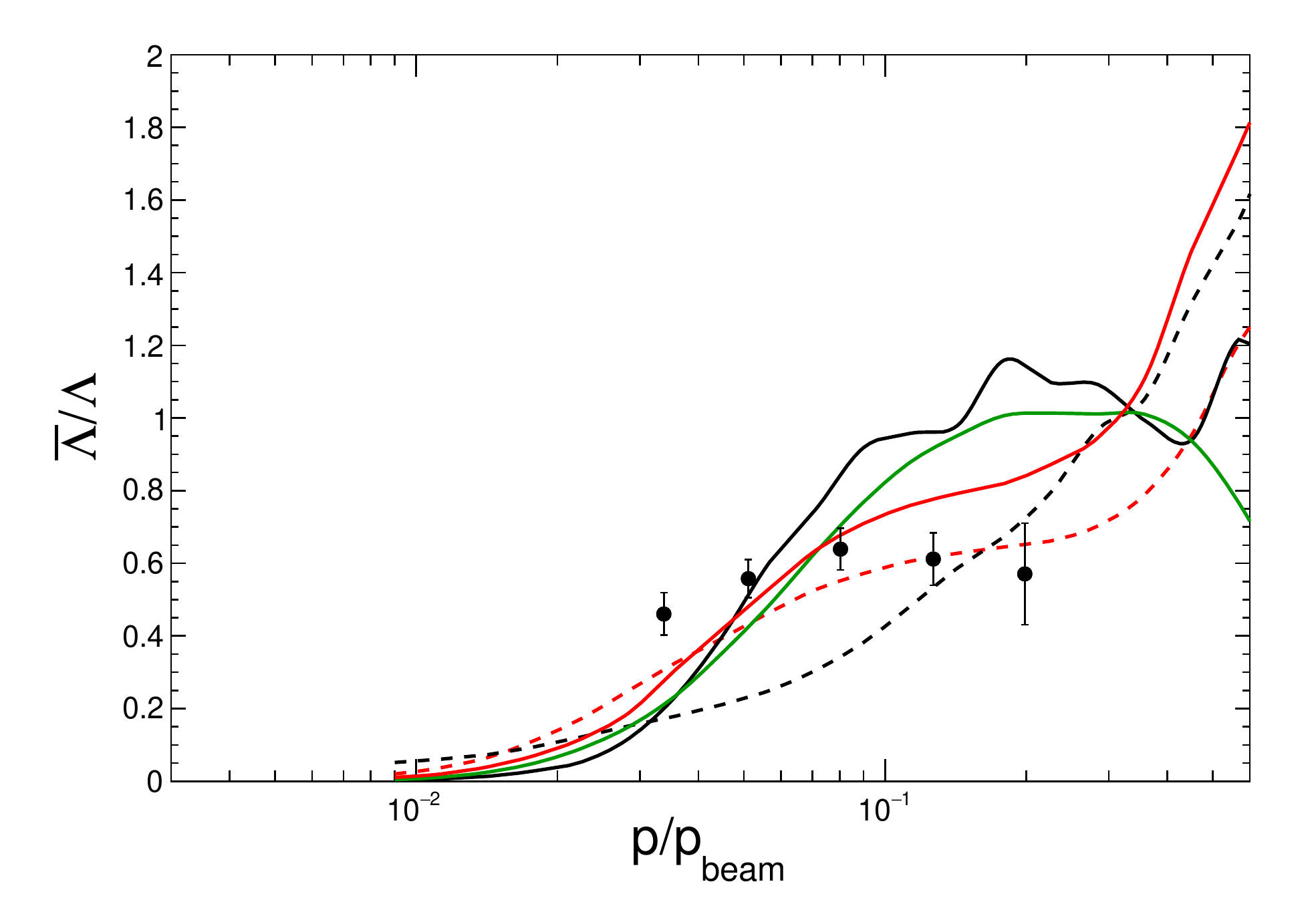}%
\includegraphics[rviewport=0.13 0 0.97 0.95,clip,width=\rffigw\linewidth]{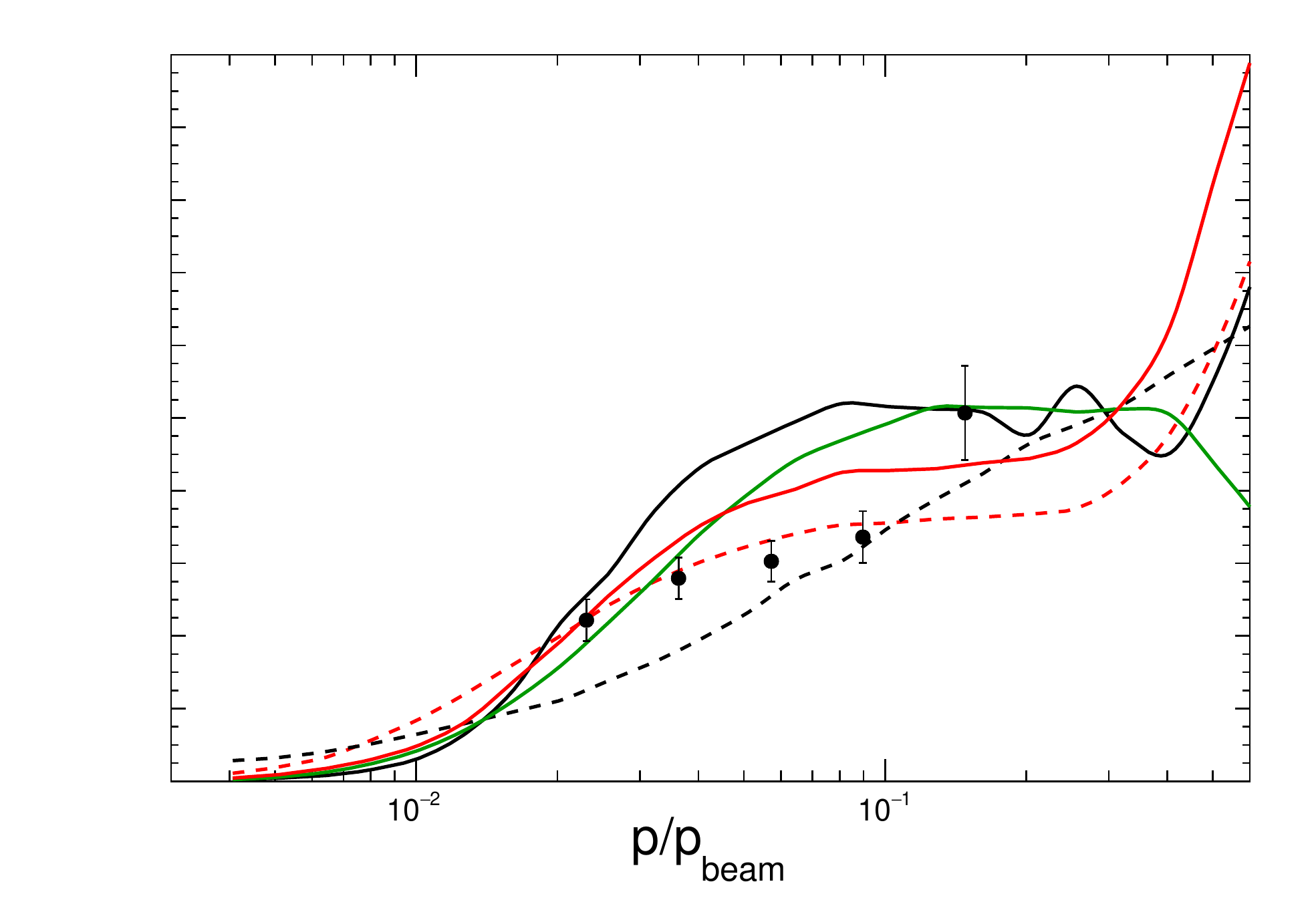}\\
\caption{Ratio of particle spectra for data (points with statistical error bars) and models (lines). Ratios
at beam energies 158 and 350\,\GeVc are displayed on the left and right column respectively.}
\label{fig:rel}
\end{figure}

For a first assessment of the validity of hadronic interaction models
used in air-shower simulations, we compare the ratio of measured
particle spectra to the model predictions in \cref{fig:rel}. Solid
lines denote recent
model-tunes~\cite{Ostapchenko2010,Ostapchenko:2013pia,Pierog:2013ria,Fedynitch:2018cbl,sib23d}
based on LHC and fixed target data, whereas dashed lines show previous
versions of these models~\cite{Ahn:2009wx,Pierog:2006qv}.  Most of the
models agree reasonably well with the measured pion charge ratio shown
in the first row of \cref{fig:rel}, but both versions of the {\scshape
  Sibyll} model overpredict the ratio at high particle momenta for the
158\,\GeVc data set. The kaon charge ratio in the second row is best
described by \EposLHCLong. All models underpredict the
antiproton-to-proton ratio at low momenta shown in the third row and
the shape of the momentum-dependence of this ratio with two inflection
points is best reproduced by \EposLong. We compare the production of
charged to neutral kaons in the fourth row by computing the ratio of
$\frac{1}{2}(\text{K}^+ +\text{K}^-)$ to \kzeros, where $\text{K}$ is
the shorthand for the production spectrum of the particles.  This
ratio is expected to be unity from simple arguments based on the
counting of valence quarks of the beam and target. Indeed all models
but the old version of {\scshape Sibyll} predict a value close to 1
whereas our data suggest values in the range of 1.2 to 1.3. The
difference to the expectation of 1 is about three times the systematic
uncertainty assigned to the integrated kaon spectra, see
\cref{fig:syst158,fig:syst350}. Finally, the ratio of \antilamb to
\lamb baryons is best described by the \EposLong model, whereas all
recent model re-tunes slightly overpredict the production of \antilamb
at intermediate momenta of $p/p_\text{beam}\sim 0.1$.

\def\ptfigw{1}
\def\ptfigh{0.174}
\begin{figure}[!p]
\centering
\includegraphics[page=10,width=\ptfigw\linewidth]{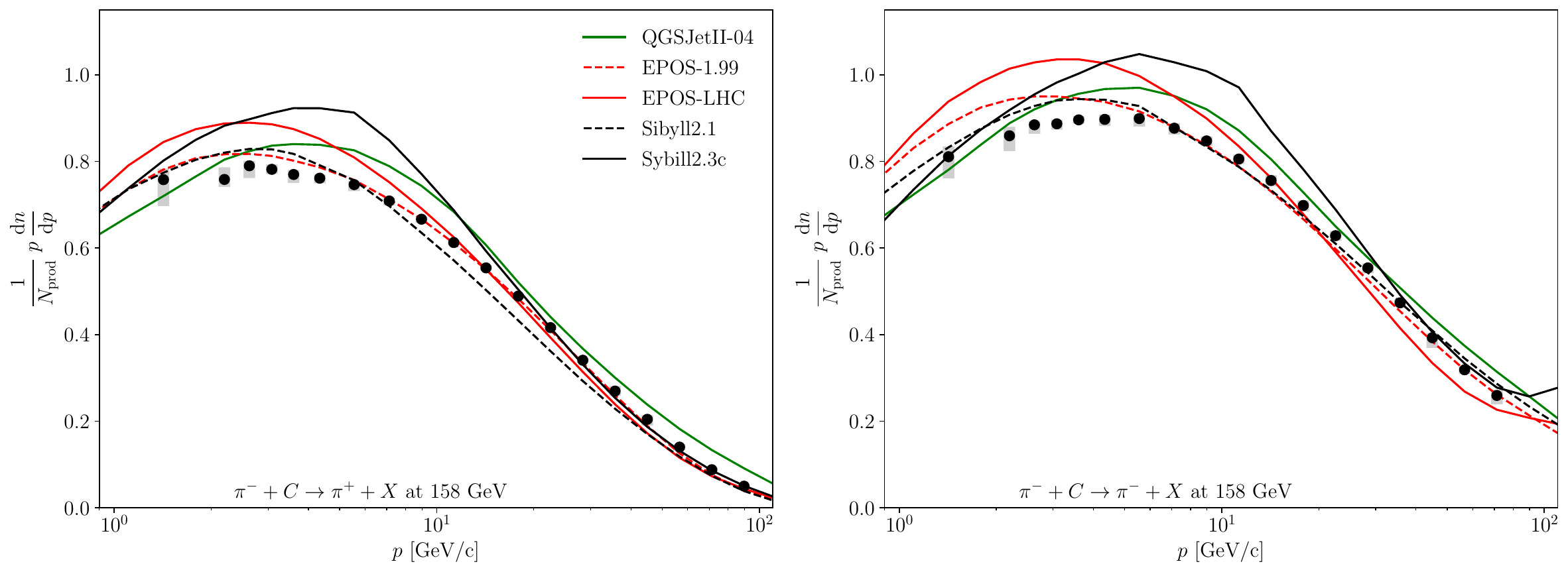}\\
\includegraphics[page=11,width=\ptfigw\linewidth]{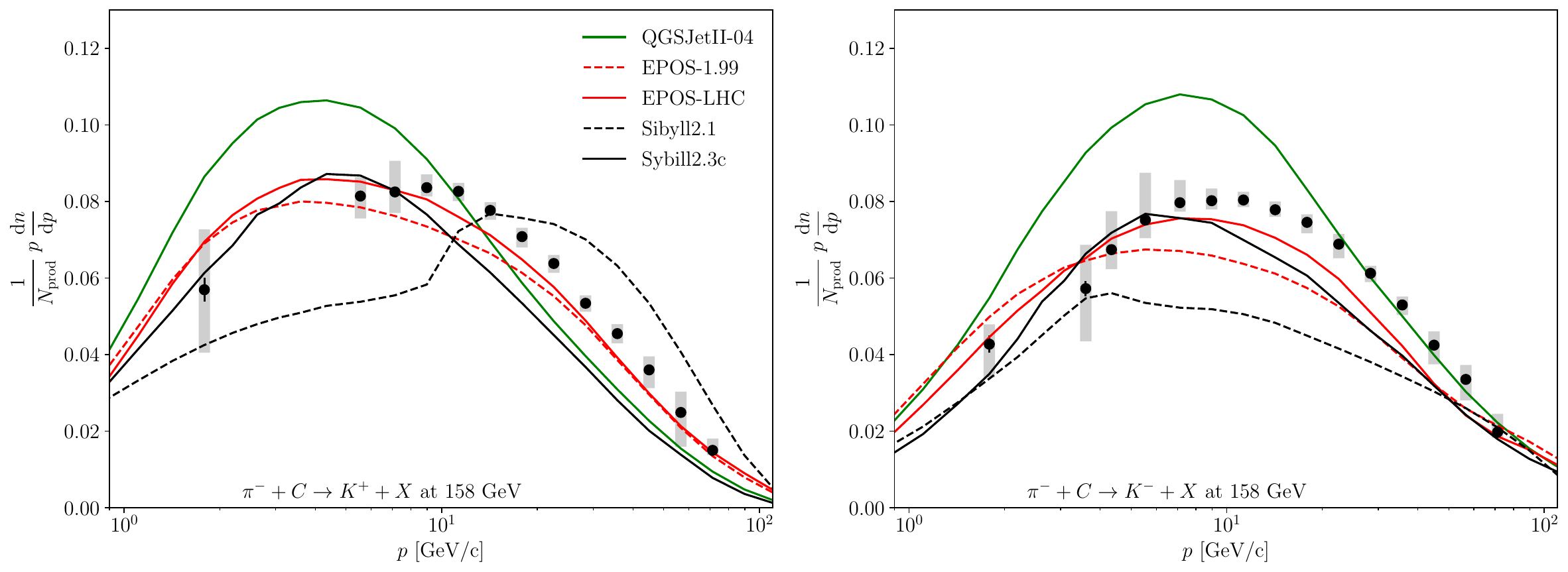}\\
\includegraphics[page=12,width=\ptfigw\linewidth]{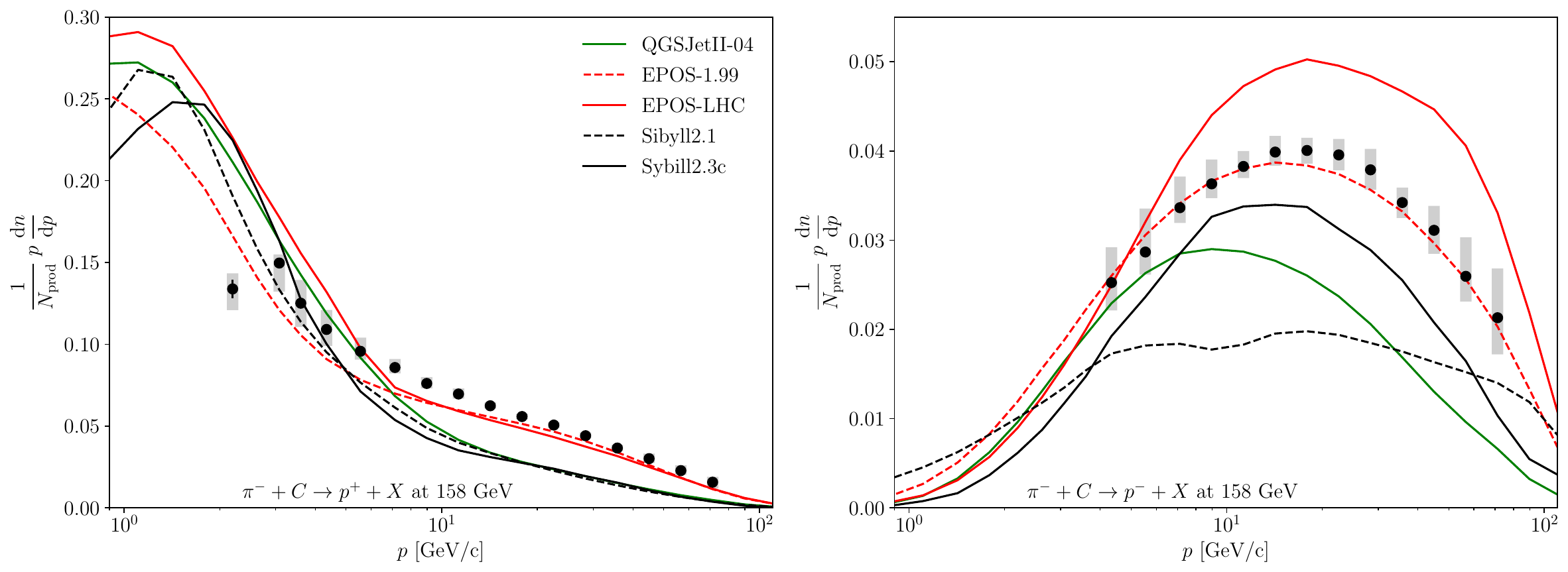}\\
\caption{Comparison of the \pT-integrated particle production spectra
  of \pions, \kaons and \protons at 158\,\GeVc with predictions of hadronic interaction models. The data is shown as black circles with statistical error bars. Systematic uncertainties are displayed by gray rectangles.}
\label{fig:pt1}
\end{figure}

\begin{figure}[!p]
\centering
\includegraphics[page=13,width=\ptfigw\linewidth]{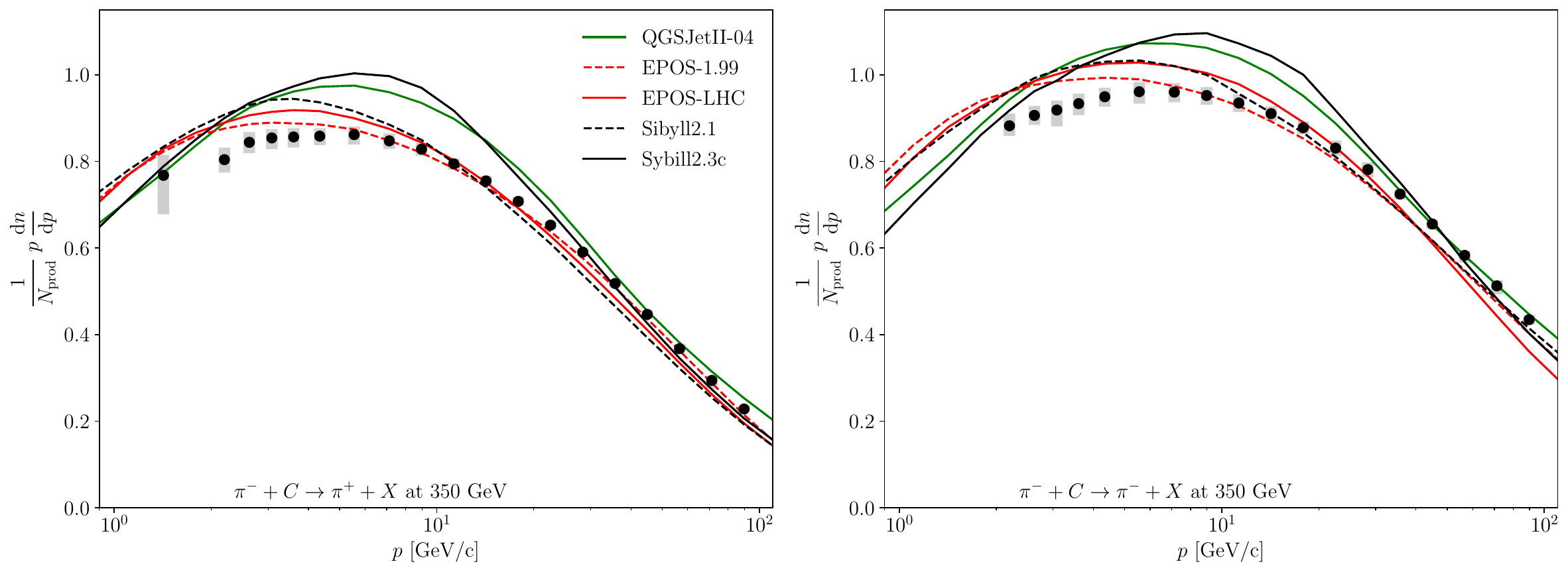}\\
\includegraphics[page=14,width=\ptfigw\linewidth]{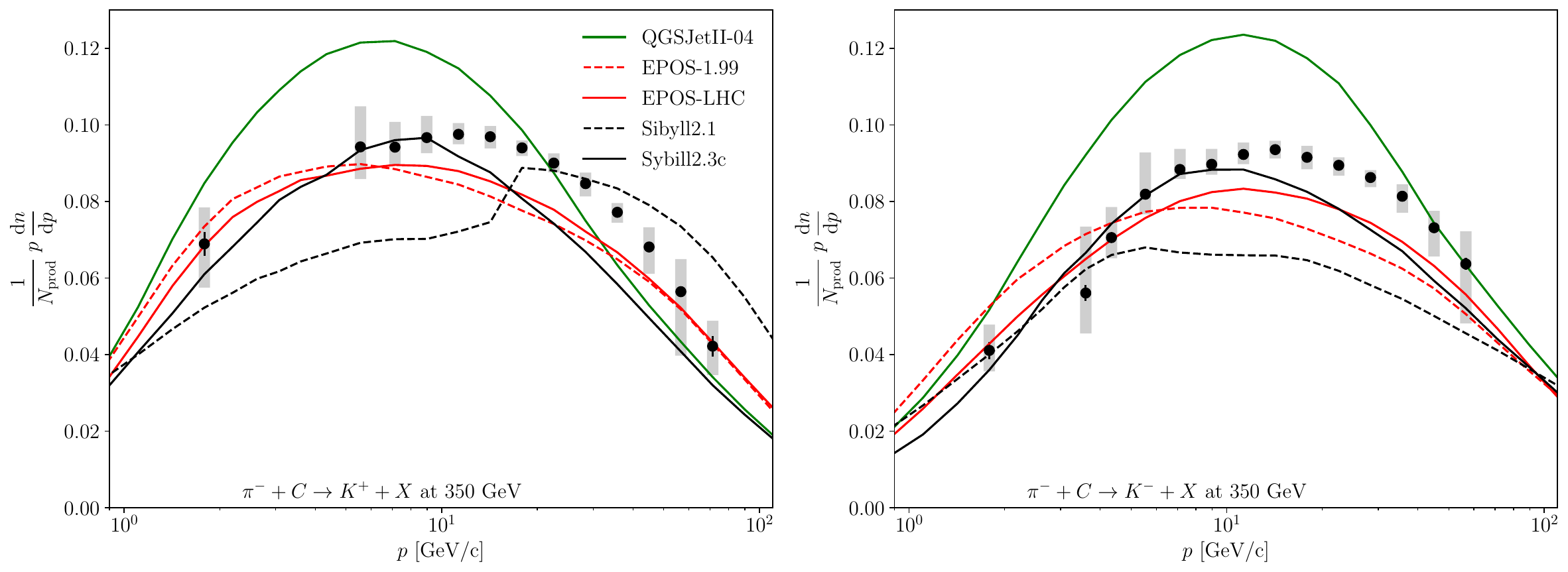}\\
\includegraphics[page=15,width=\ptfigw\linewidth]{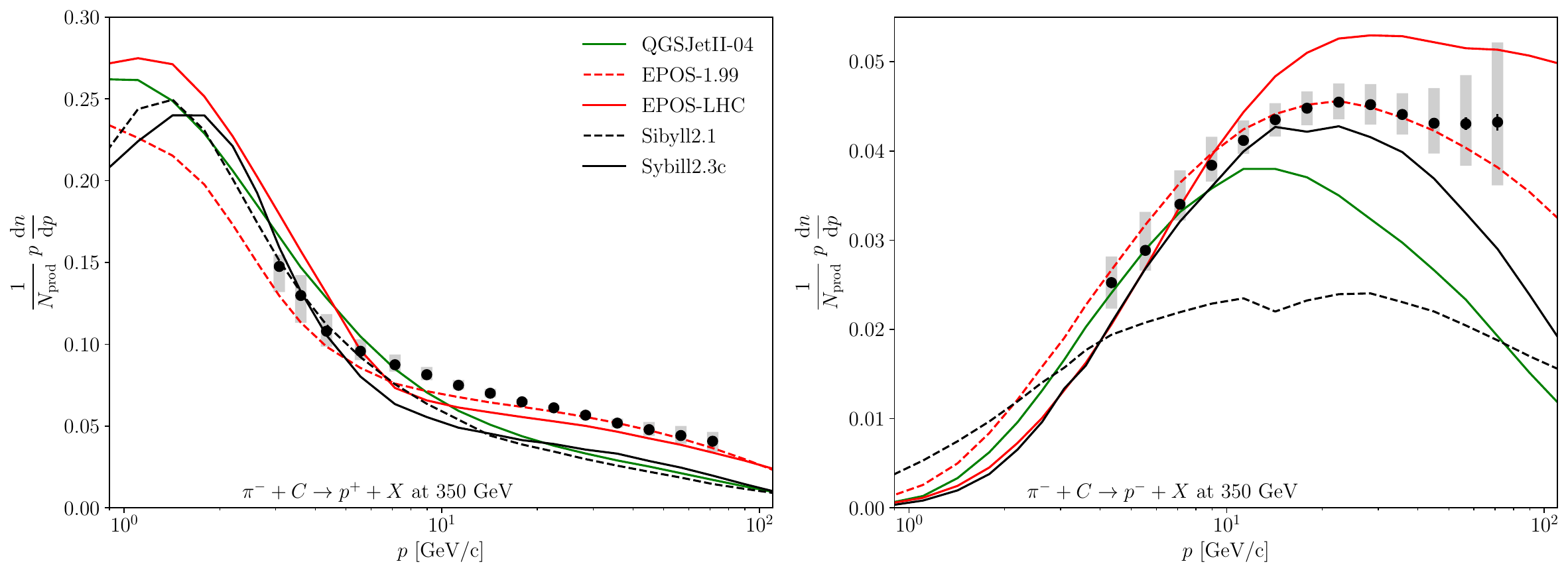}\\
\caption{Comparison of the \pT-integrated particle production spectra
  of \pions, \kaons and \protons at 350\,\GeVc with predictions of hadronic interaction models. The data is shown as black circles with statistical error bars. Systematic uncertainties are displayed by gray rectangles.}
\label{fig:pt2}
\end{figure}

\begin{figure}[!p]
\centering
\includegraphics[page=16,width=\ptfigw\linewidth]{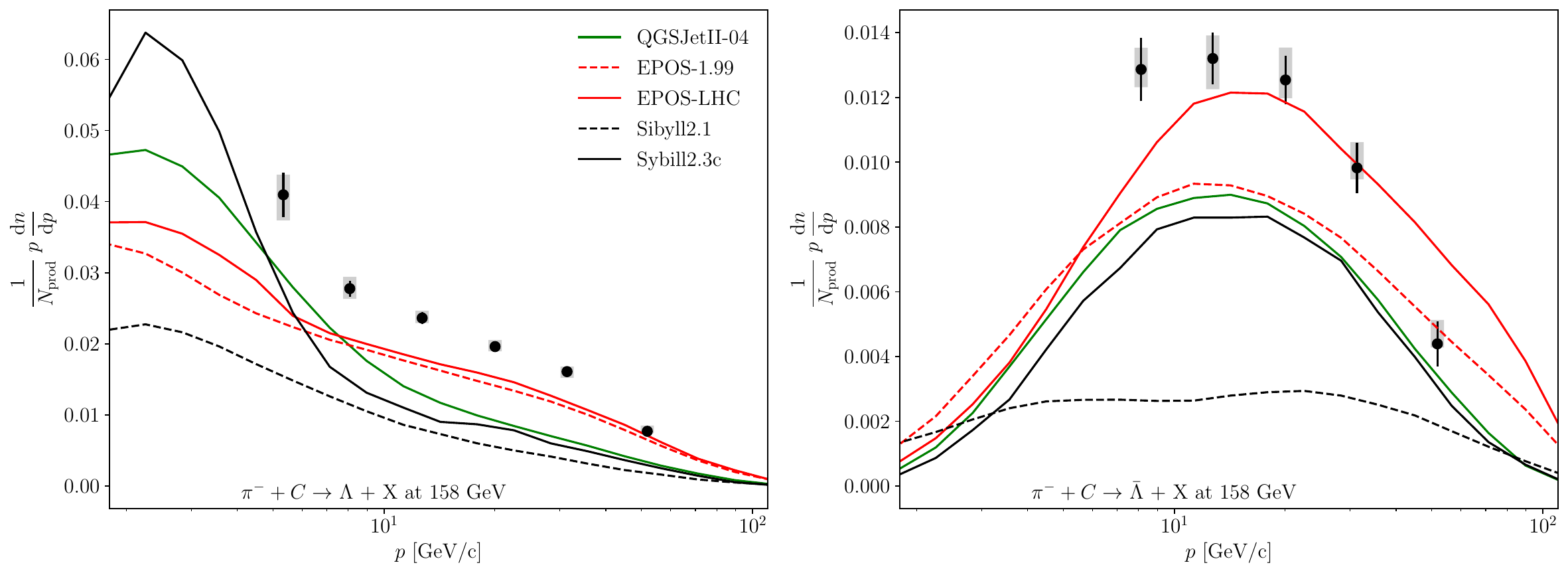}\\
\includegraphics[page=17,width=\ptfigw\linewidth]{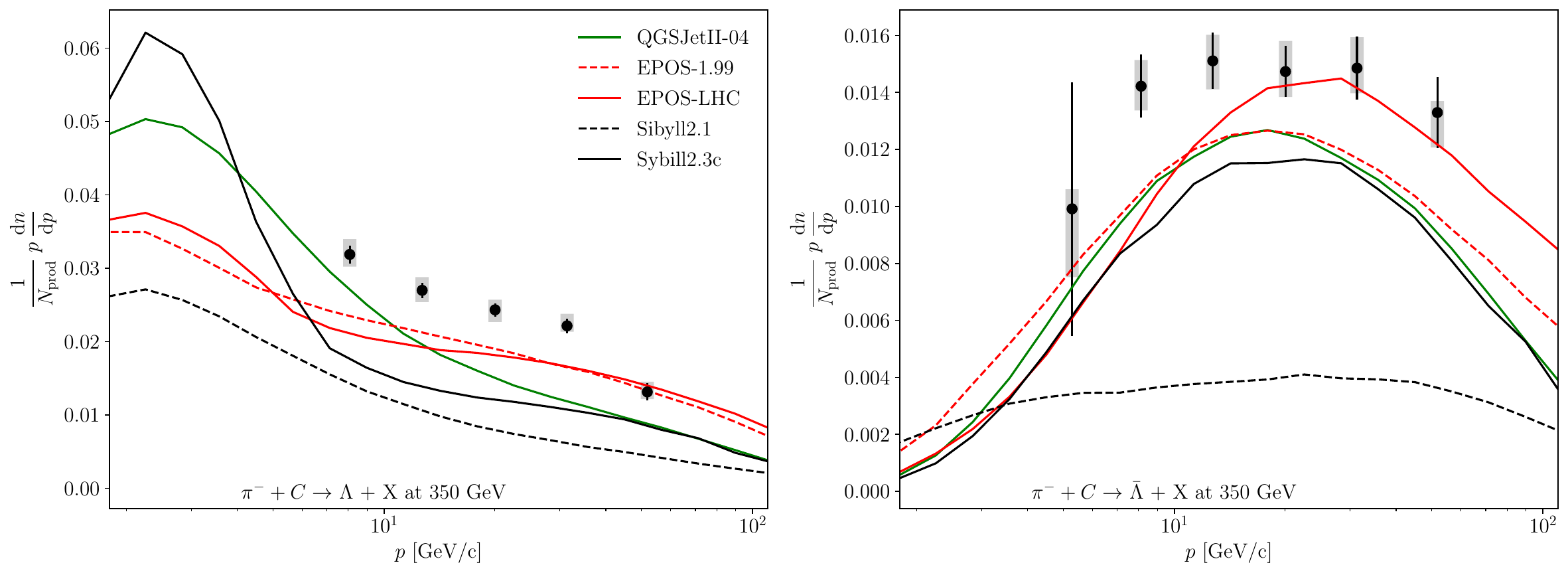}\\
\includegraphics[page=18,width=\ptfigw\linewidth]{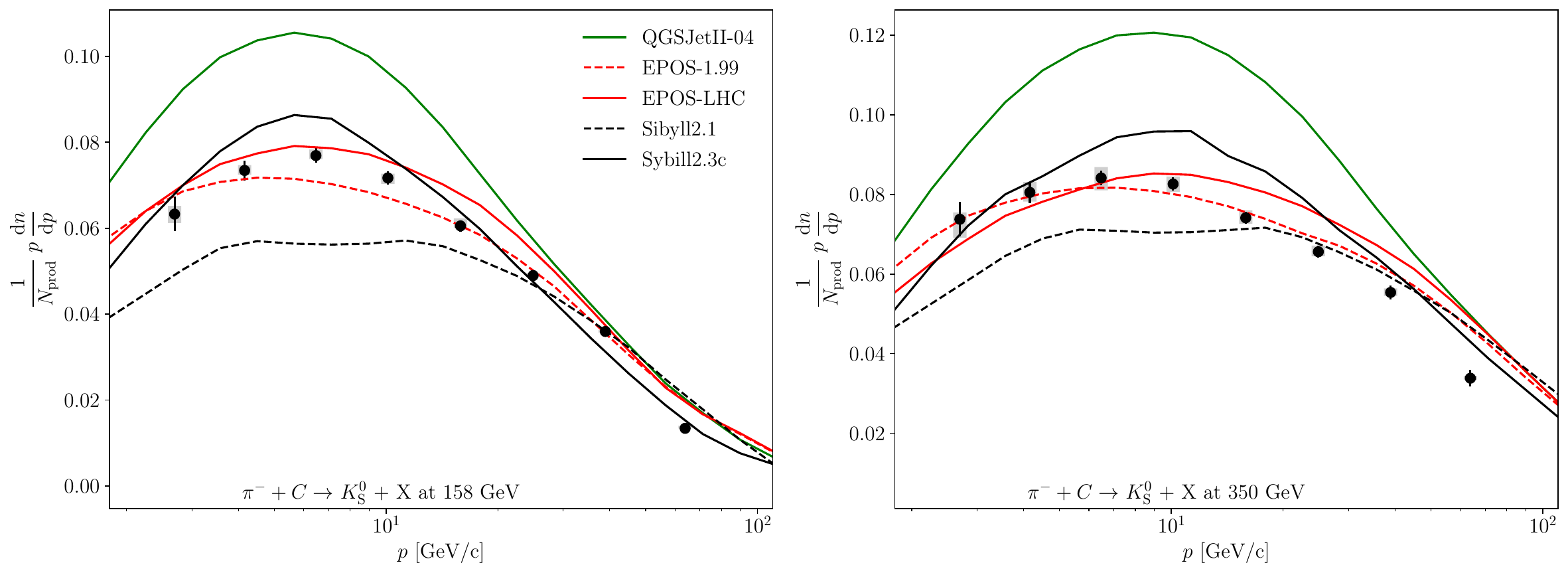}\\
\caption{Comparison of the \pT-integrated particle production spectra
  of  \lamb, \antilamb and \kzeros at 158 and 350\,\GeVc with predictions of hadronic interaction models. The data is shown as black circles with statistical error bars. Systematic uncertainties are displayed by gray rectangles.}
\label{fig:pt3}
\end{figure}

Furthermore, we compare the \pT-integrated data to
predictions from hadronic interaction models in \cref{fig:pt1,fig:pt2,fig:pt3}.  As can be
seen, none of the models provides a satisfactory description of our
data and each of the recent re-tunes has its own deficiencies.  Of all
the models, \SibyllNewLong gives the worst description of the charged
pion spectra and it under-predicts \antiproton and \antilamb
production.  \QGSJetLong fails spectacularly to reproduce
kaon-production in \pimC interactions and also produces too few
\antiproton and \antilamb particles.  In many aspects, the previous
version of the \Epos model (\EposLong) gives a better prediction of
our data than the current \EposLHCLong version.  In particular,
\EposLong provides the best description of the charged pion spectra
and a near-spot-on prediction of \antiproton production, whereas the
newer version of the model gives the best match to our \antilamb
measurements.  It will be interesting to see air shower predictions of
future versions of these models that describe all aspects of our data.

For a further qualitative analysis of the relevance of this
measurement to the ``muon puzzle'' in cosmic-ray-induced air showers,
it is useful to recall that the key to model muon production in air
showers is to correctly predict the fraction $f$ of the energy that
remains in the hadronic cascade in each interaction and is not lost to
the electromagnetic component via $\uppi^0$ production.  In a
simplified model with the production of only charged and neutral
pions, this fraction is $f=2/3$ and after $n$ interactions $(2/3)^n$
of the initial energy is left in the hadronic component.  Muons are
produced when the pions reach low energies and decay, which happens
after about $n=8$ generations of interactions for an air shower
induced by a primary of $10^{20}$\,eV~\cite{Matthews:2005sd}.  In a
more realistic scenario the energy transfer to the hadronic component
is $f = (2/3 + \Delta)$, where $\Delta$ accounts for hadronic
particles without dominant electromagnetic decay channels such as
\rhozero mesons~\cite{Drescher:2007hc,Ostapchenko:2013pia} or
baryons~\cite{Pierog:2006qv}.  Then a fraction of $(2/3 +
\Delta)^n\approx (2/3)^n\,(1 + 3/2\; n\, \Delta)$ of the initial
cosmic-ray energy can produce muons after $n$ interactions and only if
the value of $\Delta$ is accurately known throughout the whole chain
of interactions, there is hope for a precise prediction of the muon
number in air showers.

\begin{figure}[t]
\centering
\includegraphics[width=0.73\linewidth]{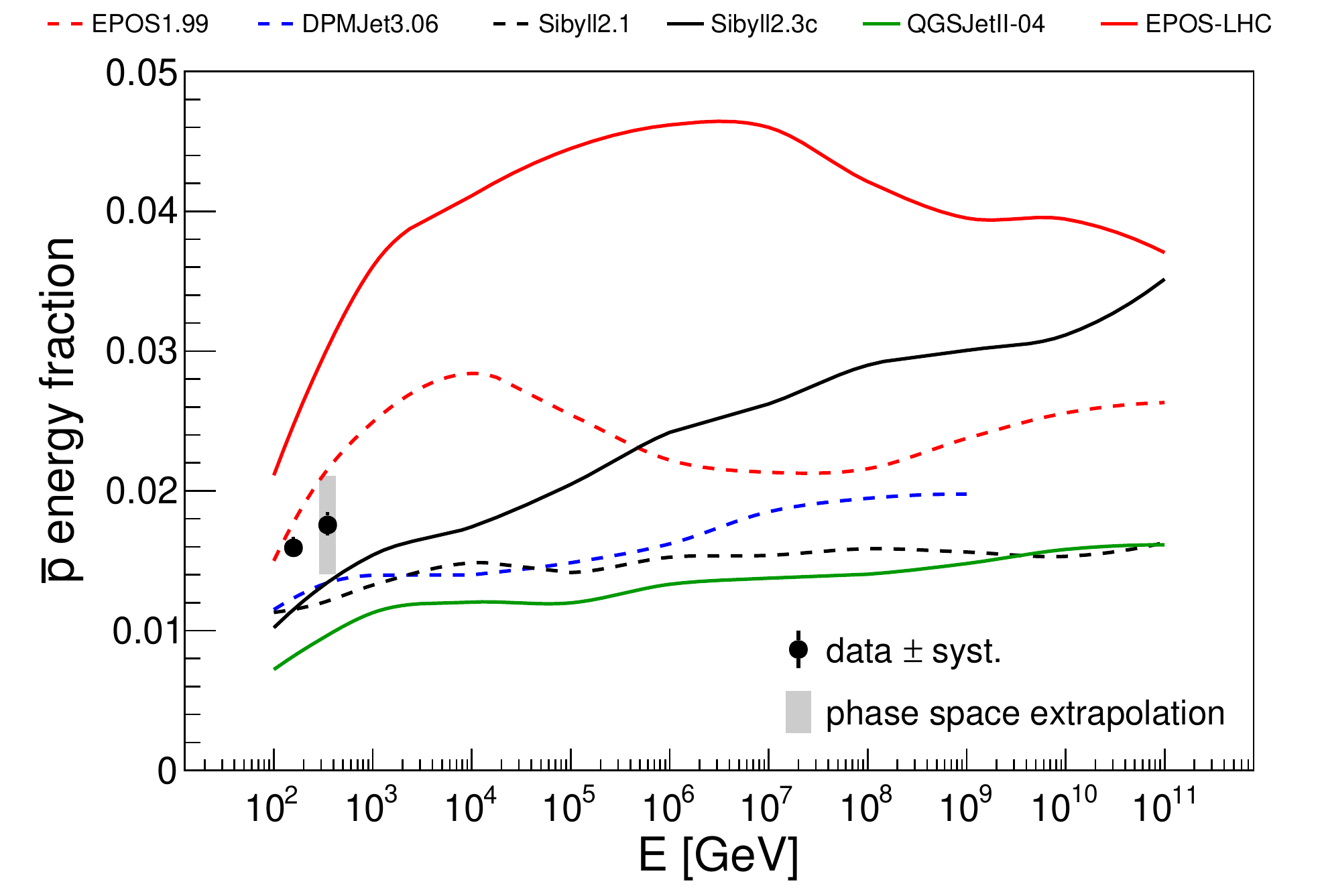}
 \caption{Energy fraction transferred to anti-protons as derived from
   the measurement presented in this article (data points) and as
   predicted by hadronic interaction models over the whole range of
   beam energies relevant for air showers.}
\label{fig:efrac}
\end{figure}

The production of \rhozero mesons in \pimC interactions has already
been addressed by \NASixtyOne in Ref.~\cite{NA61SHINE:2017vqs} and the
integrated production spectrum gives $\Delta_{\uprho^0}=\big(7.7 \pm
0.1 (\text{stat.}) \pm 0.2 (\text{syst.})\big)\%$ at 158\,\GeVc.  Here
we can comment on the baryon production for which the best proxy is
the production of anti-protons since protons can also originate from
target fragmentation.  The average energy fraction transferred to
anti-protons is displayed in \cref{fig:efrac}, and is obtained by
integrating the measured $p\,\mathrm{d}n/\mathrm{d}p$ spectra
including an extrapolation to the full beam
momentum~\cite{RaulPhD}. This gives $\Delta_{\bar{\text{p}}} = \big(
1.59 \pm 0.01 (\text{stat.})\; \pm 0.07 (\text{syst.})  \pm 0.01
(\text{mod.})\big)\%$ and $\big(1.76 \pm 0.01 (\text{stat.})\; \pm
0.08 (\text{syst.})  \pm 0.35 (\text{mod.})\big) \%$ at 158 and
350\,\GeVc, where the last of the three quoted uncertainties is due to
the model-dependence of the extrapolation to full beam momentum.  Note
that the anti-proton fraction constrains the production of p,
\antiproton, n, and $\bar{\text{n}}$, i.e.\ naively
$\Delta_\text{baryon} \sim 4 \Delta_{\bar{\text{p}}}$.  Numerically we
find $\Delta_\text{baryon} \sim \Delta_{\uprho^0}$, i.e.\ both
processes are about equally important for the evolution of air
showers.  Our measurement can be used to normalize the model
differences at low energies (cf.\ \cref{fig:efrac}), leaving then only
the energy-evolution of $\Delta_\text{baryon}$ as the remaining
uncertainty of baryon production in air showers.

\section{Summary}
\label{sec:summary}

In this article, we presented a new measurement of particle production
in interactions of negatively charged pions with carbon nuclei at beam
momenta of 158 and 350\,\GeVc. We estimated the production cross
section and determined the double-differential \p-\pT spectra of
produced \pions, \kaons, \protonpm, \lamb, \antilamb, and \kzeros.
This measurement provides a unique reference data set with
unprecedented precision and large phase-space coverage to enable
future tuning of models used for the simulation of particle production
in extensive air showers in which pions are the most numerous
projectiles. None of the current state-of-the art hadronic interaction
models describes the measured particle spectra well. A tuning of
these models to match the measurements from \NASixtyOne at SPS
energies will significantly reduce the uncertainties in predictions of
muons in air showers.

\section*{Acknowledgments}
We would like to thank the CERN EP, BE, HSE and EN Departments for the
strong support of NA61/SHINE.

This work was supported by
the Hungarian Scientific Research Fund (grant NKFIH 138136\slash138152),
the Polish Ministry of Science and Higher Education
(DIR\slash WK\slash\-2016\slash 2017\slash\-10-1, WUT ID-UB), the National Science Centre Poland (grants
2014\slash 14\slash E\slash ST2\slash 00018, 
2016\slash 21\slash D\slash ST2\slash 01983, 
2017\slash 25\slash N\slash ST2\slash 02575, 
2018\slash 29\slash N\slash ST2\slash 02595, 
2018\slash 30\slash A\slash ST2\slash 00226, 
2018\slash 31\slash G\slash ST2\slash 03910, 
2019\slash 33\slash B\slash ST9\slash 03059 and 
2020\slash 39\slash O\slash ST2/00277), 
the Norwegian Financial Mechanism 2014--2021 (grant 2019\slash 34\slash H\slash ST2\slash 00585),
the Polish Minister of Education and Science (contract No. 2021\slash WK\slash 10),
the Russian Science Foundation (grant 17-72-20045),
the Russian Academy of Science and the
Russian Foundation for Basic Research (grants 08-02-00018, 09-02-00664
and 12-02-91503-CERN),
the Russian Foundation for Basic Research (RFBR) funding within the research project no. 18-02-40086,
the Ministry of Science and Higher Education of the Russian Federation, Project "Fundamental properties of elementary particles and cosmology" No 0723-2020-0041,
the European Union's Horizon 2020 research and innovation programme under grant agreement No. 871072,
the Ministry of Education, Culture, Sports,
Science and Tech\-no\-lo\-gy, Japan, Grant-in-Aid for Sci\-en\-ti\-fic
Research (grants 18071005, 19034011, 19740162, 20740160 and 20039012),
the German Research Foundation DFG (grants GA\,1480\slash8-1 and project 426579465),
the Bulgarian Ministry of Education and Science within the National
Roadmap for Research Infrastructures 2020--2027, contract No. D01-374/18.12.2020,
Ministry of Education
and Science of the Republic of Serbia (grant OI171002), Swiss
Nationalfonds Foundation (grant 200020\-117913/1), ETH Research Grant
TH-01\,07-3 and the Fermi National Accelerator Laboratory (Fermilab), a U.S. Department of Energy, Office of Science, HEP User Facility managed by Fermi Research Alliance, LLC (FRA), acting under Contract No. DE-AC02-07CH11359 and the IN2P3-CNRS (France).

The data used in this paper were collected before February 2022.

\cleardoublepage
\bibliographystyle{na61Utphys}
\bibliography{na61References}

\cleardoublepage
\appendix
\section{Cross Section}
\label{sec:xsapp}

\begin{figure}[h!]
\centering
\includegraphics[width=\linewidth]{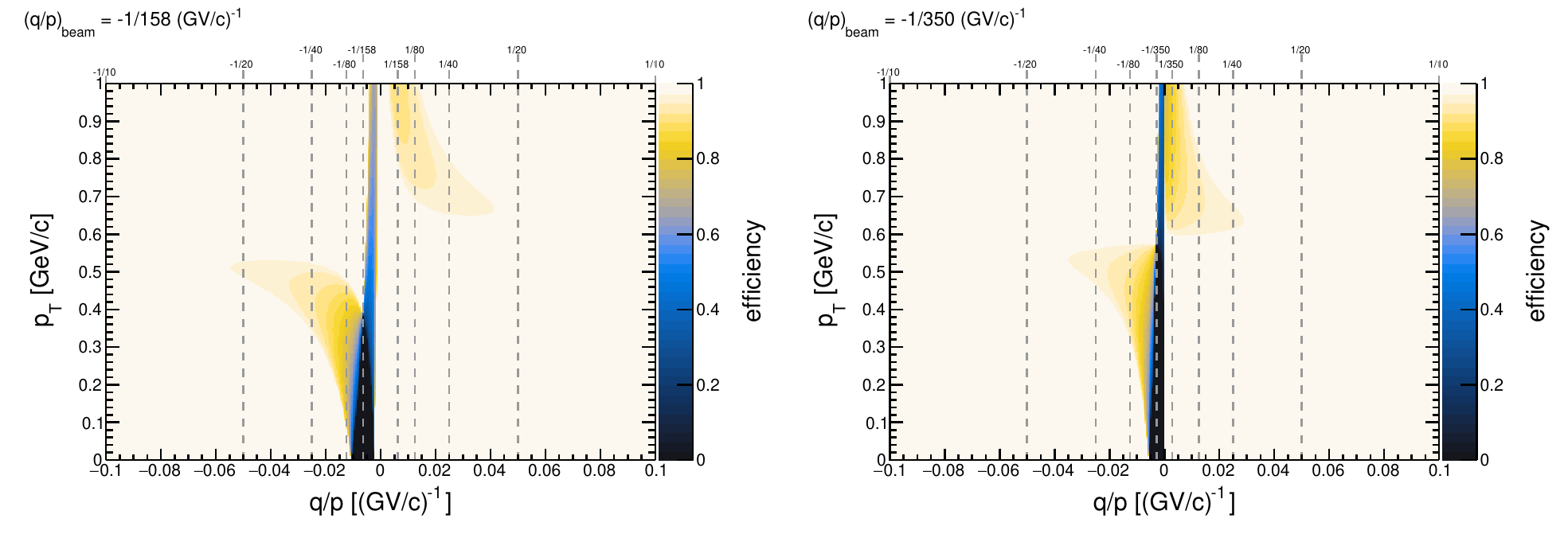}
\caption{Efficiency of the interaction trigger for particles with a certain curvature $q/p$ and transverse momentum $\pT$ for the two beam energies 158 (left, trigger radius 0.9\,cm) and 350\,\GeVc (right, trigger radius 0.6\,cm).}
\label{fig:interactionTrigger}
\end{figure}

\begin{figure}[hb!]
  \def\figw{0.33}
  \centering
\includegraphics[clip,rviewport=0 0 1.02 1,width=\figw\linewidth]{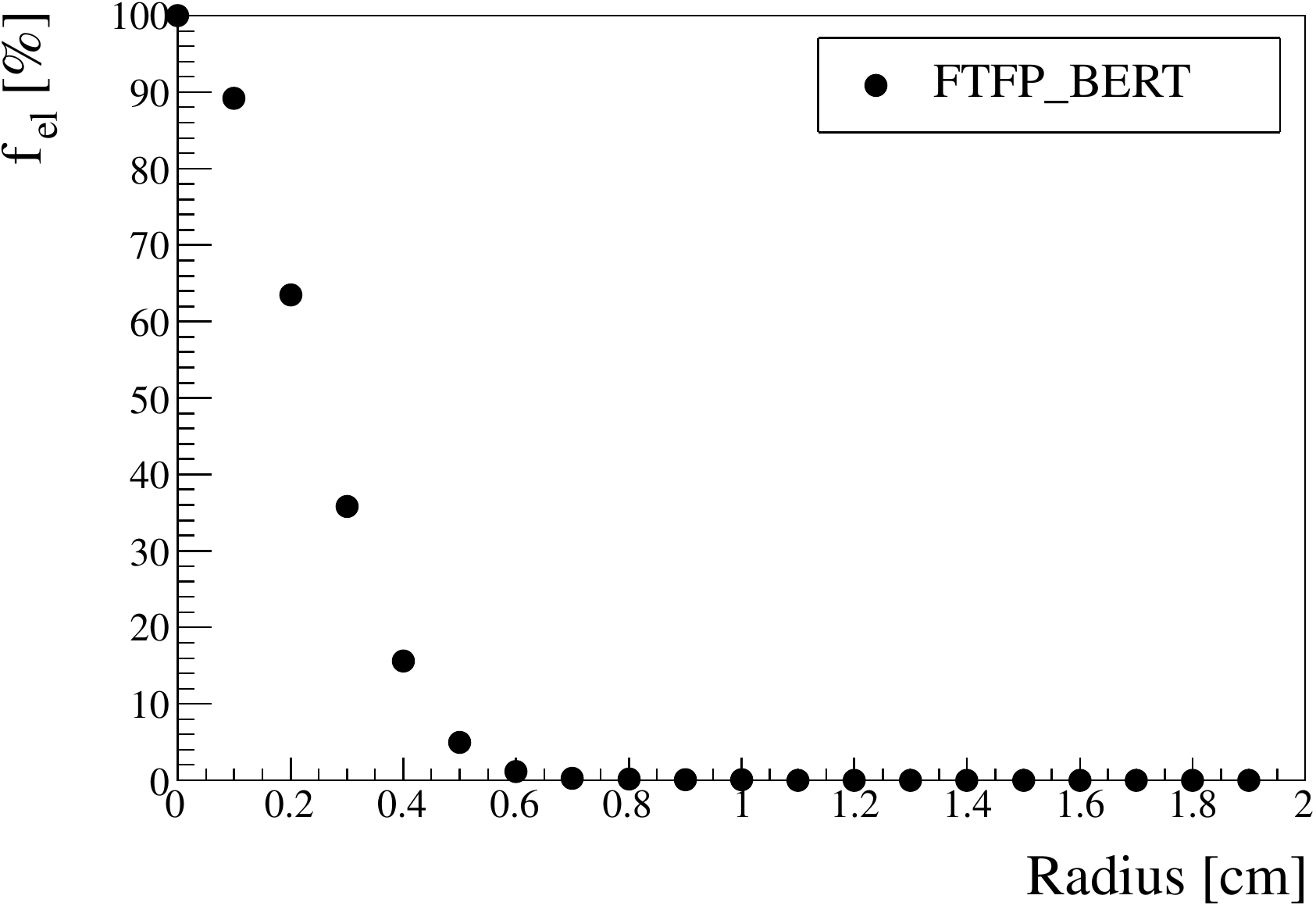}\includegraphics[clip,rviewport=0 0 1.02 1,width=\figw\linewidth]{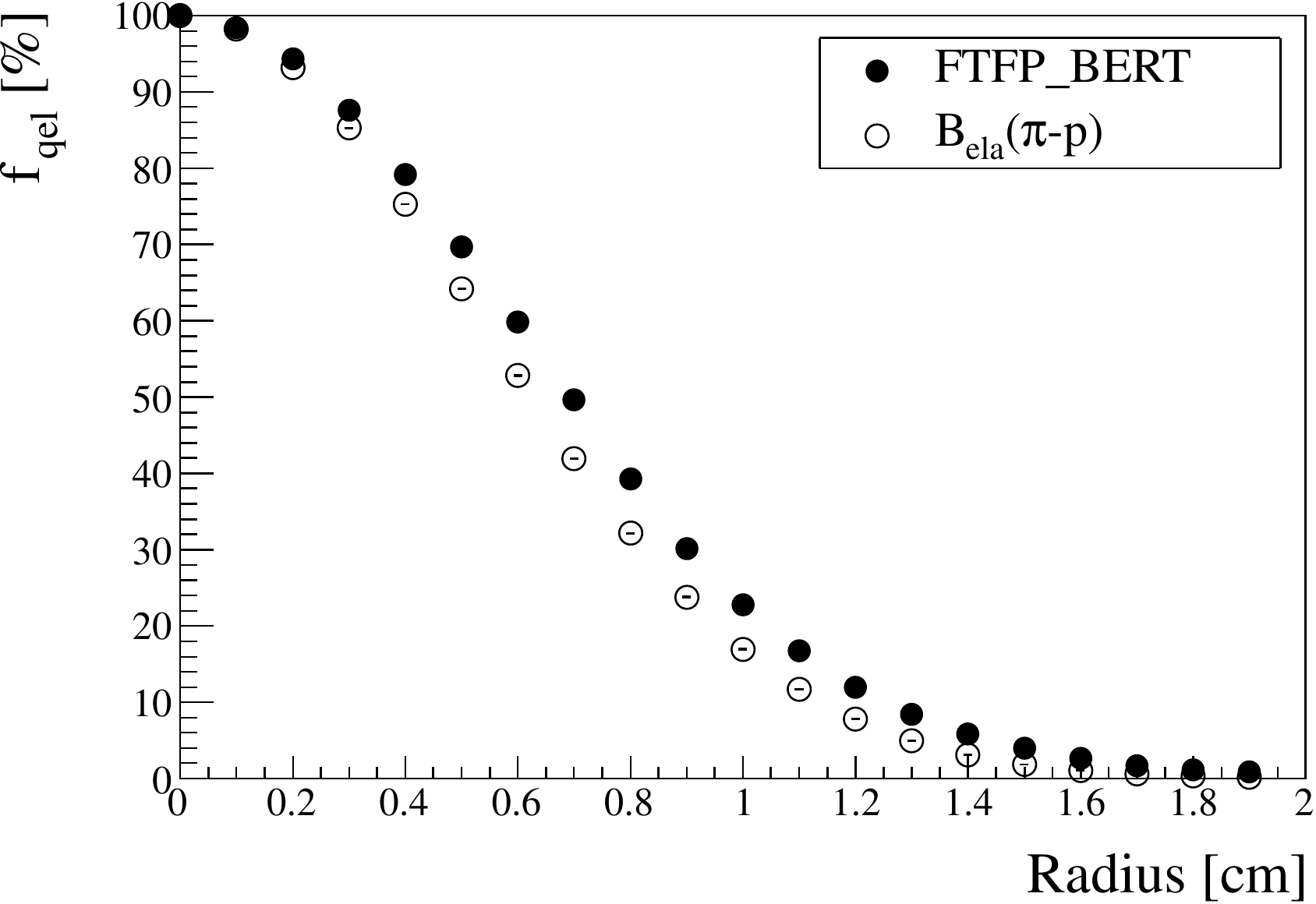}\includegraphics[clip,rviewport=0 0 1.02 1,width=\figw\linewidth]{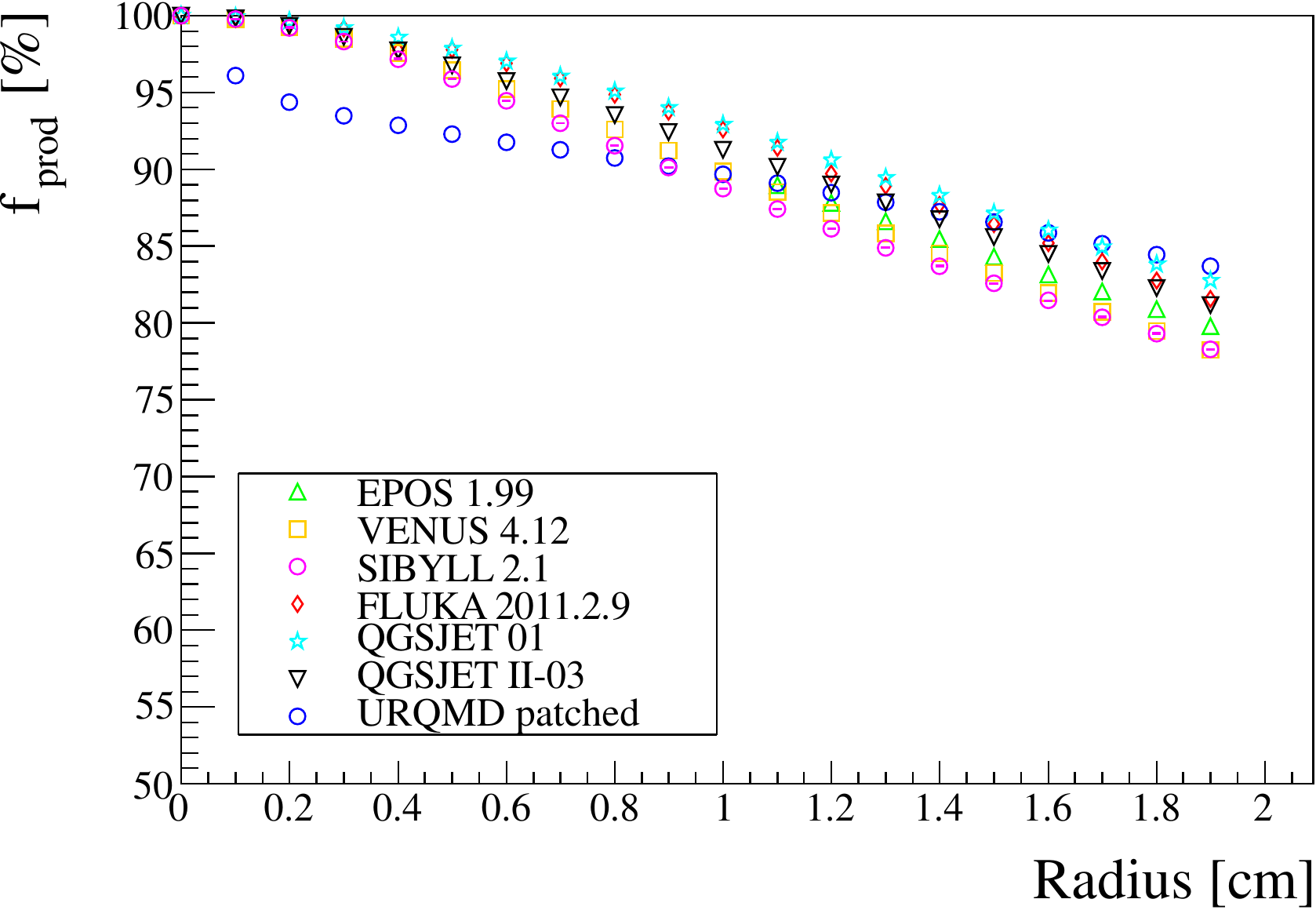}\\
\includegraphics[clip,rviewport=0 0 1.02 1,width=\figw\linewidth]{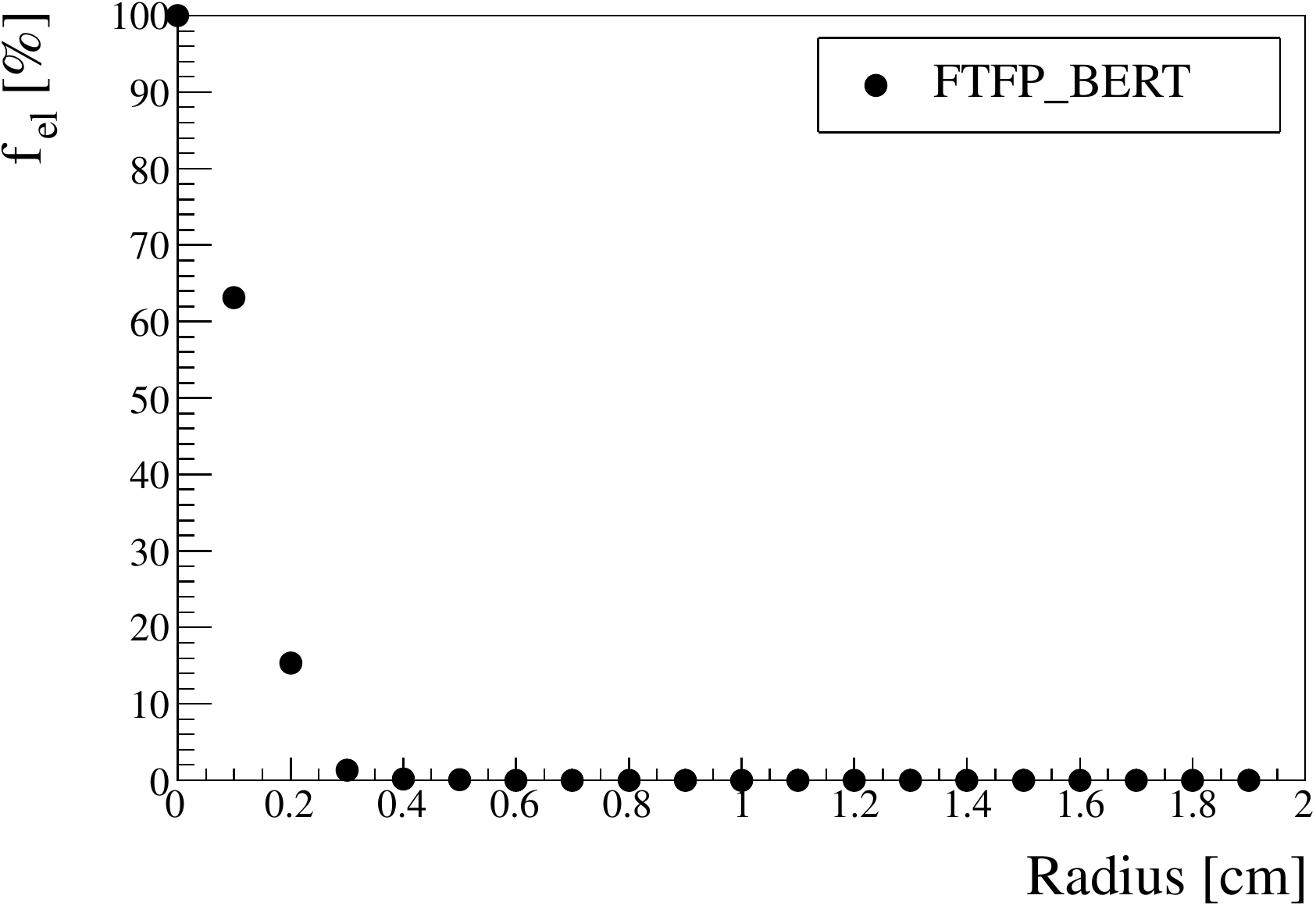}\includegraphics[clip,rviewport=0 0 1.02 1,width=\figw\linewidth]{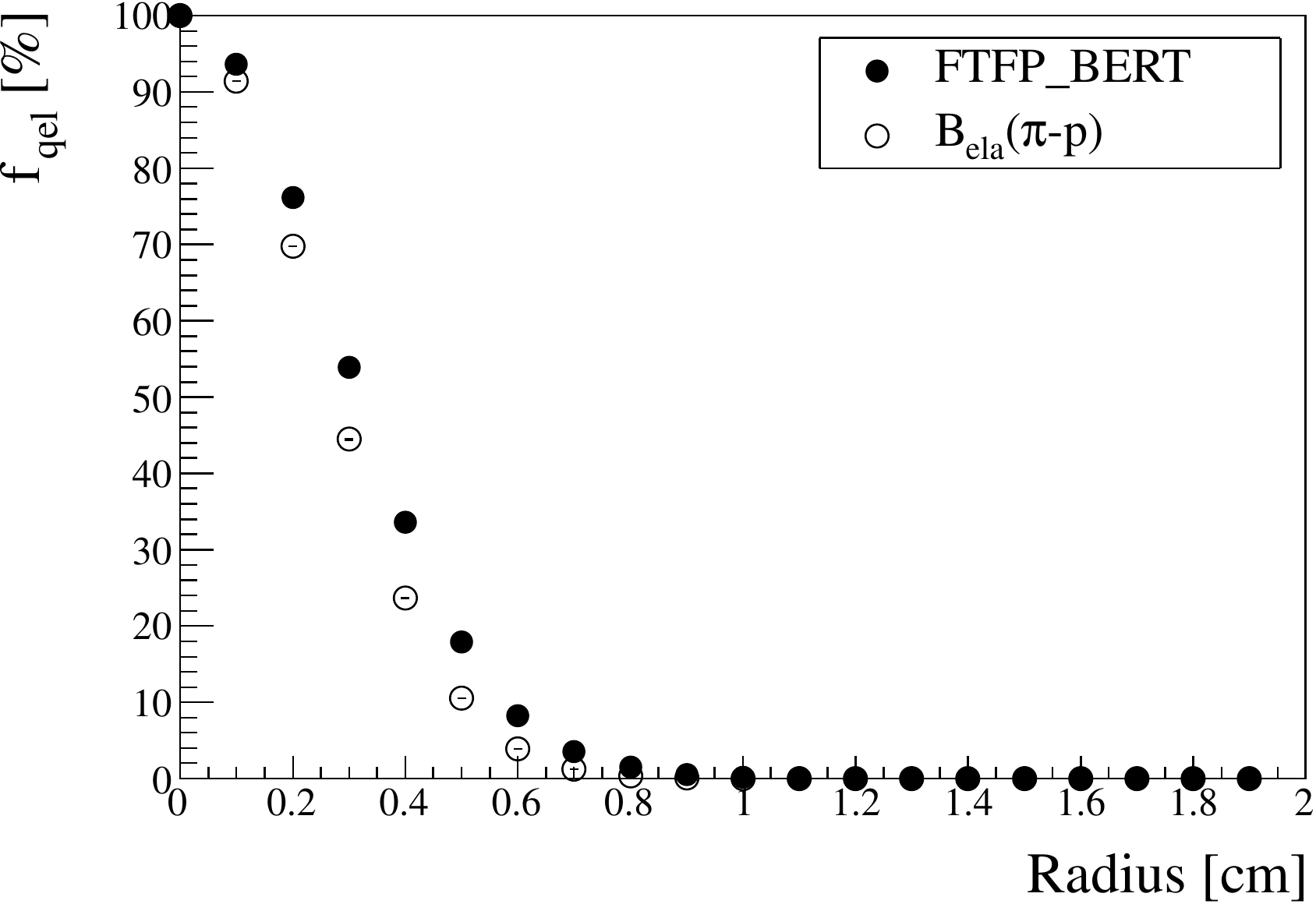}\includegraphics[clip,rviewport=0 0 1.02 1,width=\figw\linewidth]{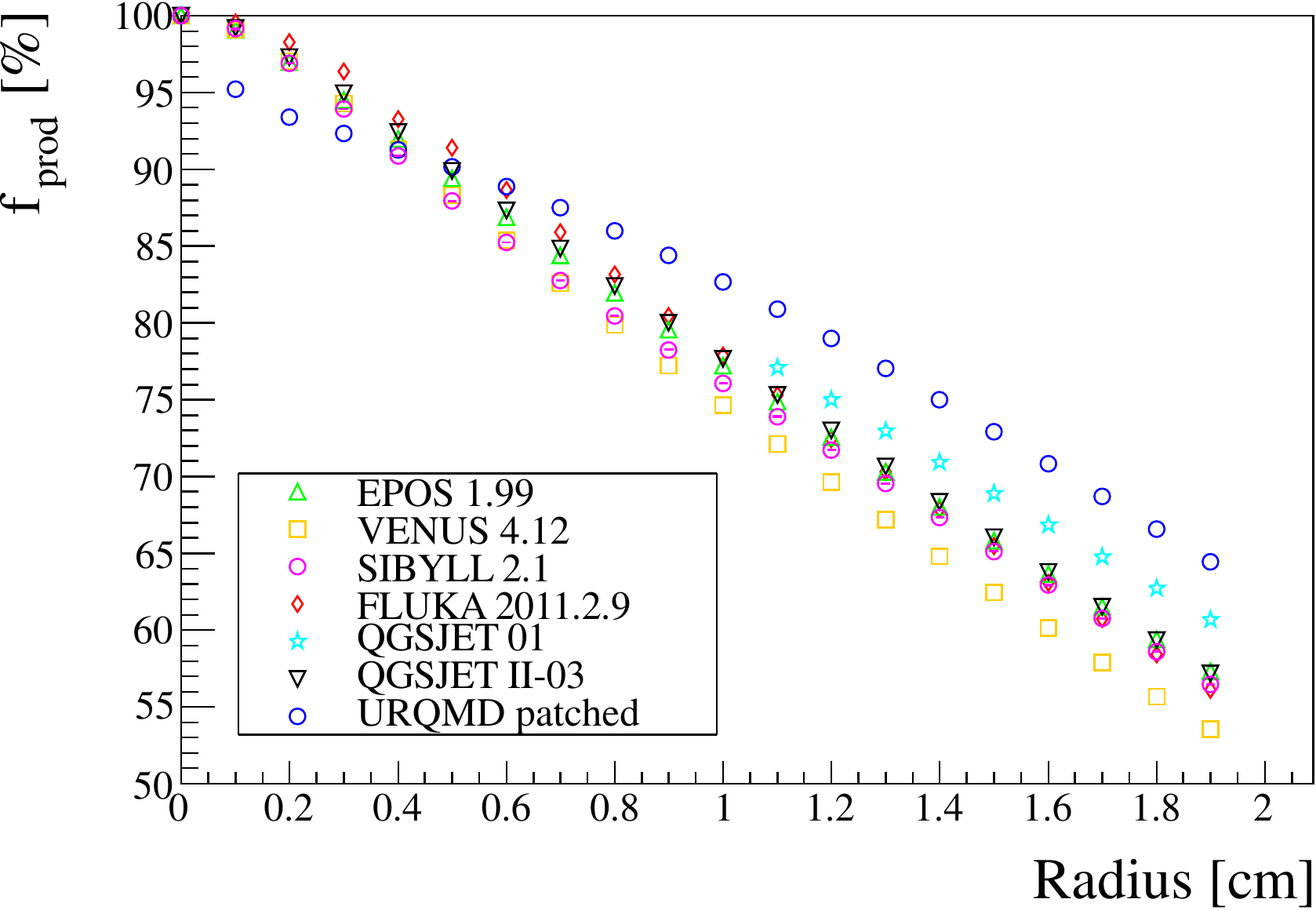}
\caption{Fraction of events leading to a forward particle within a given radial trigger distance from the beam. Columns show from left to right simulations
of elastic, quasi-elastic and production interactions. The top row shows the fractions for beam energies of 158~\GeVc and the bottom row for 350~\GeVc.}
\label{pic:fractions}
\end{figure}

The interaction trigger is set by an offline requirement on the
absence of tracks within a radius $r_\text{trig}$ with respect to the
beam particle 3.7\,m downstream of the target, i.e.\ after passing
through the magnetic field of the first superconducting dipole magnet.
To ease a possible re-analysis of the measured trigger cross section
with different choice of model corrections, we show a visual
representation of the efficiency of this requirement as a function of
curvature $q/p$ and transverse momentum $\pT$ in
\cref{fig:interactionTrigger}.  These efficiency maps are available
electronically at~\cite{dataDownload}.  The {\itshape cumulative}
efficiencies for our choice of models is shown in \cref{pic:fractions}
as a function of $r_\text{trig}$.  These efficiencies can be thought
of as the result of folding the $(q/p,\pT)$-distribution of a specific
process with the two-dimensional efficiency map.

\section{Binning}
\label{sec:binning}

\begin{figure}[!ht]
\centering
\includegraphics[clip,rviewport=0 0 1 1,width=0.5\textwidth]{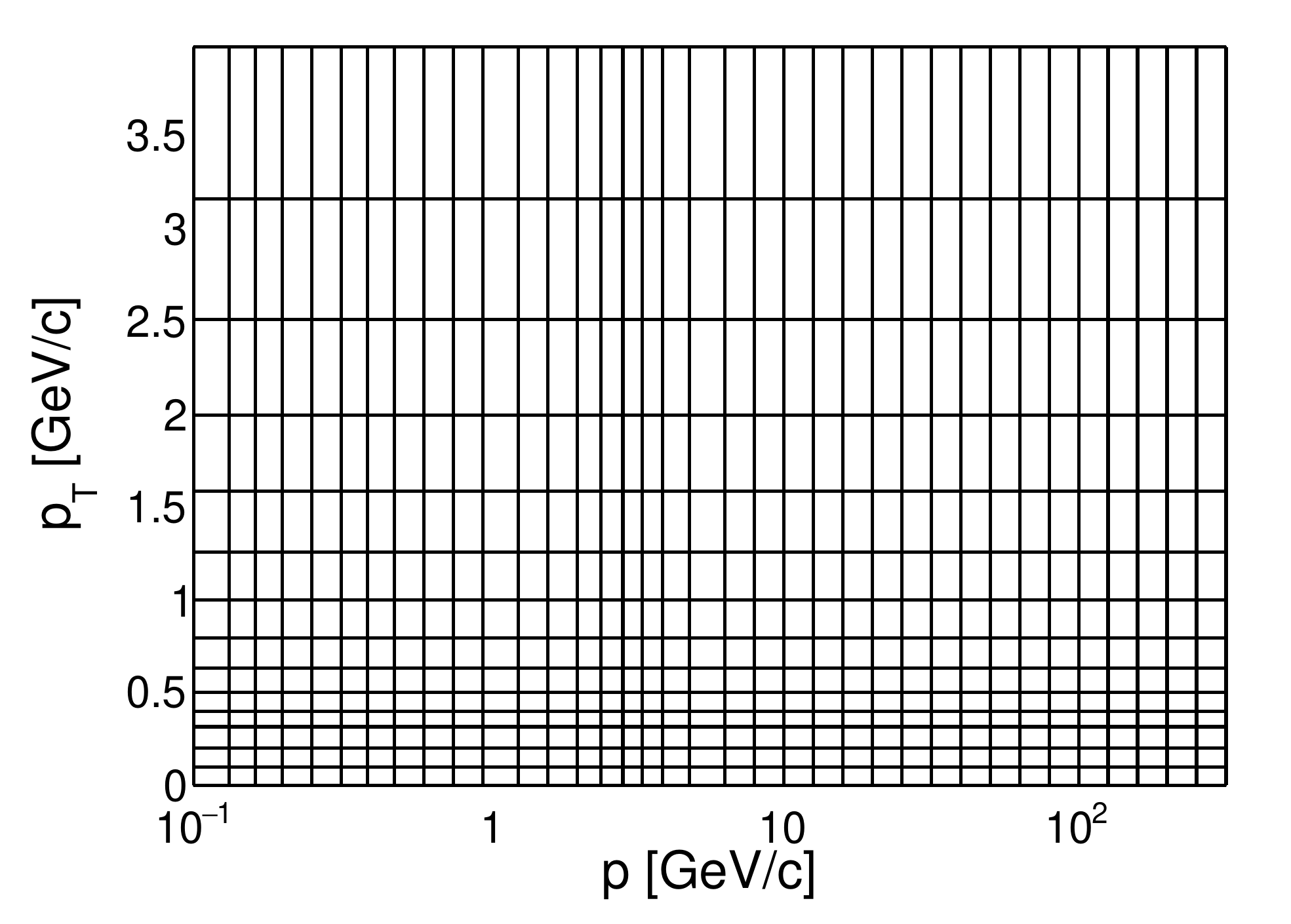}
\caption{Illustration of the phase-space binning used for the charged hadron analysis.}
\label{fig:hadron:binning:dedx}
\end{figure}

The data analysis is performed by splitting the data into 2-dimensional phase-space bins of the \pp and \pT variables.
For the charged hadron analysis a unique phase-space binning was defined.
The \pp intervals are nearly uniform in \logp.
Only small adjustments were done to move the crossing points of the energy deposit function of different particles closer to the center of the bins.
Since some of these bins in the crossing regions will be removed from the analysis, this strategy has been effective to reduce the number of removed bins.
The average width of the \logp intervals is $\Delta\lg\left(p/(\GeVc)\right)=0.1$.
Concerning the \pT intervals, the bin width increases with \pT from $\Delta \pT=0.1$ to $0.5$~\GeVc.
In~\cref{fig:hadron:binning:dedx} we show the phase-space binning used for the charged hadron analysis.

\begin{figure}[!ht]
\centering
\includegraphics[clip,rviewport=0 0 1 1,width=0.9\textwidth]{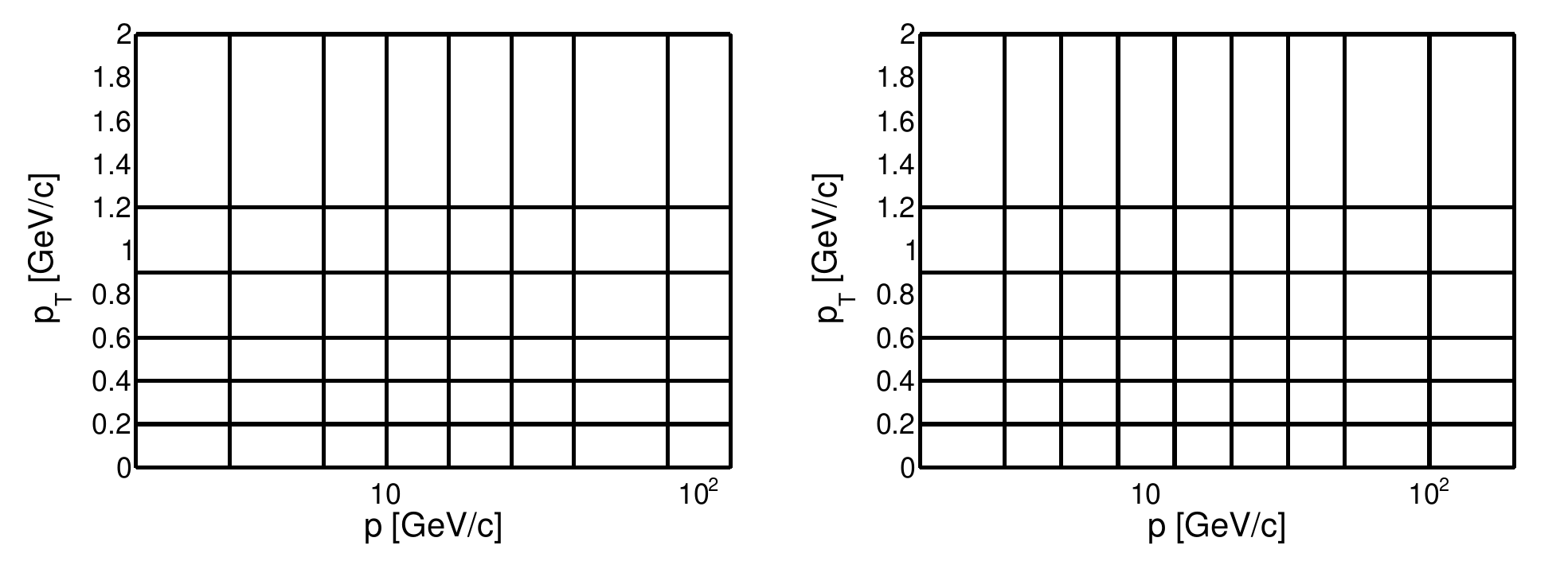}
\caption{Illustration of the phase-space binning used for the \vzero analysis.
The plot on the left show the binning used for \lamb and \antilamb, and the plot on the right shows the binning used for \kzeros.}
\label{fig:hadron:binning:vzero}
\end{figure}

Because the \vzero analysis is done independently for the \lamb, \antilamb, and \kzeros, the phase-space binning is not required to be unique.
However, because the statistics is similar for \lamb and \antilamb, the same binning was defined for these two particles.
For either \lambs or \kzeros the \pp intervals vary from $\Delta\lg\left(p/(\GeVc)\right)=0.2$ to $0.3$.
Concerning the \pT intervals, the widths vary from $\Delta\pT = 0.2$ to $0.8$.
In~\cref{fig:hadron:binning:vzero} we show the binning used for the \vzero analysis.

\section{Additional Plots for the 350~\GeVc Data Set}
\label{sec:350plots}

For a more concise flow of the main part of this article, some of the
plots are only shown for the 158\,\GeVc data set.  In this appendix,
the counterparts for the 350\,\GeVc data are presented.

\betafig{350}
\sysfig{350}
\dedxfig{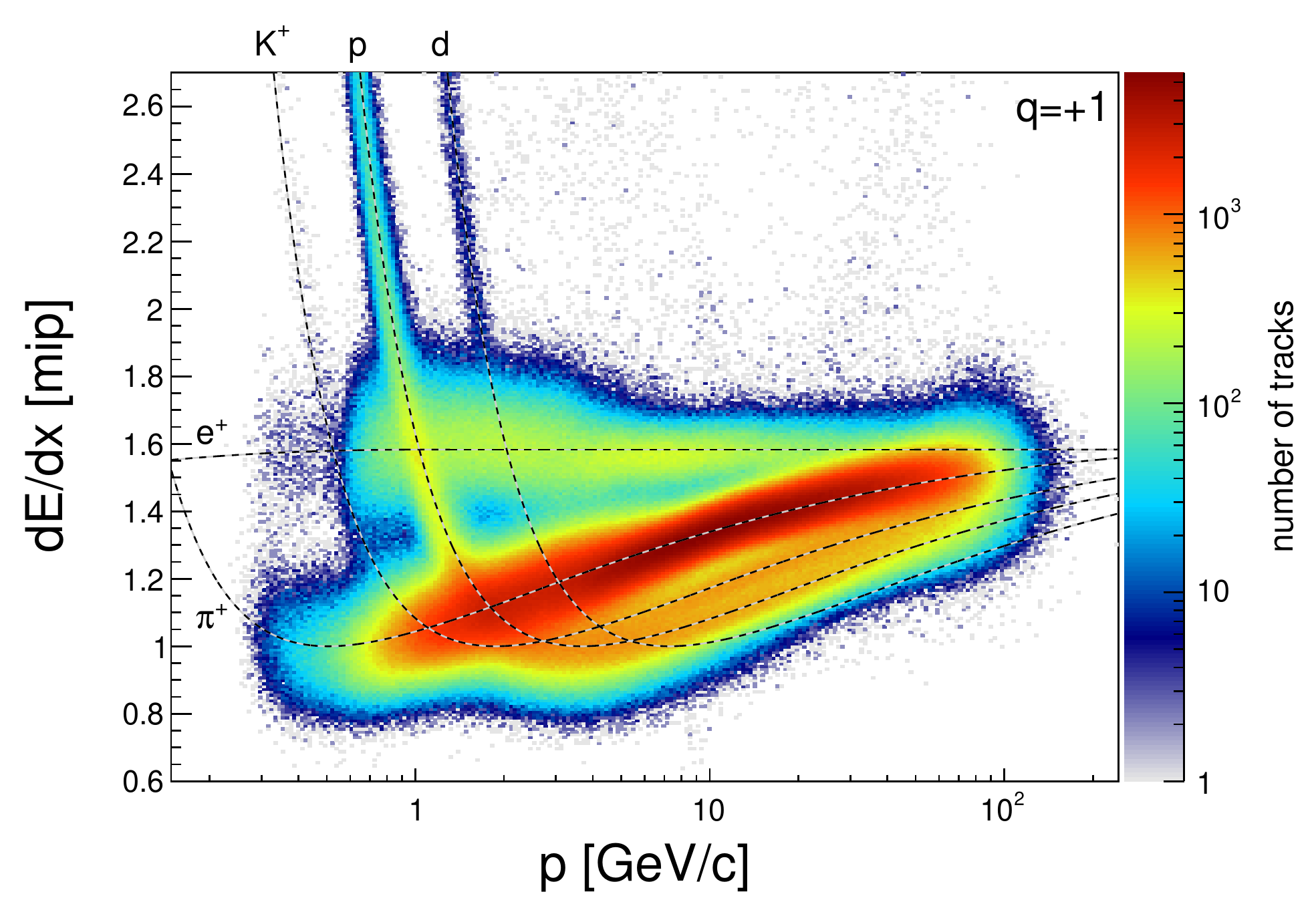}{h}

\cleardoublepage
\textbf{\Large The \NASixtyOne Collaboration}

\begin{wrapfigure}[6]{l}{0.13\linewidth}
\vspace{-2mm}
\includegraphics[width=0.98\linewidth]{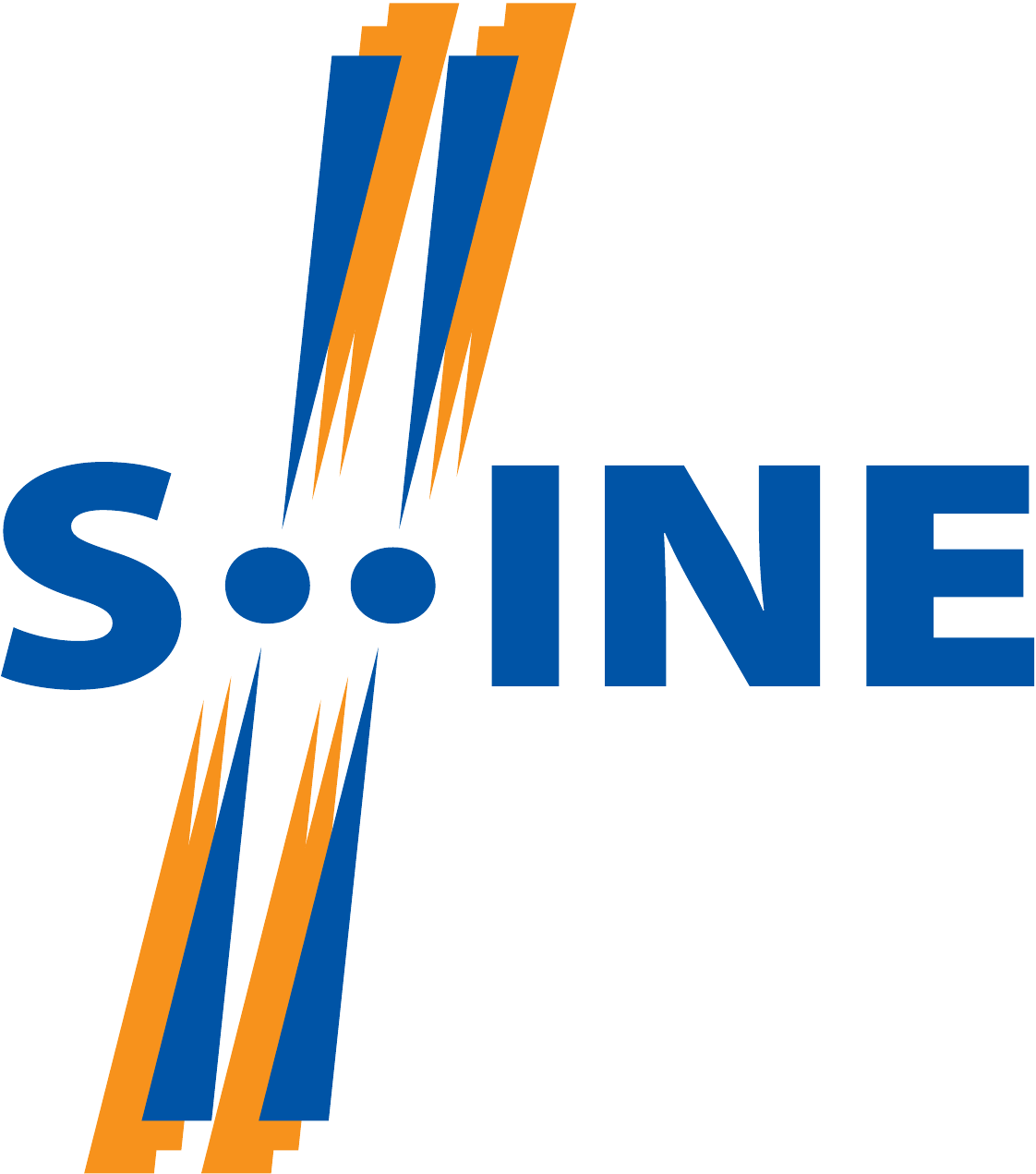}
\end{wrapfigure}
\begin{sloppypar}

\noindent
H.~Adhikary$^{\,13}$,
K.K.~Allison$^{\,30}$,
N.~Amin$^{\,5}$,
E.V.~Andronov$^{\,25}$,
T.~Anti\'ci\'c$^{\,3}$,
I.-C.~Arsene$^{\,12}$,
Y.~Balkova$^{\,18}$,
M.~Baszczyk$^{\,17}$,
D.~Battaglia$^{\,29}$,
S.~Bhosale$^{\,14}$
% ...

\end{sloppypar}
\begin{multicols}{2}
\small

\noindent
% ...
$^{\phantom{2}3}$~Ru{\dj}er Bo\v{s}kovi\'c Institute, Zagreb, Croatia\\
% ...
$^{\phantom{2}5}$~Karlsruhe Institute of Technology, Karlsruhe, \\ \phantom{$^{25}$~}Germany\\
% ...
$^{12}$~University of Oslo, Oslo, Norway\\
$^{13}$~Jan Kochanowski University in Kielce, Poland\\
$^{14}$~Institute of Nuclear Physics, Polish Academy of\\
\phantom{$^{14}$~}Sciences, Cracow, Poland\\
% ...
$^{17}$~AGH - University of Science and Technology,\\
\phantom{$^{17}$~}Cracow, Poland\\
$^{18}$~University of Silesia, Katowice, Poland\\
% ...
$^{25}$~St. Petersburg State University, St. Petersburg, Russia\\
% ...
$^{29}$~University of Notre Dame, Notre Dame , USA\\
$^{30}$~University of Colorado, Boulder, USA\\
% ...

\end{multicols}
\end{document}